\documentclass[printer]{aa}
\usepackage{amssymb}
\usepackage{amsmath}
\usepackage{natbib}
\bibpunct{(}{)}{;}{a}{}{,} 
\usepackage{graphicx}
\usepackage{txfonts}
\usepackage{color}
\usepackage{epstopdf}
\usepackage{wasysym}
\usepackage{booktabs}
\usepackage{hyperref}

\begin{document}

\title{Radio spectral index distribution of SDSS-FIRST sources across optical diagnostic diagrams\thanks{Table~\ref{tab_three_freq1} is only available in electronic form
at the CDS via anonymous ftp to cdsarc.u-strasbg.fr (130.79.128.5)
or via \href{http://cdsweb.u-strasbg.fr/cgi-bin/qcat?J/A+A/}{http://cdsweb.u-strasbg.fr/cgi-bin/qcat?J/A+A/}.}}
\author{Michal Zaja\v{c}ek\inst{1,2,3}, Gerold Busch\inst{1}, M\'onica Valencia-S.\inst{1}, Andreas Eckart\inst{1,2}, Silke Britzen\inst{2},  Lars Fuhrmann\inst{2}, Jana Schneeloch\inst{1}, Nastaran Fazeli\inst{1}, Kevin C. Harrington\inst{2,4}, \& J. Anton Zensus\inst{2}}
\institute{I. Physikalisches Institut der Universit\"at zu K\"oln, Z\"ulpicher Strasse 77, D-50937 K\"oln, Germany \and Max-Planck-Institut f\"ur Radioastronomie (MPIfR), Auf dem H\"ugel 69, D-53121 Bonn, Germany \and Center for Theoretical Physics, Polish Academy of Sciences, Al. Lotnikow 32/46, 02-668 Warsaw, Poland \and Argelander Institut für Astronomie, Auf dem Hügel 71, D-53121 Bonn, Germany}

\authorrunning{M.~Zaja\v{c}ek et al.}
\titlerunning{Radio spectral index distribution}
\date{Received .. .............. 2018; accepted .. .......... 2018}
\abstract{The empirical relations between supermassive black holes and their host spheroids point towards the crucial role of galactic nuclei in affecting the properties of their hosts. A detailed understanding of how the activity of a galactic nucleus regulates the growth of its host is still missing.}{To understand the activity and the types of accretion of supermassive black holes in different hosts, it is essential to study the radio-optical properties of a large sample of extragalactic sources. In particular, we aim to study the radio spectral index trends across the optical emission line diagnostic diagrams to search for potential (anti)correlations.}{To this goal, we combined flux densities from the radio FIRST survey at $1.4\,{\rm GHz}$ (with the flux density range $10\,{\rm mJy} \leq F_{1.4} \leq 100\,{\rm mJy}$) for 396 SDSS sources at intermediate redshift $(0.04\leq z \leq 0.4)$ with the Effelsberg radiotelescope measurements at $4.85\,{\rm GHz}$ and $10.45\,{\rm GHz}$. The information about the optical emission-line ratios is obtained from the SDSS-DR7 catalogue.}{Using the Effelsberg data, we were able to infer the two-point radio spectral index distributions for star-forming galaxies, composite galaxies (with a combined contribution to the line emission from the star formation and AGN activity), Seyferts, and low ionization narrow emission region (LINER) galaxies.}{While studying the distribution of steep, flat, and inverted sources across optical diagnostic diagrams, we found three distinct classes of radio emitters for our sample: (i) sources with a steep radio index, high ionization ratio, and high radio loudness, (ii) sources with a flat radio index, lower ionization ratio, and intermediate radio loudness, (iii) sources with an inverted radio index, low ionization ratio, and low radio loudness. The classes (i), (ii), and (iii) cluster mainly along the transition from Seyfert to LINER sources in the optical diagnostic (Baldwin, Phillips \& Telervich; BPT) diagram. We interpret these groups as a result of the recurrent nuclear-jet activity.}
\keywords{radio continuum: galaxies -- methods:observational -- techniques: spectroscopic -- Galaxies: active}
\maketitle

\section{Introduction}

In previous decades, several correlations between the mass of supermassive black holes (SMBH) and the properties of their host spheroids (spheroidal mass, luminosity, stellar velocity dispersion) point towards the co-evolution between SMBHs and host galaxies \citep[see][and references therein]{1998AJ....115.2285M,2000ApJ...539L...9F,2009ApJ...698..198G,2013ARA&A..51..511K,2016arXiv161107872B}. There are several observational studies that point towards the relationship between black hole activity and star-formation rate in the hosts of active galactic nuclei \citep[AGN, see e.g.][]{2006NewAR..50..677H}.

In addition, the observed bimodality in the colours of local galaxies \citep{2001AJ....122.1861S,2003MNRAS.346.1055K,2004ApJ...600..681B,2004ApJ...615L.101B} points towards the correlation between the star formation activity and the morphological type of the galaxy. The more massive elliptical and lenticular galaxies are preferentially located in the massive red cloud, while less massive spiral and irregular galaxies are located in the blue cloud. The intermediate region, the so-called ``green valley'', is far less occupied \citep{2007ApJ...665..265F} and the ``green'' sources belong to mixed types, that is, red spirals. This implies that the AGN are involved in the process of star formation quenching, which is commonly referred to as AGN feedback in galaxy evolution theories and simulations.

\begin{figure}[tbh]
   \centering
   \includegraphics[width=0.5\textwidth]{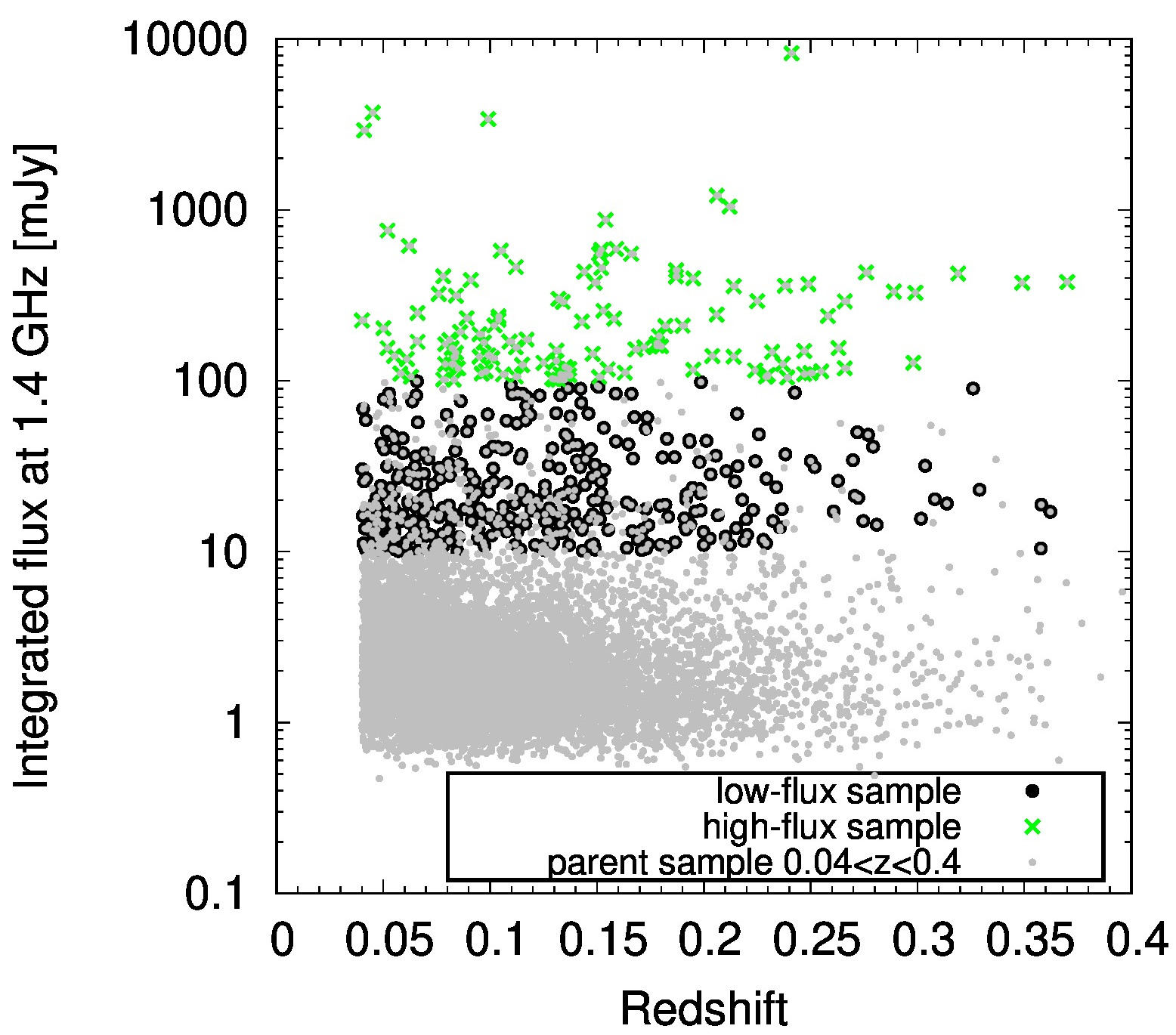}\\
    \includegraphics[width=0.5\textwidth]{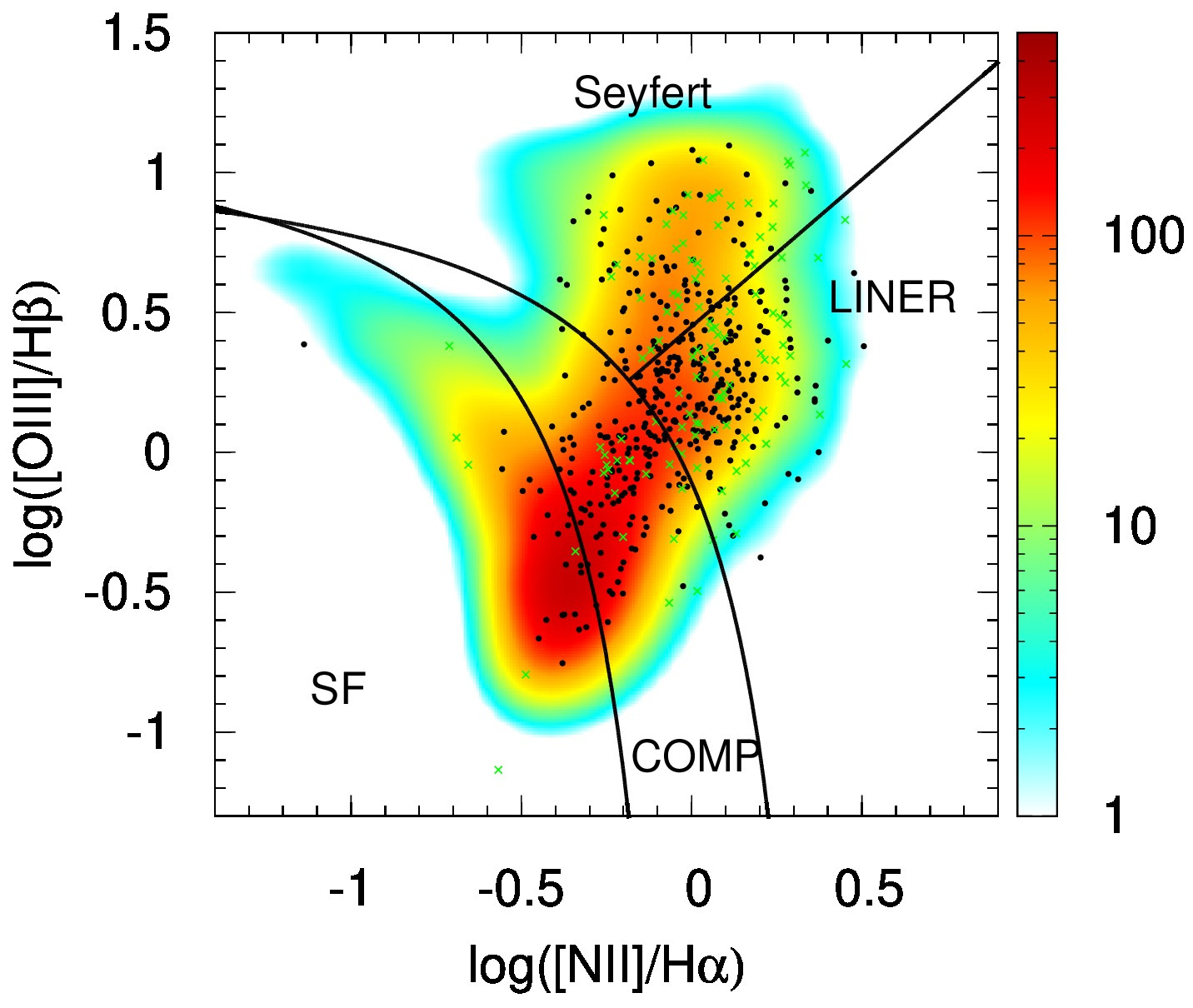}    
   \caption{Distribution of the low-flux and high-flux samples in the redshift-flux plane and in the BPT diagram. \textbf{Top panel:} Two samples of radio galaxies in the redshift-flux density plane: a high-flux sample of 119 sources previously analysed by \citet{2015A&A...573A..93V} and a sample extended towards lower radio flux densities at $1.4\,{\rm GHz}$, $10\,{\rm mJy}\leq F_{1.4} \leq 100\,{\rm mJy}$ (black points). The grey points represent a subset of the parent sample in the redshift range of $0.04\leq z \leq 0.4$. \textbf{Bottom panel:} Selected radio galaxies for the Effelsberg observations in the [NII]-based BPT diagnostic diagram (black points). The sample of radio galaxies with $F_{1.4}\geq 100\,{\rm mJy}$ analysed by \citet{2015A&A...573A..93V} is denoted by green crosses. The parent sample is colour coded depending on the density per bin in the [OIII]-[NII] plane (bin size is set to $0.1$ along both ratios).}
   \label{fig_redshift_flux_surplot_parent}
\end{figure}

A useful tool for distinguishing the galaxies with different prevailing photoionization sources is the Baldwin, Phillips, and Terlevich diagram \citep[BPT,][]{1981PASP...93....5B}, in which the source location is determined by a pair of low-ionization, emission-line intensity ratios. A commonly used pair is the ratio [\ion{N}{ii}]/H$\alpha$ and [\ion{O}{iii}]/H$\beta$ (see Fig.~\ref{fig_redshift_flux_surplot_parent}, right panel), but ratios of other line intensities of a similar wavelength can be used as well. BPT diagrams include forbidden lines, which are for AGN sources associated with the narrow line region (NLR) and hence type 2 AGN. 

The emission-line ratios depend on the strength and shape of
the ionizing radiation field, as well as the physical properties
of the line-emitting gas including gas density, metal abundances,
dust, and
cloud thickness \citep[e.g.][]{1987ApJS...63..295V,1997A&A...323...31K,1997A&A...327..909B,2004ApJS..153....9G,2004ApJS..153...75G,2016MNRAS.458..988R}.
Systematic trends and correlations in
different sections of the diagnostic diagrams have been traced back to
systematic changes in the ionization parameter, the shape of the ionizing
continuum, the fraction of matter-bound clouds, and/or the role of
metal abundances. 

According to a given set of ratios, four spectral classes of galaxies are commonly distinguished. In star-forming galaxies (SF), the ionizing flux is provided mostly by hot, massive, young stars and associated supernovae that are surrounded by HII regions. They have lower [OIII]/$H\beta$ and [NII]/$H\alpha$ ratios than pure AGN sources (see also Figs.~\ref{fig_redshift_flux_surplot_parent} and \ref{fig_effelsberg_sample}). In between SF and AGN sources are composite (COMP) galaxies, with a mixed contribution from star formation (HII regions) and AGN. The AGN spectral class was further subdivided into Seyfert 2 sources (Sy) and low-ionization nuclear emission regions (LINERs). LINERs are characterized by a lower [OIII]/$H\beta$ ratio in comparison with Seyfert 2 AGN sources and a higher [NII]/$H\alpha$ ratio with respect to star-forming sources (both SF and COMP). 

Systematic changes
of line ratios across different regimes, from the HII
regime across the composite region and into the AGN regime are seen in
spatially-resolved spectroscopy of nearby Seyfert galaxies \citep[e.g.][]{2006A&A...459...55B,2006A&A...456..953B,2011AJ....142...43S},
while \citet{1994A&A...291..713S} speculated about an
ionization sequence from the AGN into the LINER regime, based on systematic
continuum-dilution processes.

Radio galaxies are generally considered as active galaxies that are luminous in radio bands. \citet{1980A&A....88L..12S} argued that the radio loudness of quasars, which is typically defined as the ratio of the radio to the optical flux densities, shows a certain degree of bimodality. \citet{1989AJ.....98.1195K} confirmed the bimodality in the radio loudness in the sense that radio-quiet quasars are five to ten times more frequent than radio-loud quasars, that is, they have radio emission comparable to the optical emission. A similar result was also obtained by \citet{2016ApJ...831..168K}. Recent studies based on deep radio studies, namely the FIRST (Faint Images of the Radio Sky at Twenty centimeters) and the NVSS (National Radio Astronomy Observatory Very Large Array Sky Survey -- NVSS) \citep{1995ApJ...450..559B,1998AJ....115.1693C}, in combination with sensitive optical surveys SDSS (Sloan Digital Sky Survey) and 2dF (Two Degree Field Survey) \citep{2000AJ....120.1579Y,2001MNRAS.322L..29C} demonstrate the large scatter in the radio loudness. They are, however, generally inconclusive about the bimodal character of the distribution \citep{2000ApJS..126..133W,2002AJ....124.2364I,2003MNRAS.346..447C,2003MNRAS.341..993C,2003ApJ...590...86L}. 

For a sample of galaxies, it was possible to determine the estimates of black hole masses and hence study the dependence of the radio loudness, $R=F_{\nu_{\rm R}}/F_{\nu_{\rm O}}$, on the Eddington ratio, $\eta \equiv L_{\rm Bol}/L_{\rm Edd}$, where $F_{\nu_{\rm R}}$ and $F_{\nu_{\rm O}}$ are monochromatic radio and optical flux densities at frequencies $\nu_{\rm R}$ and $\nu_{\rm O}$, respectively, and $L_{\rm Bol}$ and $L_{\rm Edd}$ are the bolometric luminosity and the Eddington limit of the AGN. The general trend found is that the radio loudness increases with a lower Eddington ratio, that is, there seems to be an anti-correlation between $R$ and $\eta$ with a large scatter \citep{2002ApJ...564..120H,2007ApJ...658..815S}. This trend is so far consistent with the hardness-luminosity diagram for X-ray binaries \citep{2004MNRAS.355.1105F,2017A&A...603A.127S} and leads to the conclusion that accretion at lower Eddington rates leads to smaller cooling rates, and hence to hot, geometrically thick, and optically thin radiatively inefficient accretion flows \citep[RIAFs,][]{1977ApJ...214..840I,1982Natur.295...17R}. RIAFs are often modelled as advection-dominated flows \citep[ADAFs,][]{1995ApJ...444..231N}, which can launch jets more effectively \citep{1982Natur.295...17R} than colder, optically thick, and geometrically thin accretion discs that are associated with larger Eddington ratios due to efficient cooling \citep[e.g.][]{1973A&A....24..337S}. This yields the trend of larger radio loudness for sources with lower Eddington ratios. However, since $R\propto 1/L_{\rm O}$, where $L_{\rm O}$ is the optical luminosity, and $\eta\propto L_{\rm O}$, the anti-correlation is generally expected and one should be careful when interpreting the results. Moreover,  optical emission as a tracer of accretion rate only makes sense for type I AGN, since for type II sources the optical emission is influenced by obscuration and continuum dilution \citep{1994A&A...291..713S}. Therefore, adding more parameters to the study of trends, such as spectral slopes in the radio domain, may shed more light on the radio-optical properties of galaxies and the physical processes involved, namely the formation, acceleration, and collimation of jets \citep{2007ApJ...658..815S} or the effect of mergers on the radio loudness of the AGN \citep{2015ApJ...806..147C}.        

At frequencies $\lesssim 10\,{\rm GHz}$, the radio-continuum spectra are dominated by non-thermal synchrotron emission with the characteristic power-law slope, $S_{\nu} \propto \nu^{\alpha}$, while the thermal bremsstrahlung (free-free) emission is negligible \citep{1988AJ.....96...81D}, contributing less than 10\% at 10 GHz \citep{1982A&A...116..164G}, and becomes more prominent towards mm-wavelengths. Concerning the synchrotron spectral slope for radio galaxies with jets, primary components (cores) are generally self-absorbed with positive slopes of $\alpha\sim 0.4$, while secondary components have negative spectral indices with mean values $\alpha \sim -0.7$ \citep{1986A&A...168...17E}, which is consistent with optically thin synchrotron emission. The integrated radio spectrum of radio jets is typically associated with a flat spectral slope due to the superposition of self-absorbed synchrotron spectra.


The role of radio galaxies in the galaxy evolution at low to intermediate redshifts $(z<0.7)$ is unclear \citep{2016A&ARv..24...10T}. In general, the star formation activity in their hosts is expected to be one order of magnitude smaller than for the peak of quasar and star-formation activity at $z\approx 1.9$ \citep{2014ARA&A..52..415M}. However, the AGN in these radio galaxies are expected to come through phases of intermittent accretion activity \citep{2009ApJ...698..840C}, as well as mergers that influence the overall host properties, and are expected to prevent the host gas reservoirs from cooling and forming stars.

The contribution of different sources to the overall radio luminosity of galaxies remains largely unclear. The value of the radio spectral index $\alpha$ helps to distinguish between the prevalence of optically thin and optically thick emission mechanisms. \citet{2019MNRAS.482.5513L} made use of the high-resolution Very Large Array (VLA) observations at 5 and 8.4 GHz of optically selected radio-quiet (RQ) Palomar-Green (PG) quasars. They determined the corresponding spectral slopes $\alpha_{5/8.4}$ for 25 RQ PG sources and found a significant correlation between the slope value and the Eddington ratio. Specifically, high Eddington-ratio quasars ($L/L_{\rm Edd}>0.3$) have steep spectral slopes, $\alpha_{5/8.4}<-0.5$, while lower Eddington-ratio sources ($L/L_{\rm Edd}<0.3$) have flat to inverted slopes, $\alpha_{5/8.4}>-0.5$. A correlation is also found with an Eigenvector I (EV1) set of properties \citep[\ion{Fe}{ii}/H$\beta$, H$\beta_{\rm asym}$, X-ray slope $\alpha_{\rm X}$; see ][]{1992ApJS...80..109B,2000ApJ...536L...5S,2001ApJ...558..553M}, where the flat to inverted RQ PG sources have low \ion{Fe}{ii}/H$\beta$ and a flat soft X-ray slope. \citet{2019MNRAS.482.5513L} found a dichotomy between radio-quiet PG quasars and 16 radio-loud (RL) PG quasars, which in contrast with RQ sources do not exhibit the correlations with EV1 and the radio slope is instead determined by the black hole mass, which implies a different radiation mechanism for RL sources. These findings provide a motivation for the investigation of further correlations between radio slopes and optical emission-line properties for a larger sample of radio galaxies.

Previously, we performed radio continuum observations of intermediate redshift $(0.04 \leq z \leq 0.4)$ SDSS-FIRST sources at $4.85\,{\rm GHz}$ and $10.45\,{\rm GHz}$ to determine their spectral index and curvature distributions \citep{2015A&A...573A..93V}. This sample included  star-forming, composite, Seyfert, and LINER galaxies that obeyed the flux density cut of $\geq 100\,{\rm mJy}$ at $1.4\,{\rm GHz}$ (see Fig.~\ref{fig_redshift_flux_surplot_parent}, green crosses). \citet{2015A&A...573A..93V} searched for the radio spectral index trends in BPT diagnostic diagrams as well as for the relation between optical and radio properties of the sources. For the limited sample of 119 sources, they found a rather weak trend of spectral index flattening in the [NII]-based diagnostic diagram along the star-forming--composite--AGN Seyfert branch. 

The radio spectral index flattening trend triggered the motivation to study more sources with lower radio continuum fluxes at $1.4\,{\rm GHz}$. The sample was extended by 381 additional sources towards lower radio flux densities at $1.4\,{\rm GHz}$, with integrated flux densities $10\,{\rm mJy} \leq F_{\rm 1.4}\leq 100\,{\rm mJy}$. Using the cross-scan observations conducted by the 100-m Effelsberg radio telescope, for point-like sources we determined flux densities at two frequencies, $4.85\,{\rm GHz}$ and $10.45\,{\rm GHz}$, which enabled us to determine the spectral indices $\alpha_{[1.4-4.85]}$ (for 298 sources) and $\alpha_{[4.85-10.45]}$ (for 90 sources).

In this paper, we present the findings based on the radio-optical properties of the low-flux sample along with radio-brighter sources previously reported in \citet{2015A&A...573A..93V}. We searched for any trends of the radio spectral slope in the low-ionization diagnostic diagrams. In other words, we were interested in whether the changes in emission-line ratios across the BPT diagram are systematically reflected in the radio spectral index. Furthermore, the relation between the spectral index and the radio loudness of the sources was unknown. 



The paper is structured as follows. In Sect.~\ref{radio_optical_samples} we introduce the optical (SDSS survey) and radio (FIRST) samples used in our study. Subsequently, in Sect.~\ref{effelsberg_sample_observation} we present the selection of radio sources for follow-up observations with the Effelsberg 100-m telescope at two higher frequencies. The basic statistical properties of spectral index distributions are presented in Sect.~\ref{analysis_results} in combination with the optical properties of the sources. Subsequently, we describe the fundamental trends of the radio spectral slope in optical diagnostic diagrams as well as with respect to the radio loudness in Sect.~\ref{sec_trends}.  We continue with the interpretation of the results in Sect.~\ref{interpretation}. Finally, we summarize the main results in Sect.~\ref{conclusions}. Additional materials, specifically radio flux densities for new low-flux sources, are included in Appendix~\ref{appa}.

\section{Radio and optical samples}
\label{radio_optical_samples}

\subsection{Sloan Digital Sky Survey}

The Sloan Digital Sky Survey (SDSS) is a photometric and spectroscopic survey of celestial sources that covers one quarter of the north Galactic hemisphere \citep{2000AJ....120.1579Y,2002AJ....123..485S}. The spectra and magnitudes have been obtained by a $2.5$-m wide field-of-view (FOV) telescope at Apache Point in New Mexico, USA. The spectra have an instrumental resolution of $\sim 65\,{\rm km\,s^{-1}}$ in the wavelength range of $380$--$920\,{\rm nm}$. The identified galaxies have a median redshift of $0.1$. The spectra were obtained by fibers with 3'' diameter (the linear scale of $5.7\,{\rm kpc}$ at $z=0.1$), which makes the sample sensitive to aperture effects, that is, low-redshift galaxies are dominated by nuclear emission \citep[see e.g.][]{2015A&A...580A.113T}. 

The seventh data release of SDSS \citep[SDSS DR7, ][]{2009ApJS..182..543A} contains parameters of $\sim 10^6$ galaxies inferred from the spectral properties based on the Max Planck Institute for Astrophysics (MPIA) and Johns Hopkins University (JHU) emission-line analysis. The stellar synthesis continuum spectra \citep{2003MNRAS.344.1000B} were applied for the continuum subtraction, after which emission line characteristics were derived. In particular, SDSS DR7 contains the emission-line characteristics of low-ionization lines that are used to distinguish star-forming galaxies from AGN (Seyfert galaxies and LINERs or high-excitation and low-excitation systems, respectively) in the diagnostic diagrams. DR7 also contains the source images as well as stellar masses inferred from the broad-band fitting of spectral energy distributions by stellar population models. 

\subsection{Faint Images of the Radio Sky at Twenty-centimeters Survey}

The Faint Images of the Radio Sky at Twenty-centimeters Survey \citep[FIRST][]{1995ApJ...450..559B} was performed by the Very-Large-Array (VLA) in its B-configuration at $1.4\,{\rm GHz}$. The FIRST survey covers $\sim 10\,000\,{\rm deg^2}$ in the north Galactic cap, partially overlapping the region mapped by SDSS. The sky brightness was measured with a beam-size of $5.4''$ and an rms sensitivity of $\sim 0.15\,{\rm mJy/beam}$. At the sensitivity level of $\sim 1 {\rm mJy}$, the FIRST survey contains $\sim 10^6$ sources, of which about a third are resolved with structures on the angular scale of $2''$--$30''$ \citep{2002AJ....124.2364I}. The survey contains both the peak and the integrated flux densities, which allows us to distinguish resolved and unresolved sources. The flux density measurements have uncertainties smaller than $8\%$. The images for each source are provided on the website.

\subsection{SDSS-FIRST cross-identification}

As described in \citet{2015A&A...573A..93V}, we performed a cross-identification of SDSS DR7 and FIRST source catalogues using SDSS DR7 CasJobs, with a matching radius set to 1'' \citep{2005cs........2072O}. This results in a total of $37\,488$ radio-optical emitters as a basis for further studies. This initial sample constitutes $\sim 4\%$ of SDSS sources and contains mostly active, metal-rich galaxies \citep[see also][for details]{2012A&A...546A..17V}.   

As was already done in \citet{2015A&A...573A..93V} for a high-flux sample, we apply the following selection criteria:
\begin{itemize}
\item the redshift limits, $0.04 \leq z \leq 0.4$,
\item the signal-to-noise lower limit of $S/N>3$ on the equivalent width $EW$ of the emission lines used in the low-ionization optical diagnostic diagrams.
\end{itemize}
The lower redshift limit of $0.04$ is due to the fact that  nearby sources have angular sizes larger than the optical fiber used for the SDSS survey. Hence they are dominated by their nuclear emission \citep[see e.g.][]{2003AAS...20311901K}. The upper redshift limit is imposed to make sure that the emission-line diagnostics concerning [\ion{N}{ii}] and H$\alpha$ lines is reliable, meaning that they fall into the observable spectral window. By imposing the redshift constraints, we obtain our parent sample with 9951 sources, which are shown as grey points in the redshift-flux plot (see Fig.~\ref{fig_redshift_flux_surplot_parent}). There is no evident dependency of the flux density on the redshift, apart from the expected tendency of having more radio galaxies towards lower radio flux densities. 

Out of nearly $10\,000$ radio sources, only $\sim 1\%$ of the sources has the integrated flux density at $1.4\,{\rm GHz}$ above $100\,{\rm mJy}$. The combined radio-optical properties of these brightest sources were investigated in \citet{2015A&A...573A..93V}. By decreasing the lower boundary of the flux density cut by one order of magnitude to $10\,{\rm mJy}$, the number of sources increases to $5.6\%$, that is, by a factor of five. 

Most of the sources of the parent sample, $93.5\%$, have flux densities at $1.4\,{\rm GHz}$ below $10\,{\rm mJy}$. Under the assumption that many of these sources have steep to flat spectra, $S_{\nu} \propto \nu^\alpha$, where $\alpha \leq 0$, the detection of their flux density at frequencies larger than $1.4\,{\rm GHz}$ would be beyond the detection limit of the Effelsberg telescope, which is $\sim 5\,{\rm mJy}$.

In Fig.~\ref{fig_redshift_flux_surplot_parent}, the subset of the parent sample in the redshift range $0.04<z<0.4$ is plotted in the [\ion{N}{ii}]-diagnostic diagram with the aim of showing the source density in the $[\ion{O}{iii}]-[\ion{N}{ii}]$ plane. We see that most of the sources are located in the star-forming composite branch, where AGN are supposed to turn on.

\begin{figure*}[tbh]
  \centering
  \begin{tabular}{ccc} 
     \includegraphics[width=0.33\textwidth]{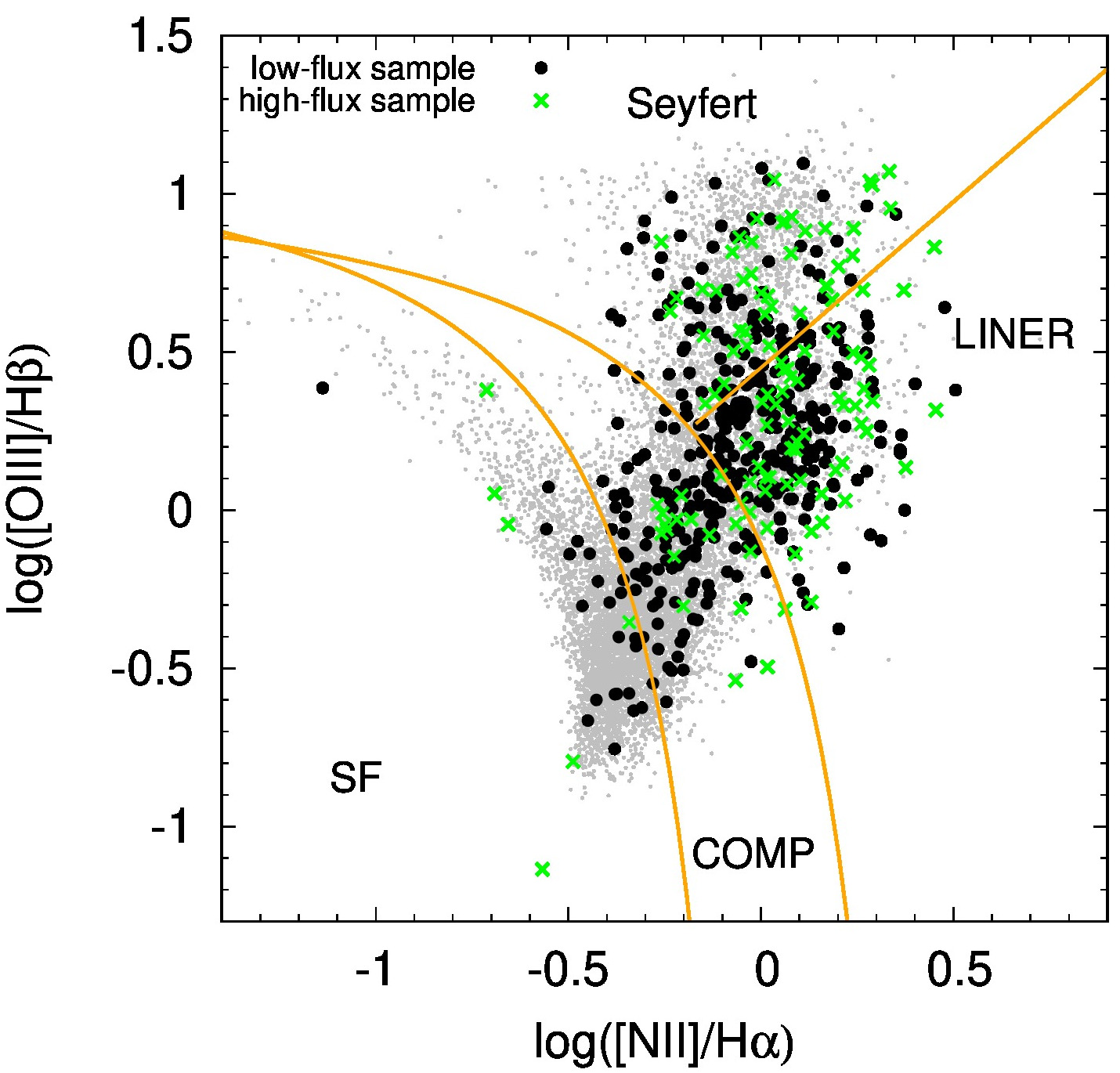} & \hspace{-0.5cm}
       \includegraphics[width=0.33\textwidth]{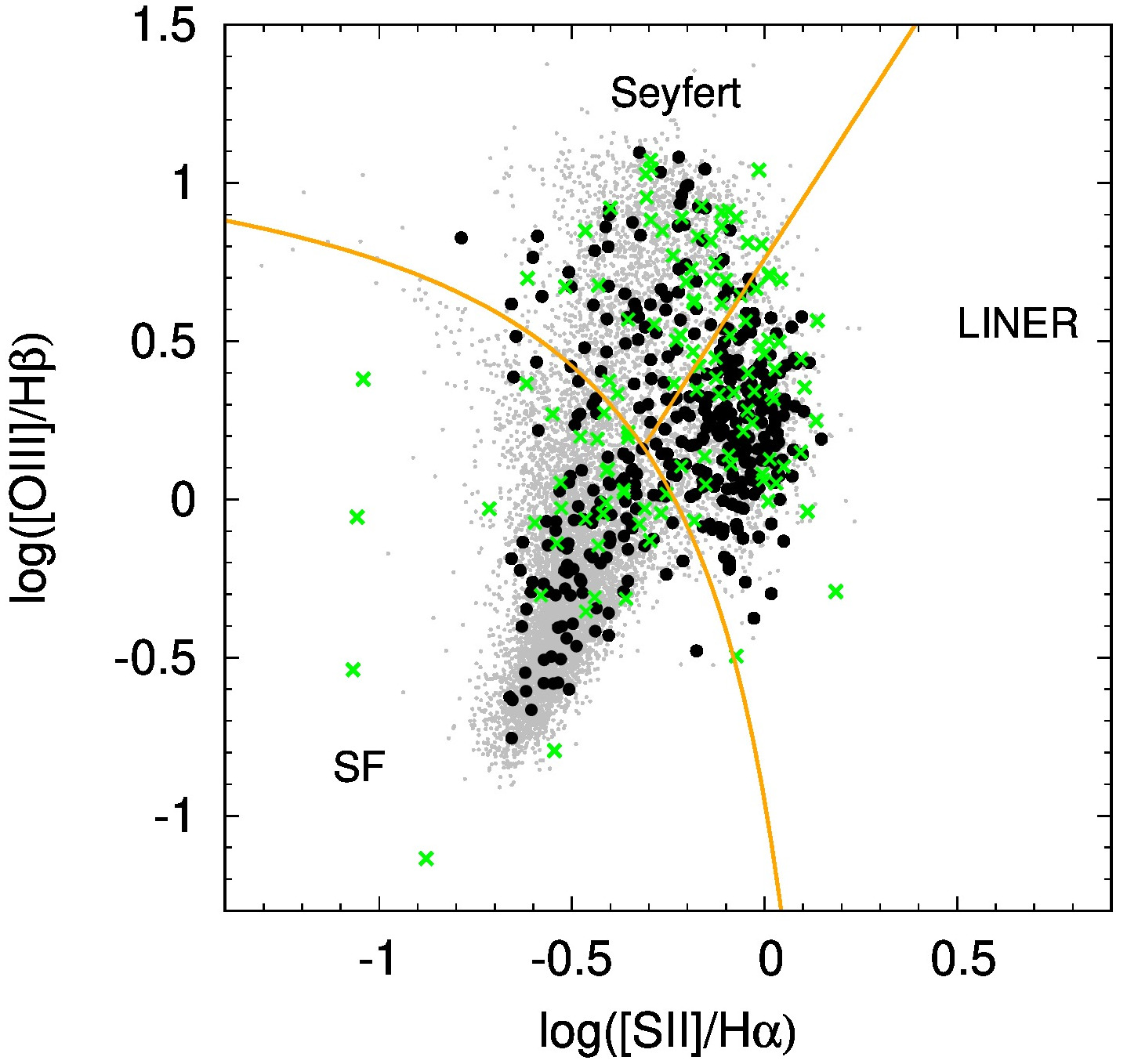}& \hspace{-0.5cm}
       \includegraphics[width=0.33\textwidth]{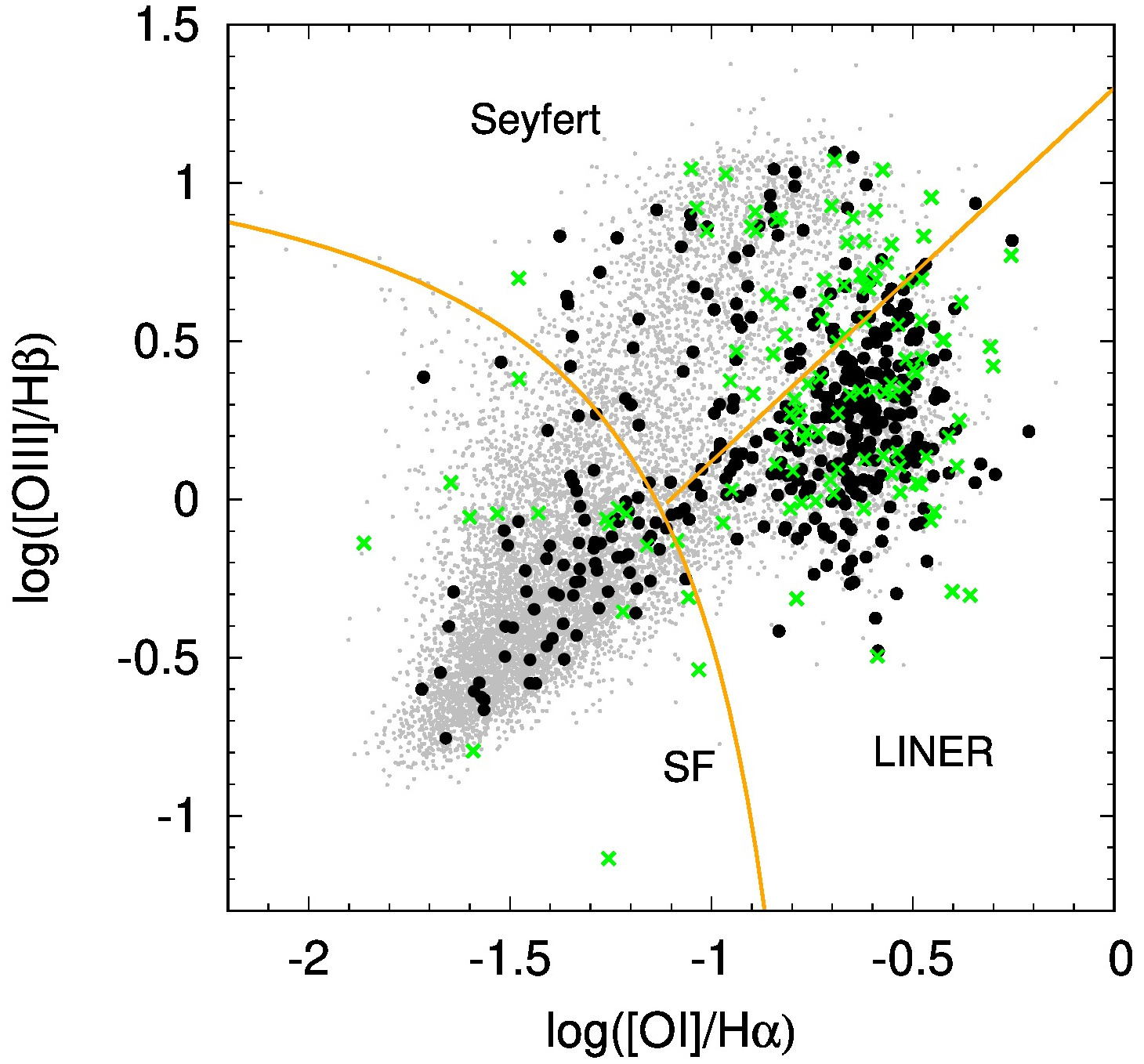}  
  \end{tabular}    
   \caption{The low- and high-flux sample distribution in different optical diagnostic (BPT) diagrams. \textbf{Left:} [NII]-based  diagnostic diagrams  of  the  parent  (grey)  and
Effelsberg samples: a low-flux sample represented by black points and a high-flux sample denoted by green crosses. Demarcation  lines  were  derived  by \citet{2001ApJ...556..121K} to set an upper limit for the star-forming galaxies and  by \citet{2003MNRAS.346.1055K} to  distinguish  purely  star-forming galaxies. The  dividing  line  between Seyferts and LINERs was derived by \citet{2007MNRAS.382.1415S}. The new Effelsberg  sample (black points), extended towards lower radio fluxes, covers the whole diagnostic diagram. \textbf{Middle:} The same samples as in the left panel in [SII]-based diagnostic diagram. \textbf{Right:} High- and low-flux samples in [OI]-based diagram.}
  \label{fig_effelsberg_sample}
\end{figure*}  

\section{Effelsberg sample and observations}
\label{effelsberg_sample_observation}

The aim of the previous study by \citet{2015A&A...573A..93V} was to analyse the radio-optical properties across the parent sample. The radio information was complemented by the radio flux density measurements at two higher frequencies -- $4.85$ and $10.45\,{\rm GHz}$ -- using the 100-m Effelsberg radio telescope. \citet{2015A&A...573A..93V} selected the sources from the parent sample using the lower limit for the integrated flux density of $F_{1.4}\geq 100\,{\rm mJy}$ at $1.4\,{\rm GHz}$. 
In total, 119 sources selected according to the criteria above were observed by the Effelsberg radio telescope at two additional frequencies. Thus, for these sources it was possible to determine the spectral slopes $\alpha_{[1.4-4.85]}$ and $\alpha_{[4.85-10.45]}$. This sample is denoted as a high-flux sample and the main properties of the sources are summarized in Table A.1 of \citet{2015A&A...573A..93V}. 

In Fig.~\ref{fig_redshift_flux_surplot_parent}, the sample occupies the upper part of the redshift-integrated flux plot. Due to the large radio flux, the sources are dominated by galaxies with an AGN. In the [\ion{N}{ii}]-based optical diagnostic diagram (BPT diagram) the high-flux radio-optical sample is dominated by composite sources, Seyfert, and LINER galaxies (see green points in Fig.~\ref{fig_redshift_flux_surplot_parent}). There are only few metal-rich star-forming sources whose radio emission generally originates in the shocks from supernovae and in (re)ignited AGN and jet activity.

Therefore the high-flux sample is certainly not complete in a statistical sense and the results published in \citet{2015A&A...573A..93V} are not representative of the whole radio-optical parent sample. This was the main motivation for the extension of the Effelsberg sample towards lower radio flux densities; the integrated radio flux density at $1.4\,{\rm GHz}$ was considered in the interval $10\,{\rm mJy} < F_{1.4} < 100\,{\rm mJy}$, that is, we decreased the upper and the lower flux limit of the high-flux sample by one order of magnitude. As  shown in Figs.~\ref{fig_redshift_flux_surplot_parent} and \ref{fig_effelsberg_sample}, the low-flux sample (black points) covers the whole [\ion{N}{ii}]-based diagnostic diagram and its coverage is also more uniform. It should therefore better represent the radio-optical properties of the parent sample. However, the bias towards the AGN and LINER sources is partially maintained, as can be inferred from the [\ion{N}{ii}]-, [\ion{S}{ii}]-, and [\ion{O}{i}]-based diagrams in Fig.~\ref{fig_effelsberg_sample}.

By imposing the redshift limits $0.04\leq z \leq 0.4$, as well as the signal-to-noise criterion on the equivalent width of the optical emission lines, $S/N>3$, the low-flux sample initially consisted of $381$ galaxies with the integrated flux densities $10\,{\rm mJy} \leq F_{1.4} \leq 100\,{\rm mJy}$. These sources were first observed at $4.85\,{\rm GHz}$ with the 100-m telescope in Effelsberg. Observations were performed between April 2014 and June 2015. The receiver at $4.85\,{\rm GHz}$ is mounted on the secondary focus of the Effelsberg antenna. It has multi-feed capabilities with two horns, which allow real-time sky subtraction in every subscan measurement. The total intensity of each source was determined via scans in the azimuth and the elevation. According to the brightness of the source, several subscans were used, ranging from 6 up to 24. In the data reduction process, subscans were averaged to produce final subscans used for further processing. Each scan had a length equal to 3.5 of that of the beam size at the corresponding frequency to ensure the correct subtraction of linear baselines. 

Before combining the subscans, we checked each for possible radio interference, bad weather effects, or detector instabilities. During each observational run, we observed standard bright calibration sources, such as 3C286, 3C295, and NGC7027, which were used for correcting gain instabilities and elevation-dependent antenna sensitivity. Finally, we used these sources for the absolute flux calibration. The whole data reduction was performed using a set of Python and Fortran scripts. The flux density was obtained by fitting Gaussian functions to the signal in the averaged single-dish cross scans. Further details on the data reduction can be found in \citet{2015A&A...573A..93V}, who applied the same routines for brighter sources. 

From 381 sources, we managed to determine the reliable flux density at $4.85\,{\rm GHz}$ for 298 sources. The flux densities range between $350\,{\rm mJy}$ and $4\,{\rm mJy,}$ with the mean and median values of $30\,{\rm mJy}$ and $17\,{\rm mJy}$, respectively. Other sources were too faint or extended at least in one direction and therefore it was not possible to reliably determine an integrated flux density. For $10.45\,{\rm GHz}$ observations, 256 sources out of 298 were scheduled for observing based on the extrapolated flux density based on the non-simultaneous $1.4\,{\rm GHz}$ and $4.85\,{\rm GHz}$ flux densities. At $10.45\,{\rm GHz}$ some sources were too faint, were extended at least in one direction, or the reliable flux density determination was not possible due to a higher sensitivity to weather effects at $10.45\,{\rm GHz}$. In the end, we obtained flux densities at $10.45\,{\rm GHz}$ for 90 sources. The maximum and minimum flux densities are $206\,{\rm mJy}$ and $6\,{\rm mJy}$, respectively. The mean and median values are $32\,{\rm mJy}$ and $19\,{\rm mJy}$, respectively. The three non-simultaneous flux densities for 90 sources in the low-flux sample, $F_{1.4}$, $F_{4.85}$, and $F_{10.45}$, are listed in Table~\ref{tab_three_freq1} in Appendix~\ref{appa}, along with the radio spectra for each source as well as the mean radio spectrum for each galaxy spectral class (star-forming, composite, AGN Seyfert, and LINER galaxies). 

In the following analysis, unless otherwise stated, we use the low-flux sample in combination with the high-flux sample of \citet{2015A&A...573A..93V}. For the study of radio continuum properties between $1.4$ and $4.85\,{\rm GHz}$, we have $298$ low-flux sources and $119$ high-flux sources available. Between $4.85$ and $10.45\,{\rm GHz}$, there are $90$ low-flux and $119$ high-flux sources.

\section{Spectral index properties}
\label{analysis_results}

\subsection{General properties of radio spectral index distributions}

We present the radio flux densities at $1.4$ (FIRST), $4.85$, and $10.45\,{\rm GHz}$ (both Effelsberg) for $90$ low-flux sources in the table in Appendix~\ref{appa}. The flux densities at $1.4\,{\rm GHz}$ (FIRST) and $4.85\,{\rm GHz}$ (Effelsberg) are non-simultaneous (more than one year apart from each other), while the Effelsberg observations at $4.85\,{\rm GHz}$ and $10.45\,{\rm GHz}$ were performed within one year of each other. 

For high-flux sources, the analysis as well as the optical and the radio images were presented in \citet{2015A&A...573A..93V}. The catalogue of the high-flux sources is available in \citet{2015yCat..35730093V}, where the (quasi)-simultaneous flux densities (obtained during a single observing session) are listed.


For the flux density in the radio domain, we assume the power-law dependency on frequency, using the notation $F(\nu) \propto \nu^{+\alpha}$, where $\alpha$ is the spectral index. Based on the non-simultaneous measurements of flux densities $F_{1.4}$ (FIRST) and $F_{4.85}$ (Effelsberg), we calculated the spectral index $\alpha_{[1.4-4.85]}$ using

\begin{equation}
  \alpha_{[1.4-4.85]}=\frac{\log{(F_{1.4}/F_{4.85})}}{\log{(1.4/4.85)}}\,.
  \label{eq_spec_index_low}
\end{equation}

We note that there is a large beam-size difference between the VLA in the B-configuration and the Effelsberg telescope at $4.85\,{\rm GHz}$: the half-power beam-width (HPBW) at $20\,{\rm cm}$ for the VLA is $\theta_{\rm HPBW}^{1.4}\approx 4.3''$, whereas for the Effelsberg telescope at $4.85\,{\rm GHz}$, $\theta_{\rm HPBW}^{4.85}\approx 2.4'$. This could have led to the exclusion of extended structures for the VLA measurements for approximately one third of the sources that are clearly extended on the scales of $2''-30''$ \citep{2002AJ....124.2364I}  and thus, for such extended sources, it influences the integrated flux densities and spectral indices $\alpha_{[1.4-4.85]}$ as well.  Moreover, the observations were more than one year apart from each other, thus possibly contaminated by the variability of the sources. We include the distribution of $\alpha_{[1.4-4.85]}$ for completeness, however, due to the potential beam-size effect, we exclude it from further analysis.

For the spectral index at higher frequencies, $\alpha_{[4.85-10.45]}$, which is determined analogously to Eq.~\eqref{eq_spec_index_low} as
\begin{equation}
  \alpha_{[4.85-10.45]}=\frac{\log{(F_{4.85}/F_{10.45})}}{\log{(4.85/10.45)}}\,,
  \label{eq_spec_index_high}
\end{equation}
the effect of excluding extended structures is largely diminished, since for the analysis in this paper we considered only point sources that were consistent with HPBWs of $2.4'$ and $1.1'$ at $4.85\,{\rm GHz}$ and $10.45\,{\rm GHz}$, respectively. Moreover, the observations at $4.85\,{\rm GHz}$ and $10.45\,{\rm GHz}$ were performed within one year of each other, hence the spectral index $\alpha_{[4.85-10.45]}$ captures the integrated radio continuum better than the spectral index $\alpha_{[1.4-4.85]}$, which can be influenced by the core emission due to the small VLA beam width.

\begin{figure}[h!]
  \centering
  \includegraphics[width=0.5\textwidth]{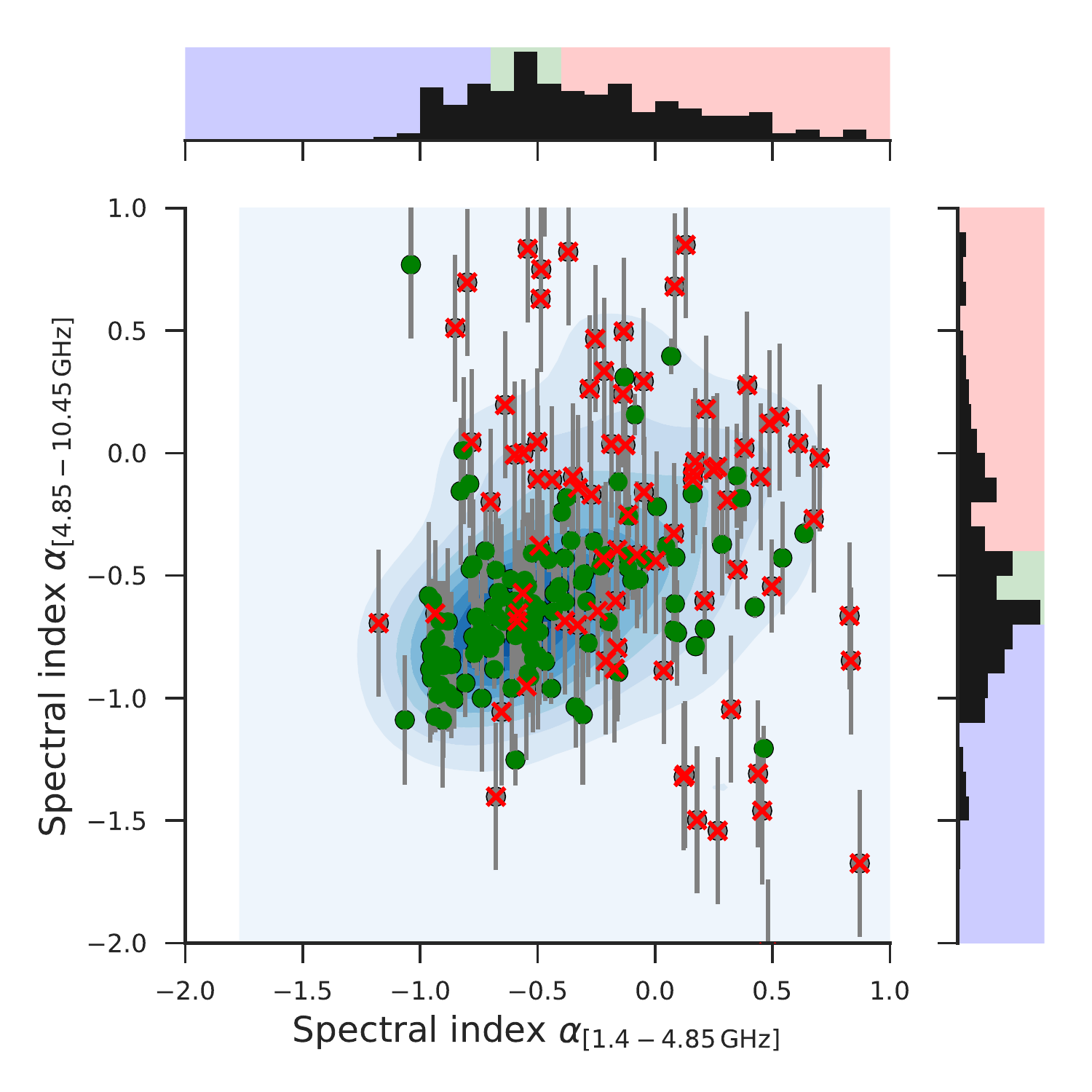}
  \caption{Two-dimensional distribution of spectral indices at lower ($x$-axis) and higher frequencies ($y$-axis). The green points mark the high-flux sources, while the red crosses depict the low-flux sample. Some of the sources have error bars for $\alpha_{[4.85-10.45]}$ to show typical uncertainties of the spectral index.}
  \label{fig_2D_spectrindex}  
\end{figure}

The uncertainty of the spectral index $\sigma_{\alpha}$ was calculated by propagating the measurement errors of flux densities at the corresponding frequencies,

\begin{equation}
  \sigma_{\alpha}=\frac{1}{|\log{(4.85/10.45)}|}\sqrt{(\sigma_{4.85}/F_{4.85})^2+(\sigma_{10.45}/F_{10.45})^2}\,,
\end{equation}
where $\sigma_{4.85}$ and $\sigma_{10.45}$ are the measurement uncertainties of flux densities at the corresponding frequencies. We show exemplary error bars in Fig.~\ref{fig_2D_spectrindex}. The median value of $\sigma_{\alpha}$ at higher frequencies for the joint sample is $\sigma_{\alpha}=0.1$.

\begin{figure*}[tbh]
  \centering
     \includegraphics[width=0.45\textwidth]{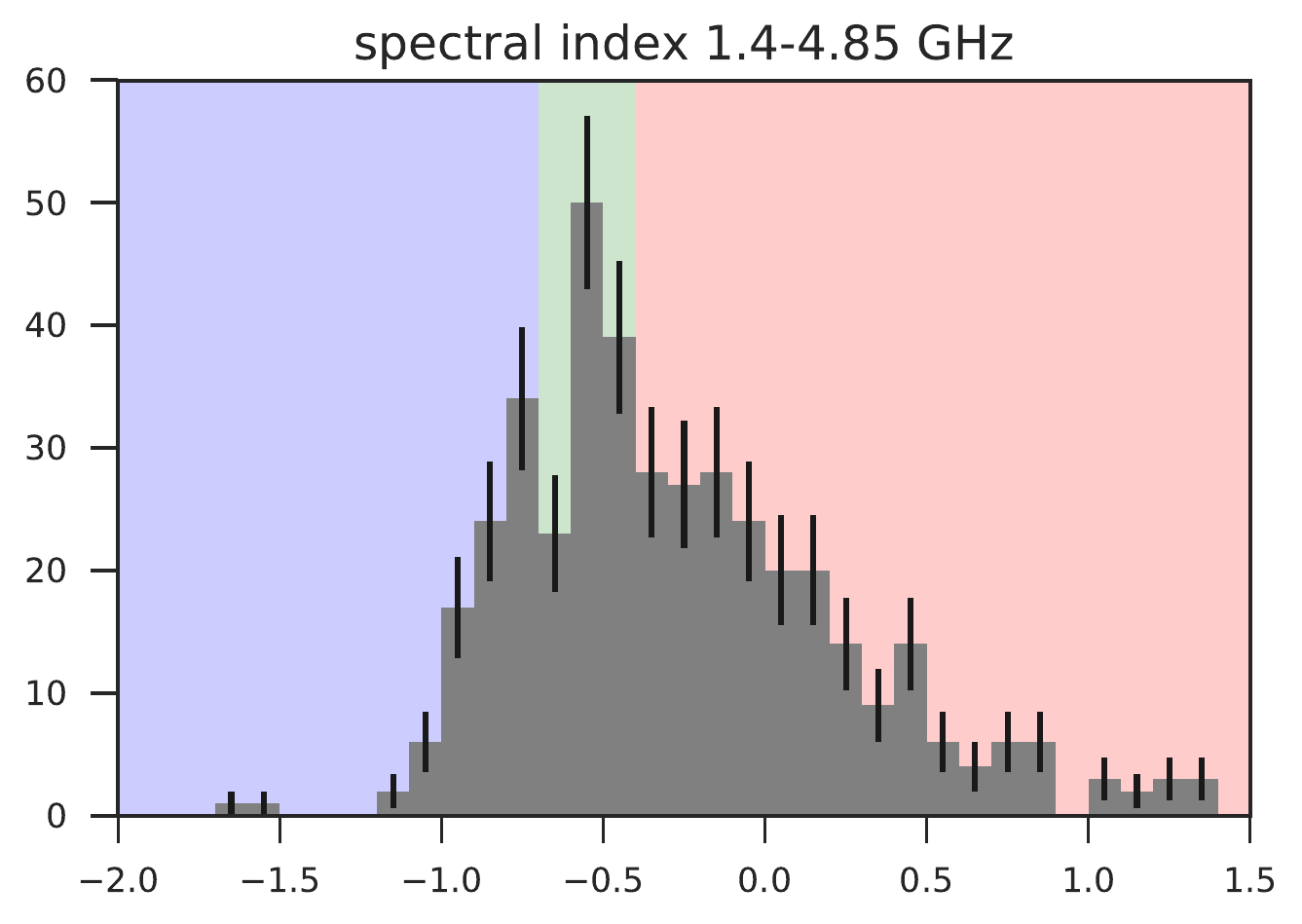} 
     \includegraphics[width=0.45\textwidth]{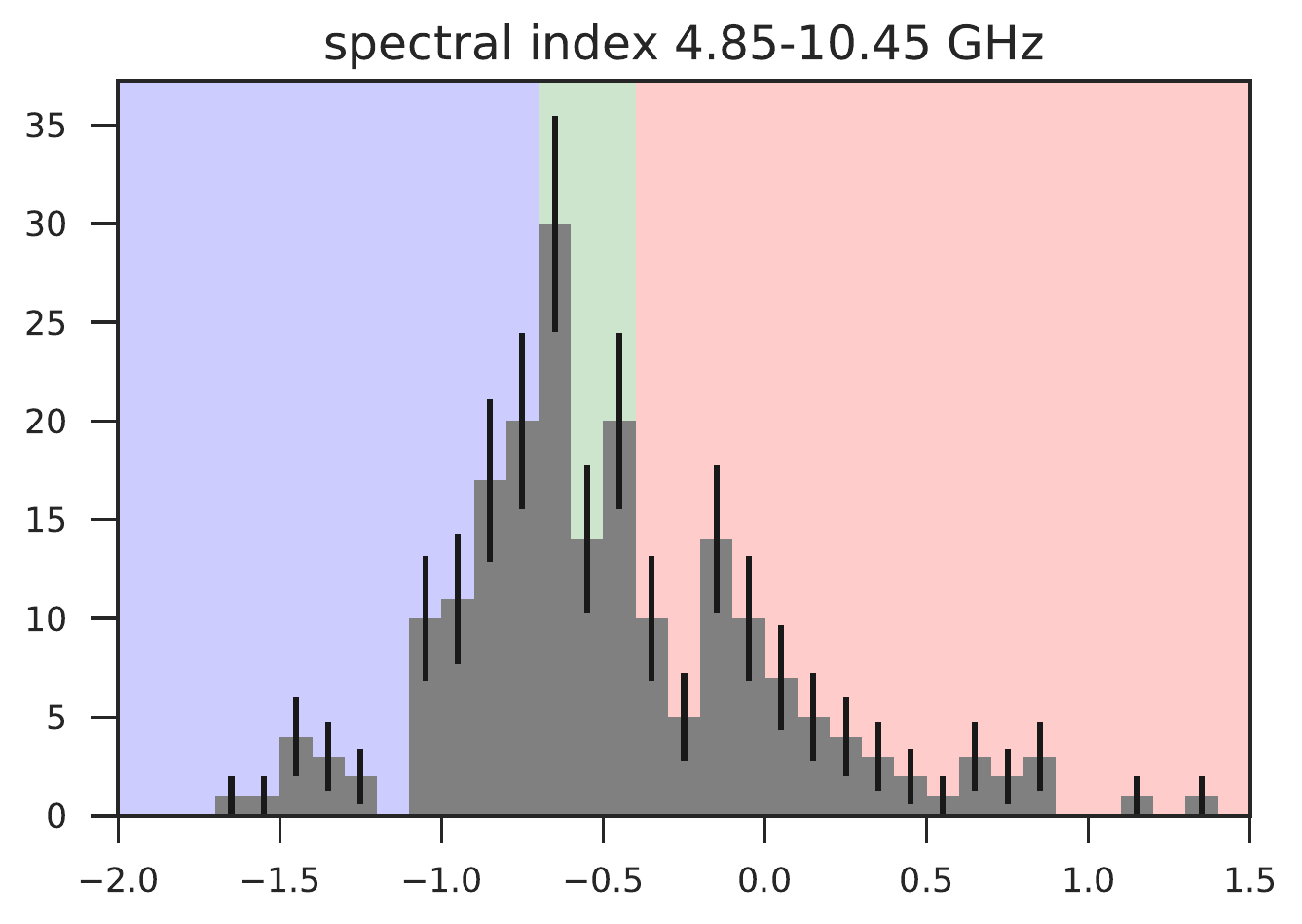}   
   \caption{Distributions of spectral indices $\alpha_{[1.4-4.85]}$ and $\alpha_{[4.85-10.45]}$.   \textbf{Left panel:} A two-point spectral index ($\alpha_{[1.4-4.85]}$) distribution for the combined low-flux and high-flux sample (in total 417 sources). \textbf{Right panel:} A two-point spectral index ($\alpha_{[4.85-10.45]}$) distribution for the combined low-flux and high-flux sample (in total 209 sources).}
  \label{fig_specindex_histograms_allsources}
\end{figure*}

The distributions of spectral indices  $\alpha_{[1.4-4.85]}$ and $\alpha_{[4.85-10.45]}$ for all observed sources are plotted in Fig.~\ref{fig_specindex_histograms_allsources} in the left and the right panel, respectively. For the lower frequencies, the mean spectral index is $\overline{\alpha}_{[1.4-4.85]}=-0.25\pm 0.54$ (median $-0.36$). For the higher frequencies, the mean spectral index is $\overline{\alpha}_{[4.85-10.45]}=-0.51\pm 0.63$ (median $-0.58$). The two-dimensional distribution of spectral indices at both lower and higher frequencies is in Fig.~\ref{fig_2D_spectrindex}.

For the general classification of radio spectra with the power-law shape $F(\nu) \propto \nu^{+\alpha}$, we use the following categories based on the spectral slope $\alpha$:
\begin{itemize}
 \item[(i)] $\alpha<-0.7$, for short denoted as steep, which are typical for optically thin synchrotron structures, where electrons have cooled off, such as radio lobes; 
 \item[(ii)] $-0.7 \leq \alpha \leq -0.4$, denoted as flat, with the mixed contribution of optically thin and self-absorbed synchrotron emission, typical for jet emission; 
 \item[(iii)] $ \alpha > -0.4$, denoted as inverted, which are characteristic for sources where synchrotron self-absorption becomes important, such as in AGN core components.
\end{itemize}
This division reflects the distributions of the spectral index for lower and higher frequencies as found for samples of radio-loud galaxies, such as in the S5 polar-cap sample \citep{1986A&A...168...17E}. We adopt it for all the histograms of the radio spectral index, from Fig.~\ref{fig_specindex_histograms_allsources} onwards.

\begin{figure}[h!]
  \centering
  \includegraphics[width=0.5\textwidth]{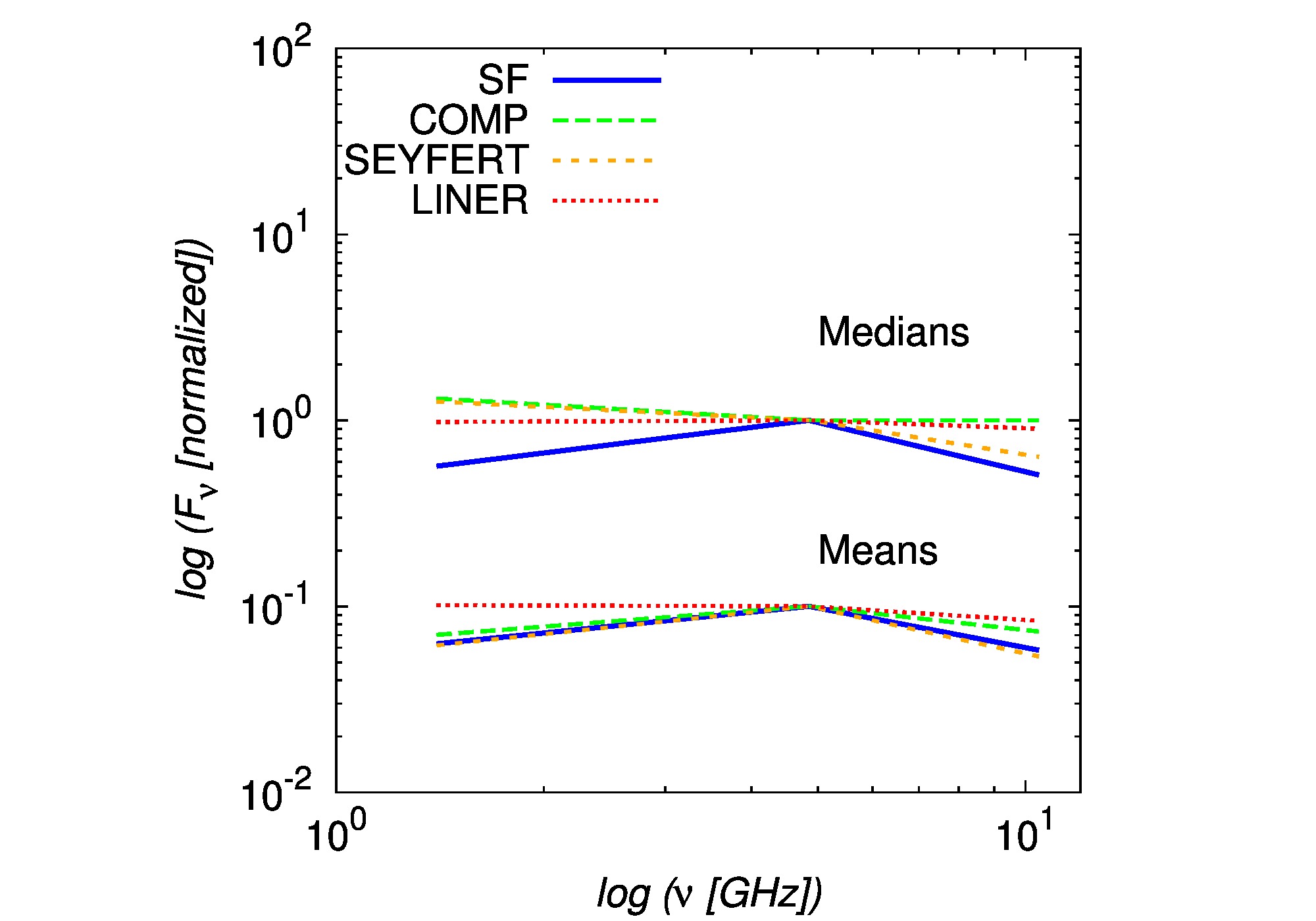}
  \caption{Median and mean spectra for different spectral classes scaled for the common value at the mid-frequency of $4.85\,{\rm GHz}$.}
  \label{fig_mean_radio_spectra}  
\end{figure}

The mean and median radio spectra calculated for different spectral classes according to the BPT diagram -- star-forming, composites, Seyferts, LINERs -- are scaled for comparison in Fig.~\ref{fig_mean_radio_spectra}. The values of spectral indices for mean spectra are $0.37$, $0.28$, $0.39$, $-0.01$  at smaller frequencies $1.4\,{\rm GHz}-4.85\,{\rm GHz}$  and $-0.71$, $-0.41$, $-0.81$, $-0.23$ at higher frequencies $4.85\,{\rm GHz}-10.45\,{\rm GHz}$ for star-forming, composite, Seyfert, and LINER sources, respectively. Spectral indices for median spectra, which were calculated from the spectra of individual sources after the normalization with respect to the mid-frequency, are $0.46$, $-0.22$, $-0.19$, $0.02$ at smaller frequencies $1.4\,{\rm GHz}-4.85\,{\rm GHz}$  and $-0.88$, $0.0$, $-0.59$, $0.13$ at higher frequencies $4.85\,{\rm GHz}-10.45\,{\rm GHz}$ for star-forming, composite, Seyfert, and LINER sources, respectively.     

The distribution of spectral indices is very reminiscent of the
distribution that one finds for higher redshift quasars. The S5-survey
\citep{1981A&AS...45..367K, 1984AJ.....89..323G}
shows the transition between the lobe and jet-dominated
structures, which are more prominent at lower radio frequencies, and the flat to
inverted cores that are dominant at higher radio frequencies.
Unfortunately, the spectral slope expected from the synchrotron emission of
supernova remnants, which are a clear tracer of high-mass star formation,
has a similar slope as the radio lobe-jet structures.
The radio emission from star-forming regions can be expected to be
relevant for objects on the division lines between star-forming and Seyfert/LINER objects.
In the Seyfert/LINER domain of the diagnostic diagrams, objects with starburst-AGN mixing
can be found \citep[e.g.][]{2006MNRAS.372..961K, 2016MNRAS.462.1616D}.
However, a detailed and sensitive structural investigation at high angular
resolution is required to differentiate between the presumably extended radio contribution
from star formation and that of the core-jet-nucleus structure of a radio-active AGN.
This will be possible with instruments dedicated for low-surface brightness investigations
like the Square Kilometer Array \citep[SKA; see e.g.][]{2015aska.confE..93A}.

\subsection{Radio spectral index between $1.4$ and $4.85$ GHz}

\begin{figure*}[tbh]
 \centering
 \includegraphics[width=0.45\textwidth]{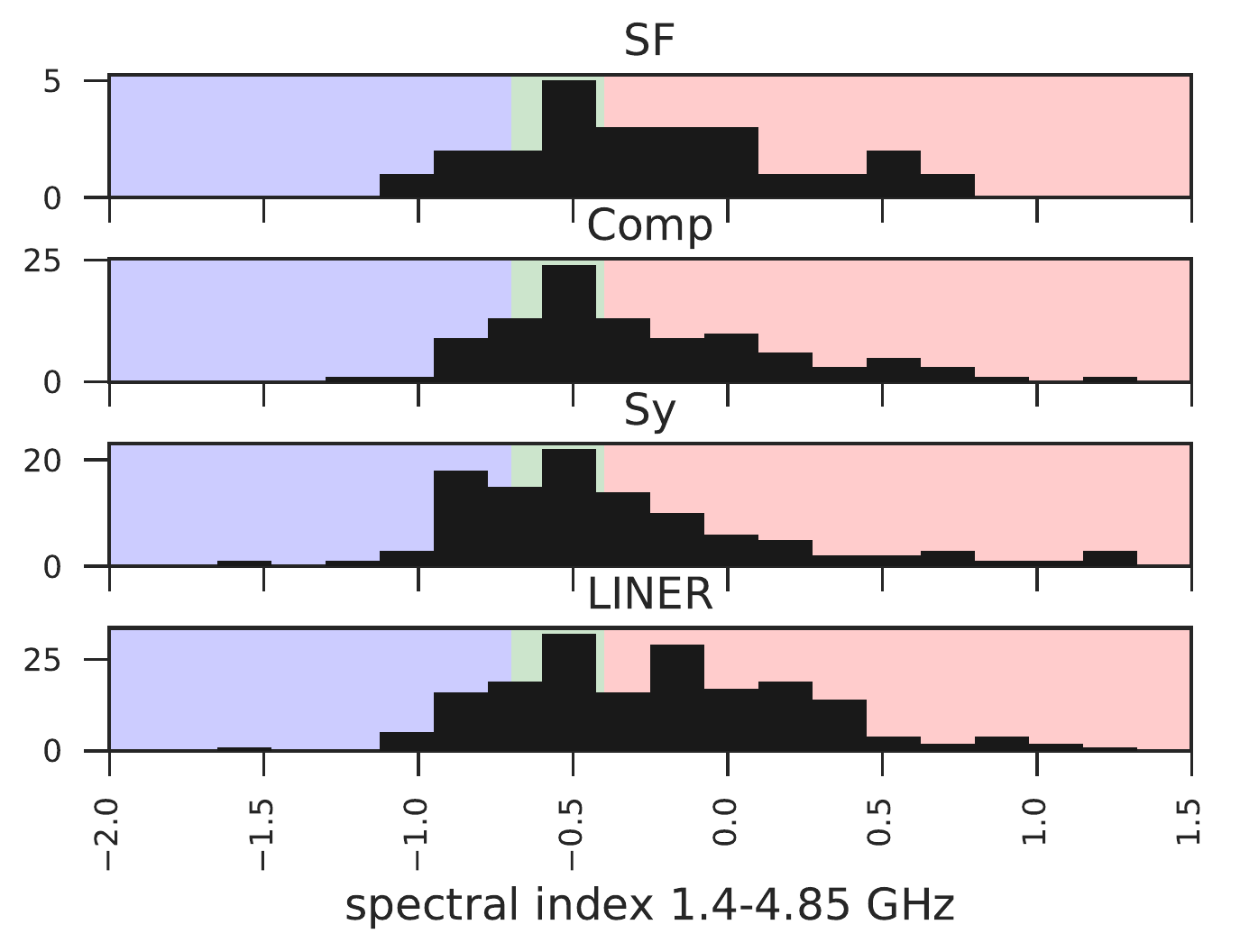}
 \includegraphics[width=0.45\textwidth]{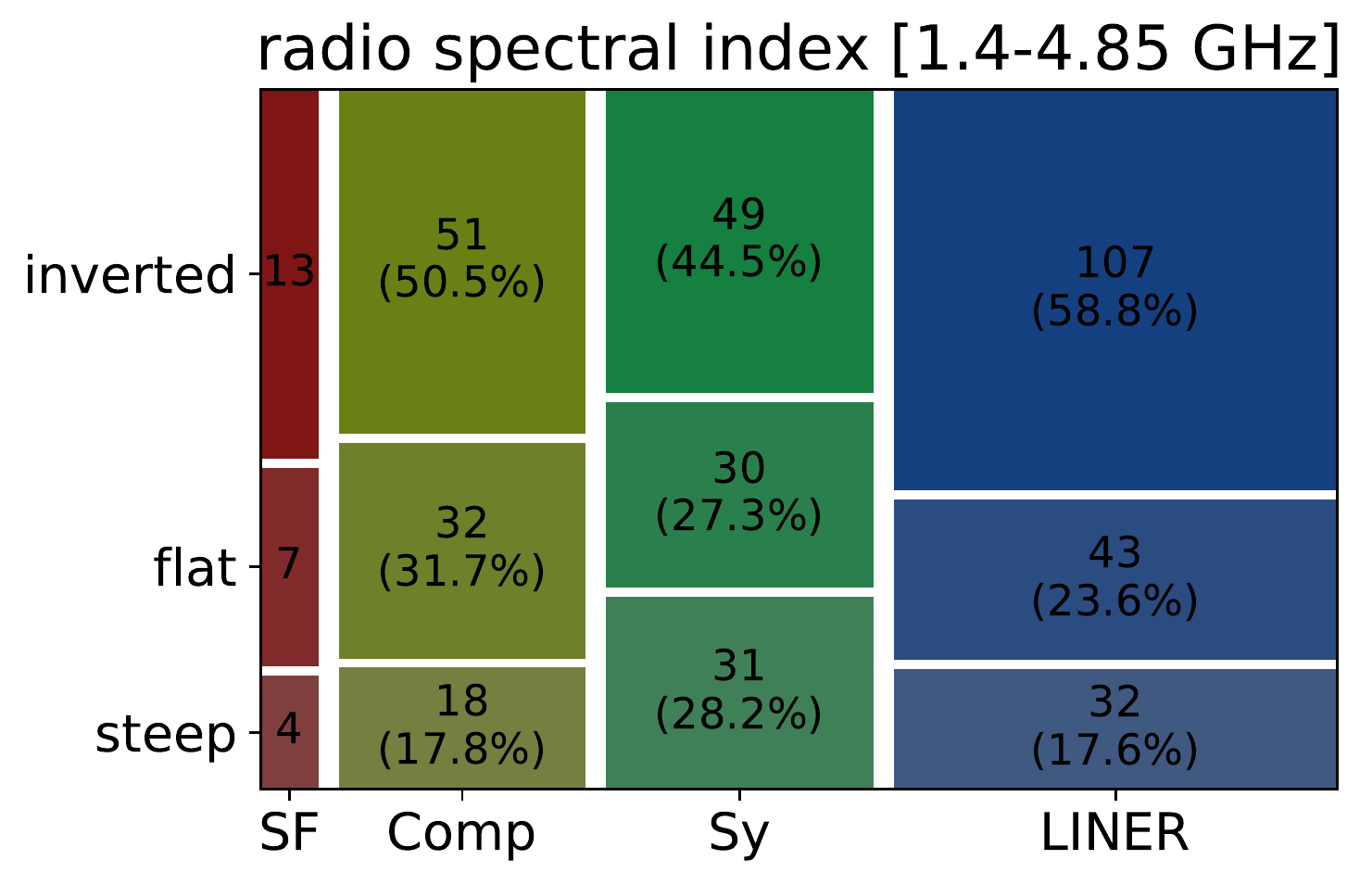}
 \caption{Distribution of the spectral slope $\alpha_{[1.4-4.85]}$ at lower frequencies among different spectral classes of galaxies. \textbf{Left:} The two-point spectral index distribution $\alpha_{[1.4-4.85]}$ for different optical spectral classes of galaxies (SF, composites, Seyferts, and LINERs). The mean and median values are listed in Table~\ref{tab_spectralindex_1.4_4.85}. \textbf{Right:} Fractional distribution of steep, flat, and inverted sources among star-forming, composite, Seyfert, and LINER galaxies.}
 \label{fig_histogram_1.4-4.85GHz}
\end{figure*}

The mean and median values of the spectral index $\alpha_{[1.4-4.85]}$ for the whole low-flux and high-flux sample are in Table~\ref{tab_spectralindex_1.4_4.85}. LINER sources have the flattest median and mean radio spectral indices in comparison with galaxies in other spectral classes. This is also apparent in the histogram in Fig.~\ref{fig_histogram_1.4-4.85GHz} (left panel), where we plot the two-point spectral index distribution for each spectral class. Another way of representing this trend is shown in the plot in Fig.~\ref{fig_histogram_1.4-4.85GHz} (right panel), which combines the radio classification (steep, flat, inverted) and the optical spectral classification of galaxies (SF, Comp, Sy, LINER). While among the composites $50.5\%$ are inverted sources and among the Seyferts $44.5\%$ are inverted sources, most of the LINERS ($58.8\%$) have inverted spectra. The fraction of steep sources $(\alpha<-0.7)$ among LINER sources is only $17.6\%$, which is comparable to composite sources ($17.8\%$). The fraction of steep Seyferts is $28.2\%$. In terms of the mean and median spectral slopes, LINERs are the flattest index $\alpha_{[1.4-4.85]}$ with a mean of $-0.22$ and a median of $-0.24$, followed by the composites with a mean and median of $-0.25$ and $-0.40$, respectively, and the Seyferts with mean and median values of $-0.31$ and $-0.49$, respectively.

\begin{table}[h!]
  \centering
  \resizebox{\linewidth}{!}{
  \begin{tabular}{cccccc}
    \hline
    \hline
    Spectral class & Mean $\alpha_{[1.4-4.85]}$ & $\sigma$ & Median $\alpha_{[1.4-4.85]}$ & $16\%\,P$ & $84\%\,P$\\
    \hline
    SF & $-0.25$ & $0.45$ & $-0.33$ & $-0.66$ & $0.24$\\
    COMP & $-0.25$ & $0.54$ & $-0.40$ & $-0.73$ & $0.22$\\
    SY & $-0.31$ & $0.61$ & $-0.49$ & $-0.83$ & $0.21$\\
    LINER & $-0.22$ & $0.50$ & $-0.24$ & $-0.72$ & $0.26$\\
    \hline 
    Total & $-0.25$ & $0.54$ & $-0.36$ & $-0.76$ & $0.24$\\
    \hline    
  \end{tabular}}
  \caption{Mean, standard deviation, median, $16\%$-, and $84\%$- values of the radio spectral index $\alpha_{[1.4-4.85]}$, respectively, for each optical spectral class of galaxies and the overall sample.}
  \label{tab_spectralindex_1.4_4.85}
\end{table}

\subsection{Radio spectral index between $4.85$ and $10.45$ GHz}



In an analogous way to frequencies $1.4-4.85\,{\rm GHz}$, we determine the spectral index $\alpha_{[4.85-10.45]}$ between the non-simultaneous Effelsberg measurements at $4.85\,{\rm GHz}$ and $10.45\,{\rm GHz}$. In this case, the primary beam size is comparable, hence the resolution effects should not be so significant as for $1.4\,{\rm GHz}$ obtained from the FIRST survey. Those sources whose emission profiles in cross-scans were clearly extended were excluded from further analysis. 

\begin{figure*}[tbh]
 \centering
 \includegraphics[width=0.45\textwidth]{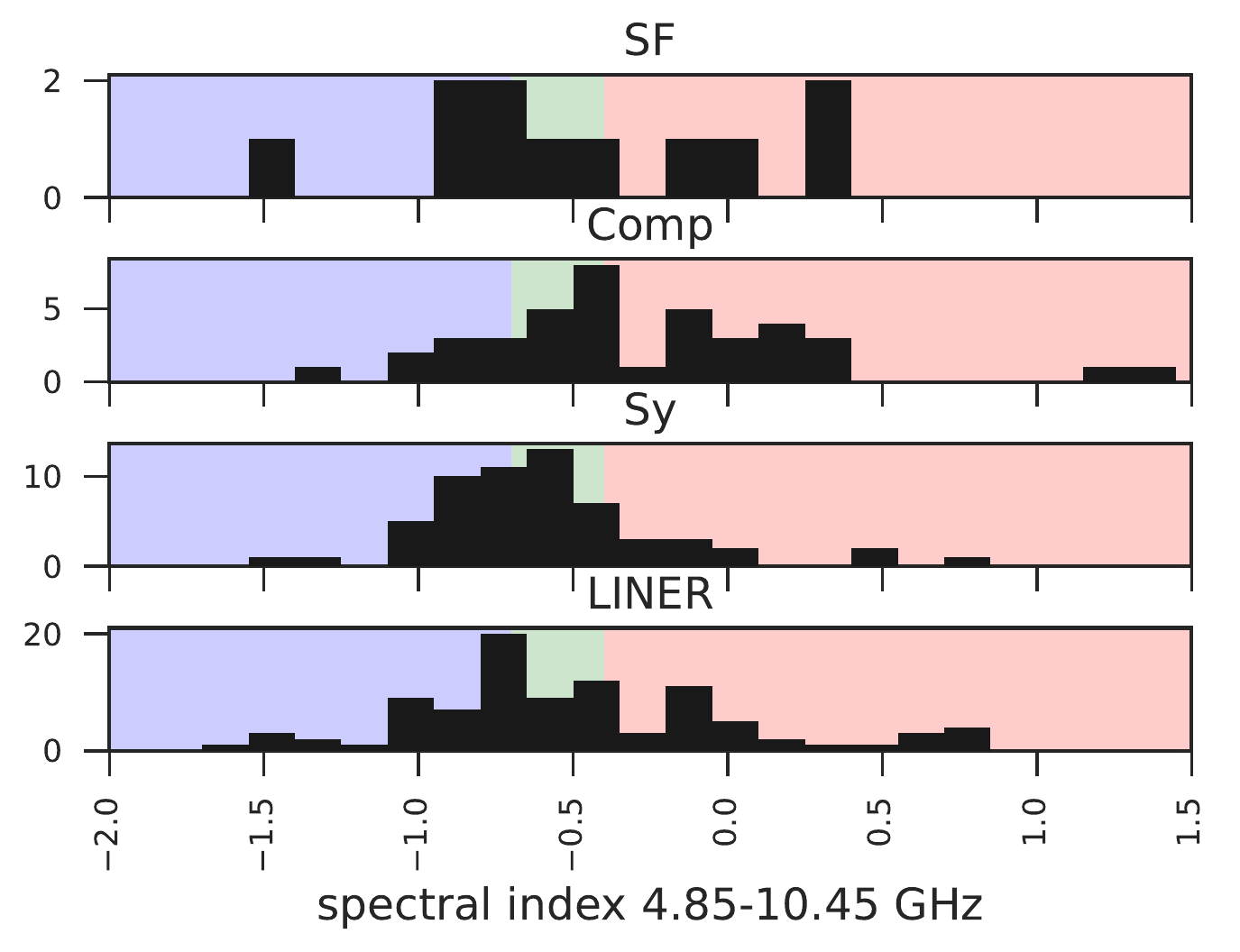}
 \includegraphics[width=0.45\textwidth]{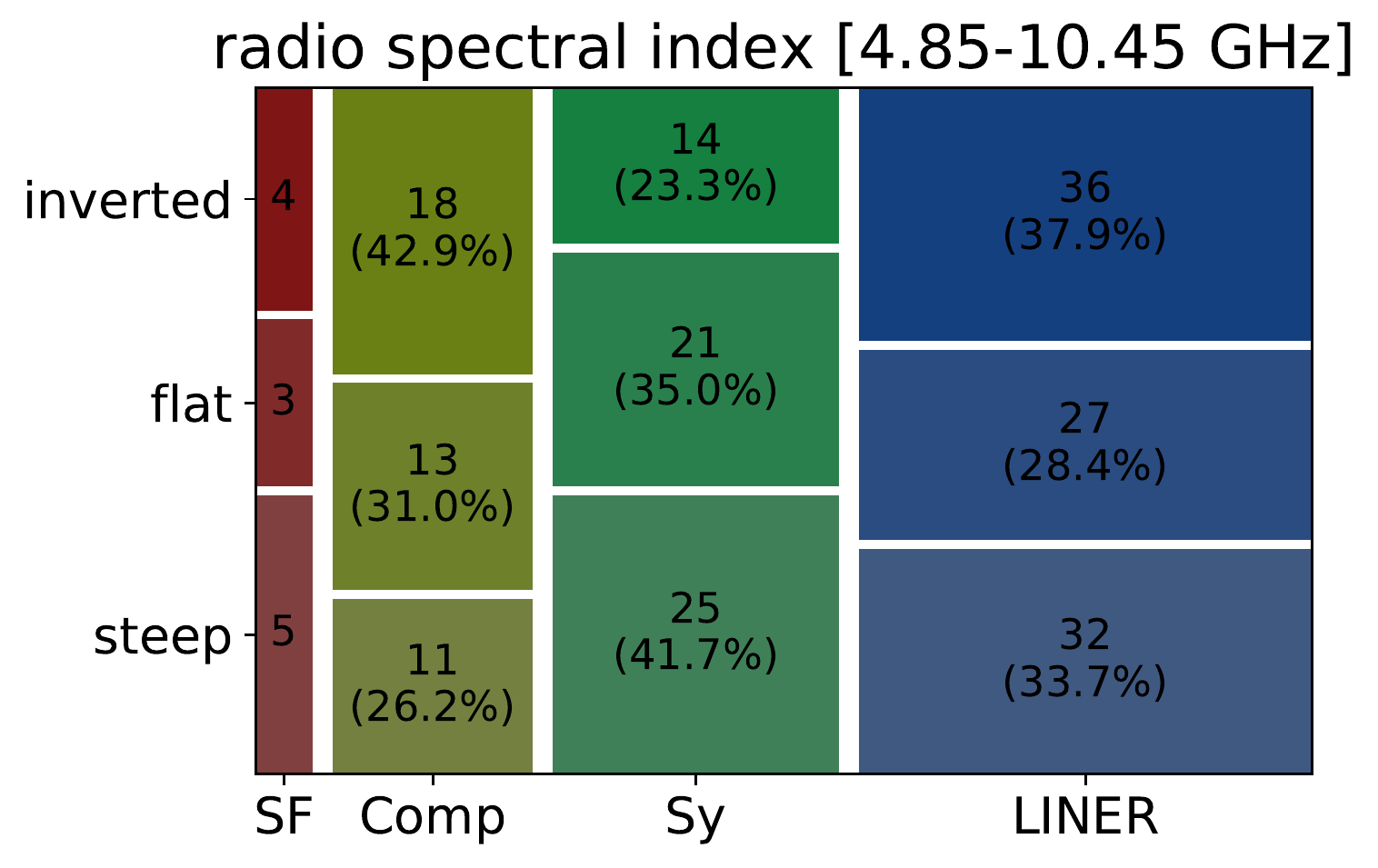}
 \caption{Distribution of the spectral slope $\alpha_{[4.85-10.45]}$ at higher frequencies among different spectral classes of galaxies. \textbf{Left:} The two-point spectral index distribution $\alpha_{[4.85-10.45]}$ for different optical spectral classes of galaxies (SF, composites, Seyferts, and LINERs). The mean and median values are listed in Table~\ref{tab_spectralindex_1.4_4.85}. \textbf{Right:} Fractional distribution of steep, flat, and inverted sources among star-forming, composite, Seyfert, and LINER galaxies.}
 \label{fig_histogram_4.85-10.45GHz}
\end{figure*}

\begin{figure*}[tbh]
 \centering
 \includegraphics[width=\textwidth]{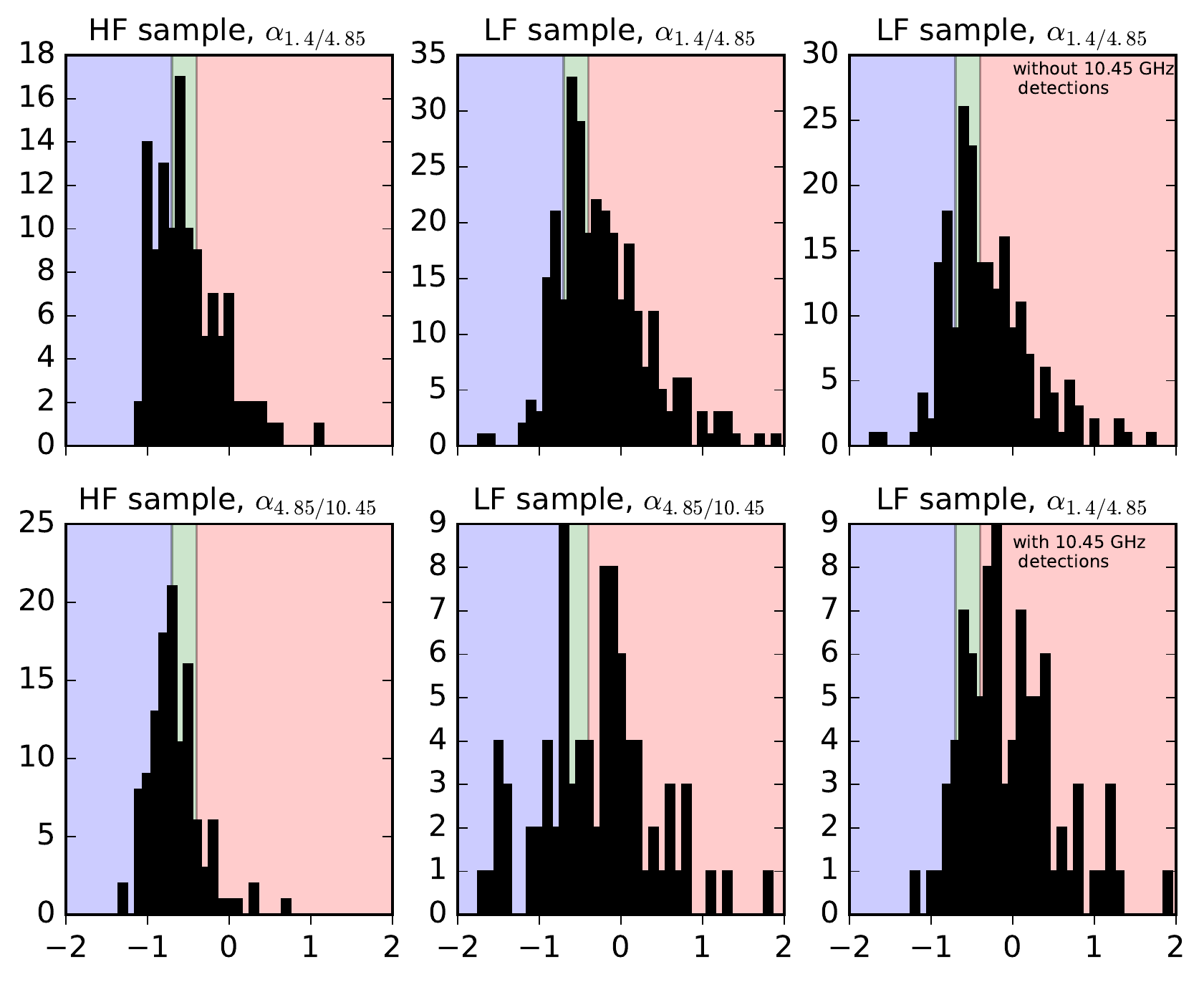}
 \caption{ 
Low-flux density sample experiences an increasing influence of flat-spectrum sources.
\textbf{Top and bottom left:}
$\alpha_{1.4/4.85}$ and $\alpha_{4.85/10.45}$ index distribution for the high-flux density sample (HF).
\textbf{Top and bottom middle:}
$\alpha_{1.4/4.85}$ and $\alpha_{4.85/10.45}$ index distribution for the low-flux density sample (LF)
(two sources - not shown here - have a $\alpha_{4.85/10.45}$ below -3).
\textbf{Top and bottom right:}
$\alpha_{1.4/4.85}$ index distribution for the low-flux sources (LF) without and with 10.45~GHz measurements.
 }
 \label{fig_new-histograms}
\end{figure*}

The distribution of the spectral index $\alpha_{[4.85-10.45]}$ for each optical spectral class is in Fig.~\ref{fig_histogram_4.85-10.45GHz} (left panel). The fractions of three radio classes -- steep, flat, and inverted -- are calculated for each spectral class in the right panel of Fig.~\ref{fig_histogram_4.85-10.45GHz}, where composites have the largest fraction of inverted sources $(\alpha>-0.4), 42.9\%$, followed by LINERs, $37.9\%$. In terms of the overall fraction of sources with a spectral index larger than $\alpha>-0.7$ (non-steep spectra), composites have the largest fraction with $73.9\%$, followed by LINERs $(66.3\%)$ and Seyferts $(58.3\%)$. In the higher frequency range $4.85-10.45\,{\rm GHz}$, the composite sources have the flattest spectral slope $\alpha_{[4.85-10.45]}$ , with a mean of $-0.42$ and a median of $-0.43$, followed by LINERs with mean and median values of $-0.46$ and $-0.59$, respectively, and Seyferts with a mean and median spectral slope of $-0.63$ and $-0.64$, respectively.   

\begin{table}[h!]
  \centering
   \resizebox{\linewidth}{!}{
  \begin{tabular}{cccccc}
    \hline
    \hline
    Spectral class & Mean $\alpha_{[4.85-10.45]}$ &  $\sigma$ & Median $\alpha_{[4.85-10.45]}$ & $16\%\,P$ & $84\%\,P$\\
    \hline
    SF & $-0.61$ &  $0.65$ & $-0.65$ & $-1.02$ & $0.05$\\
    COMP & $-0.42$ & $0.80$ & $-0.43$ & $-0.90$ & $0.19$\\
    SY & $-0.63$ & $0.52$ & $-0.64$ & $-0.89$ & $-0.29$\\
    LINER & $-0.46$ & $0.59$ & $-0.59$ & $-0.95$ & $0.02$\\
    \hline
    Total & $-0.51$ & $0.63$ & $-0.58$ & $-0.92$ & $0.00$\\
    \hline    
  \end{tabular}}
  \caption{Mean, standard deviation, median, $16\%$-, and $84\%$- values of the spectral index $\alpha_{[4.85-10.45]}$, respectively, for each optical spectral class of galaxies and the overall sample.}
  \label{tab_spectralindex_4.85_10.45}
\end{table}

\begin{figure*}[h!]
  \centering
  \includegraphics[width=\textwidth]{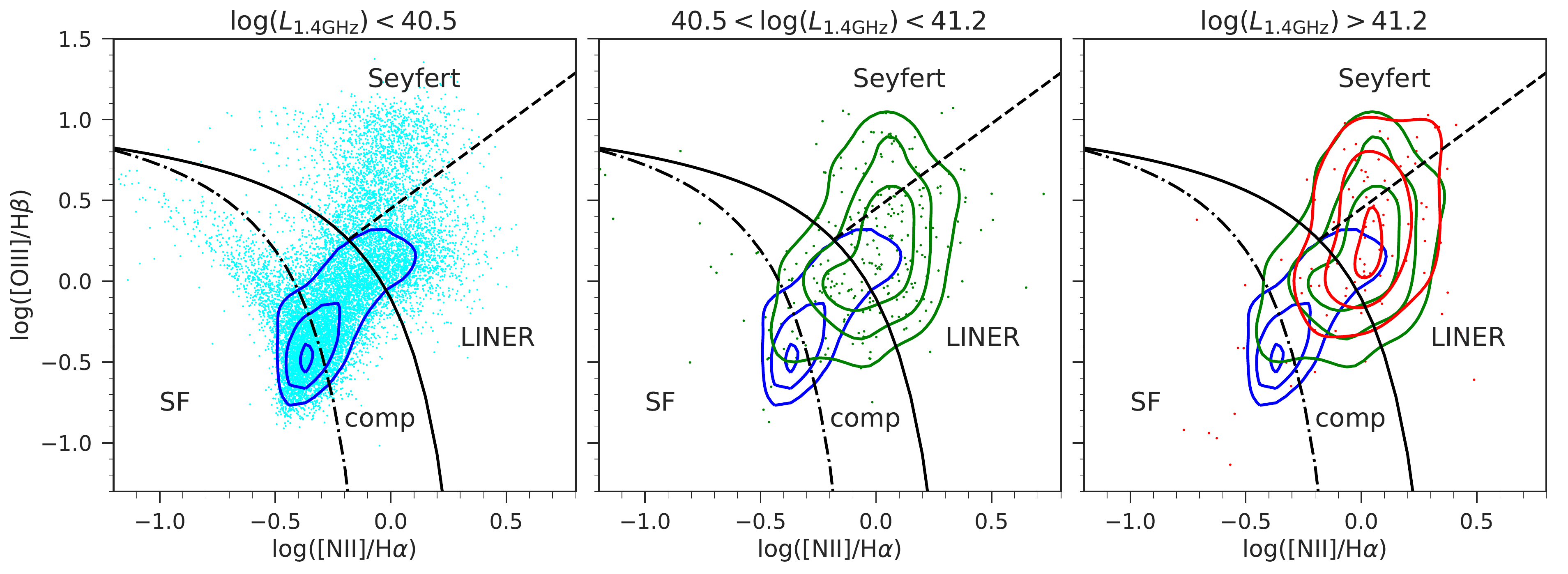}
  \includegraphics[width=\textwidth]{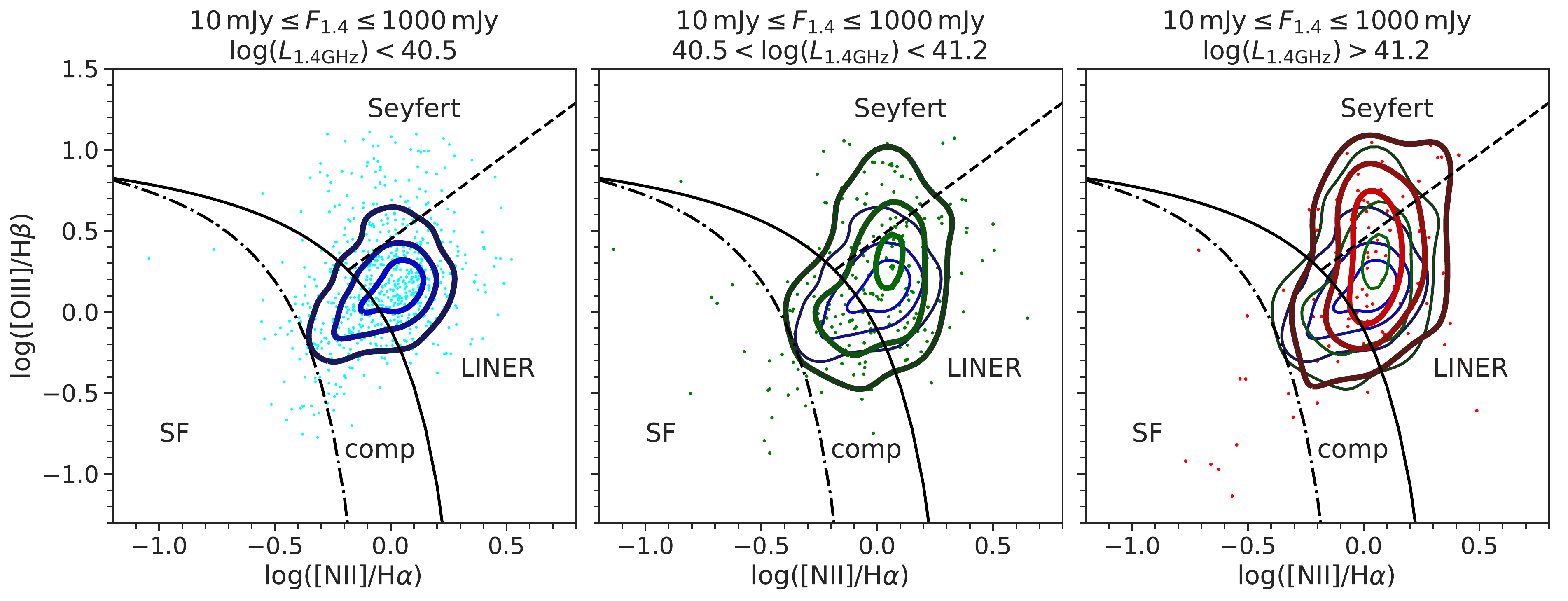}
  \includegraphics[width=\textwidth]{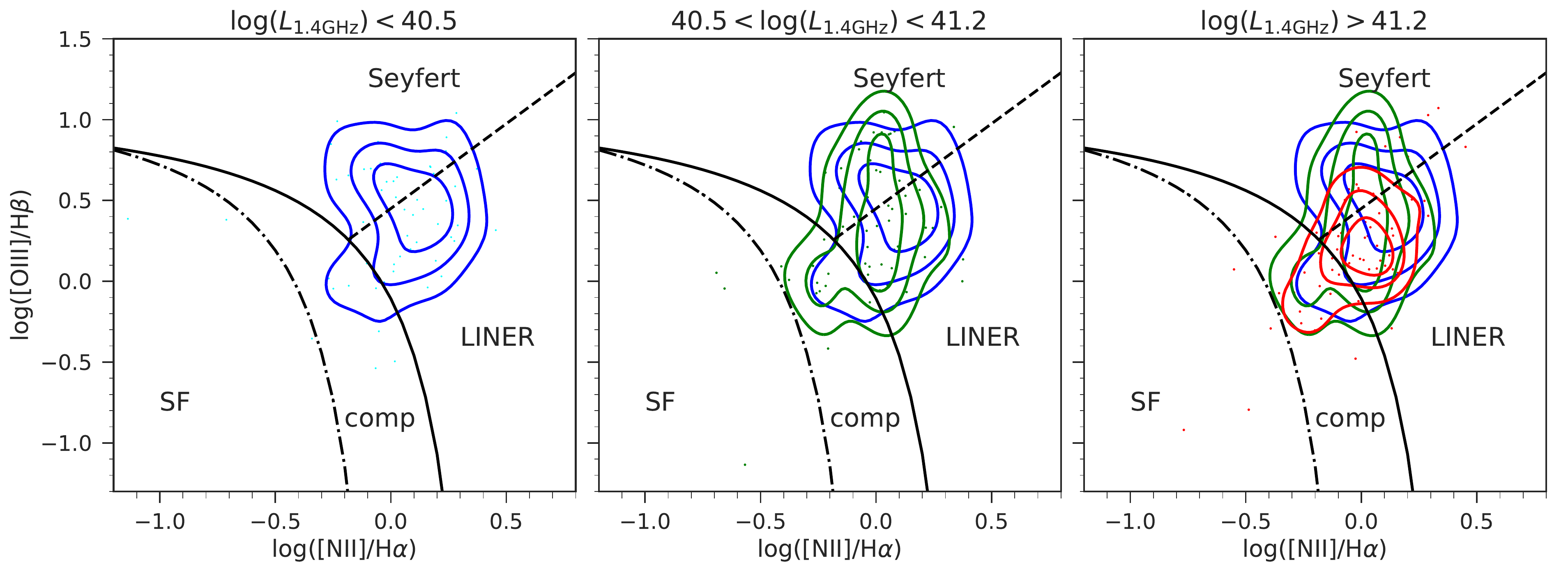}
  \caption{Distribution of the radio luminosity at $1.4$ GHz, $L_{\rm 1.4GHz}$, in optical diagnostic diagrams. \textbf{Top row:} The distribution of the radio luminosity $L_{\rm 1.4GHz}$ for the parent sample. The contours correspond to the distribution of the sources for the luminosity bins, $\log{(L_{\rm 1.4GHz})}<40.5$, $40.5<\log{(L_{\rm 1.4GHz})}<41.2$, and $\log{(L_{\rm 1.4GHz})}>41.2$, with the increasing radio luminosity from  left to right. \textbf{Middle row:} The distribution of the radio luminosity $L_{\rm 1.4GHz}$ as in the top row but for a flux-limited subsample of the parent sample, $10\,{\rm mJy} \leq F_{\rm 1.4GHz} \leq 1000\,{\rm mJy}$. The luminosity bins are the same as in the top row. \textbf{Bottom row:} The distribution of the radio luminosity $L_{\rm 1.4GHz}$ for the Effelsberg sample only (low$+$high flux sources). The luminosity bins are the same as in the top row.}
  \label{fig_1_4_luminosity}
\end{figure*}


The spectral-index distributions for the high-flux density
sample are dominated by sources with low spectral indices as can be seen in the left column of Fig.~\ref{fig_new-histograms} where we denote high-flux sources as HF and low-flux sources as LF.
The $\alpha_{1.4/4.85}$ index distribution has a median of
-0.53 and a median width of 0.50\footnote{For the
uncertainty in the median we quote the median deviation
from the median or twice the value if we refer to it as the width.}.
Towards higher frequencies the index even drops
and the distribution becomes narrower.
The $\alpha_{4.85/10.45}$ index has a median of
-0.67 and a median width of 0.20.
However, the distributions have weak tails towards
flatter spectra, indicating that several sources
contain flat spectrum components at a flux density level
lower than the fluxes for the steep components.

Therefore, the situation changes for the low-flux density sample.
The flux densities are lower now and for more sources they are
close to the typical fluxes of the flat components.
Hence, the portion of the sources with flat spectral
index increases leading to a more prominent shoulder
towards the flat side of the distributions (see the middle and right columns of Fig.~\ref{fig_new-histograms}).
This is very pronounced for the $\alpha_{1.4/4.85}$
index. For about 2/3 of the sources (i.e. 89 out of 289)
the 10.45~GHz flux drops below the detection limit.
Correspondingly, the $\alpha_{4.85/10.45}$
distribution is biased towards flat-spectrum sources.
For the low-flux density sample, the $\alpha_{1.4/4.85}$
index distribution has a median of -0.26 and a median width of 0.33.
The distribution is highly skewed to the steep side
and has a peak at -0.6.
The $\alpha_{4.85/10.45}$
index distribution has median of -0.25 and a width of 0.86, with a
similar but more pronounced 0.3 wide secondary peak
around $\alpha_{4.85/10.45}$=-0.10.
Separate plots of the $\alpha_{1.4/4.85}$
index for the sources with (median -0.11, median width 0.38)
and without (median -0.36, width 0.31) a 10.45~GHz measurement
show that the low-spectral index $\alpha_{1.4/4.85}$ of the 10.45~GHz detected
sources is also flatter by about 0.25.

The dropout in sources with  $\alpha_{4.85/10.45}$ measurements
in our low-flux sample is predominantly an effect of the flux sensitivity we
reached. From the uncertainties in Table~A1 we estimate a 3$\sigma$ flux density
limit of 12 mJy.
We find that about 70\% of all sources
without $\alpha_{4.85/10.45}$ measurements  indeed fall below this flux density limit
if one uses their low-frequency spectral index combined with the 4.85~GHz flux density
to predict their 10.45~GHz flux density.
A stronger spectral steeping towards 10.45~GHz and the effect of radio-source angular extension
may account for the remaining $30\%$. For the observations at 4.85 GHz, 64 sources out of the original 381 ($16.8\%$)  were extended at least in one direction of cross-scans (larger than the HPBW at 4.85 GHz$\sim 144''$). For the subsequent observations at 10.45 GHz of 256 selected sources, we had 50 sources ($19.5\%$) broader than the HPBW of $66''$ at least in one direction. We excluded the extended sources from  further analysis due to the fact that it was not possible to determine flux densities by Gaussian fitting to the combined cross-scan intensity profiles.

In terms of properties in the diagnostic diagrams, the sources with and without
$\alpha_{4.85/10.45}$ measurements do not differ significantly. There, median
$\log{([\ion{O}{iii}]/H\beta)}$ and $\log{([\ion{N}{ii}]/H\alpha)}$ values of 0.23$\pm$0.19 and -0.03$\pm$0.11 and
values of 0.22$\pm$0.22 and -0.09$\pm$0.14 are reached, respectively.
However, as in the high-flux density sample, the trend that the sources with
flat-spectrum components show a higher excitation remains: for the sources with
$\alpha_{4.85/10.45}$ measurements, we find median
$\log{([\ion{O}{iii}]/H\beta)}$ and $\log{([\ion{N}{ii}]/H\alpha)}$ values of 0.26$\pm$0.19 and -0.03$\pm$0.11, and for the
sources without $\alpha_{4.85/10.45}$ measurements, we find
$\log{([\ion{O}{iii}]/H\beta)}$ and $\log{([\ion{N}{ii}]/H\alpha)}$ values of 0.19$\pm$0.26 and -0.10$\pm$0.15, respectively.

\section{Trends of the spectral index in optical diagnostic diagrams}
\label{sec_trends}

While in the previous section we looked for trends in the radio spectral index between different activity classes as traced by the optical diagnostic diagram, we will pursue the reverse way and investigate how different radio spectral indices are reflected in the optical diagnostic diagram.
Considering the radio luminosity of the sources at $1.4\,{\rm GHz}$, $L_{\rm 1.4GHz}$\footnote{The integrated flux at $\nu_0 = 1.4\,\mathrm{GHz}$ in Jansky is derived from the FIRST survey. We derive the luminosity distance, $D_L$, from the redshift, using a standard cosmology with $H_0=70\,\mathrm{km}\,\mathrm{s}^{-1}\,\mathrm{Mpc}^{-1}$, $\Omega_m = 0.3$ and $\Omega_\Lambda = 0.7$. The luminosity is then given by $L_{1.4\,\mathrm{GHz}} = 4\pi D_L^2 \, \nu_0 f_{\nu_0}$.}, expressed in ${\rm erg\,s^{-1}}$ (which traces better the 
radio-core emission due to a smaller beam size of the VLA), in the parent sample the general trend is that the radio luminosity increases from the star-forming, through the composites, LINERs, and up to the Seyfert sources (Fig.~\ref{fig_1_4_luminosity}, top row). The distribution of radio luminosities implies that our source selection lacks 
metal-rich star-forming galaxies with radio luminosities below $\log{(L_{\rm 1.4GHz}/{\rm erg\,  s^{-1}})}<40.5$, and as redshift increases, systematically fainter sources in all luminosity bins.  Hence, our radio measurements and analysis mostly trace nearby luminous, active galaxies in the composite--LINER--Seyfert regions. This is demonstrated in Fig.~\ref{fig_1_4_luminosity}, where we first show the subsample of the parent sample with the integrated flux density at $1.4\,{\rm GHz}$ greater than $10\,{\rm mJy}$ and less than $1000\,{\rm mJy}$ (middle row). For the sources detected at $4.85\,{\rm GHz}$ as well as $10.45\,{\rm GHz}$ (low$+$high sample, 209 sources; see the bottom row), we see distribution peaks in the Seyfert-LINER part of the optical diagram as for the flux-limited subsample for all luminosity bins, with the apparent loss of radio sources in the composite and star-forming parts in the lowest luminosity bin, $\log{(L_{\rm 1.4GHz})}<40.5$, due to source drop-outs. 


\begin{figure*}[h!]
  \centering
  \includegraphics[width=\textwidth]{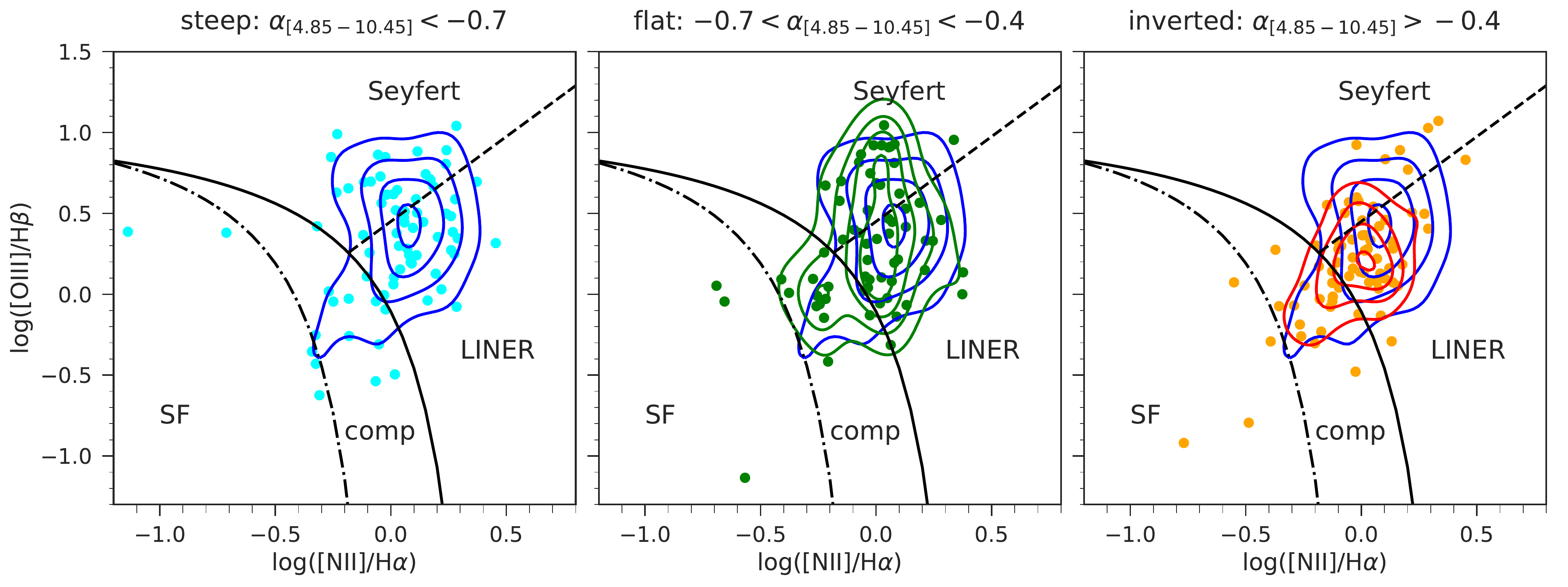}
  \includegraphics[width=\textwidth]{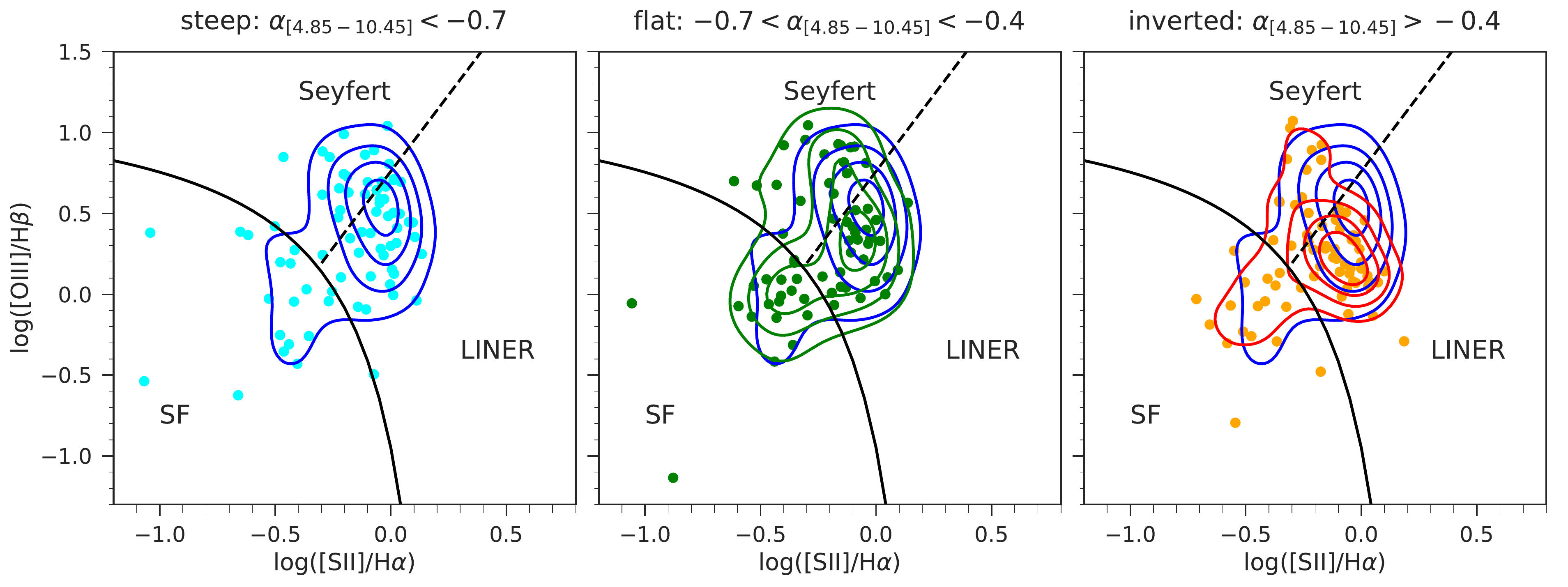}
   \includegraphics[width=\textwidth]{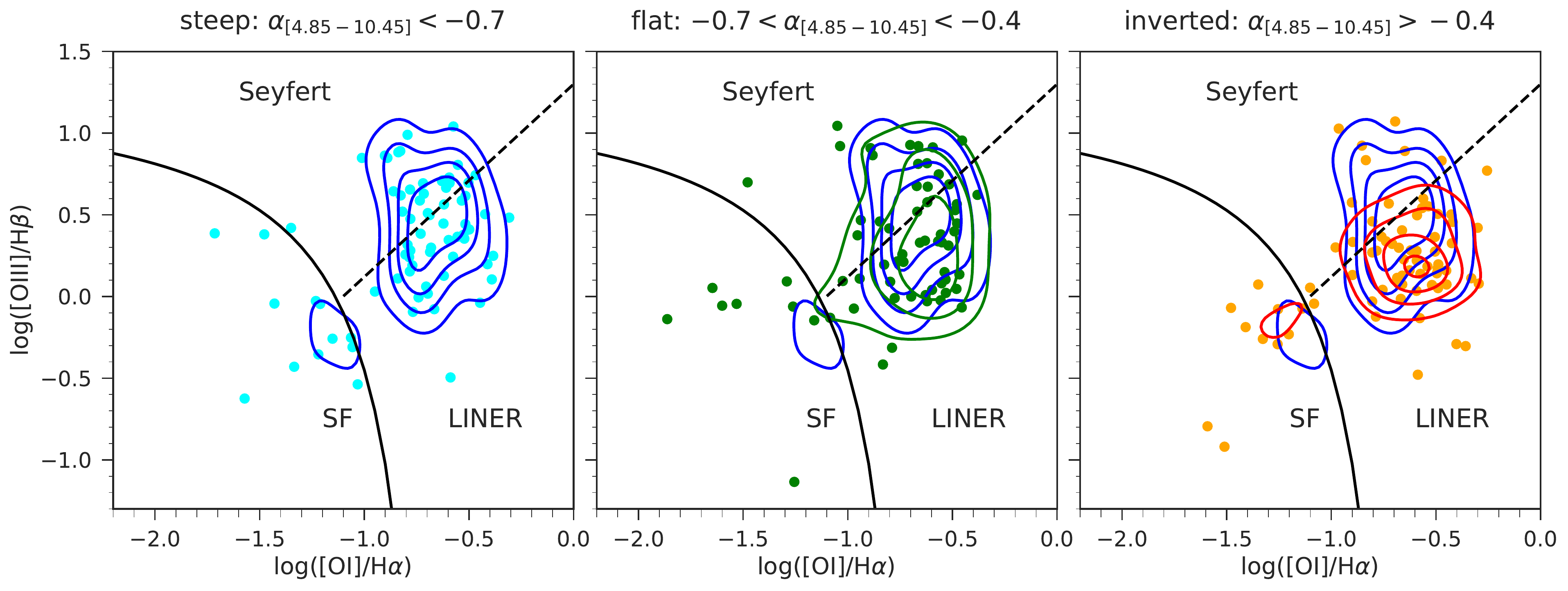}
  \caption{Spectral index trends in the optical diagnostic diagrams. From the top to the bottom panels: [NII]-, [SII]-, and [OI]-based diagrams, respectively, with the progressively increasing spectral index from the left to the right panels: $\alpha_{[4.85-10.45]}<-0.7$, $-0.7\leq\alpha_{[4.85-10.45]}\leq-0.4$, and $\alpha_{[4.85-10.45]}>-0.4$, respectively. Contours indicate Gaussian kernel density estimates.}
  \label{fig:BPT-index-reverse}
\end{figure*}



Figure~\ref{fig:BPT-index-reverse} shows the position of the sources in all three classical optical diagnostic diagrams, binned by the radio spectral index. Using the previous definitions, we distinguish between steep ($\alpha_{[4.85-10.45]} > -0.7$), flat ($-0.7 < \alpha_{[4.85-10.45]} < -0.4$), and inverted radio spectra ($\alpha_{[4.85-10.45]} < -0.4$). The plots show that when going from steep via flat to inverted spectra, the optical line ratios, in particular the ratio $\log([\ion{O}{iii}]/\mathrm{H}\beta)$, decrease. This indicates that the ionization potential of sources with inverted radio spectra is weaker than that of sources with a steep radio spectrum.

\begin{figure}[h!]
  \centering
  \includegraphics[width=0.5\textwidth]{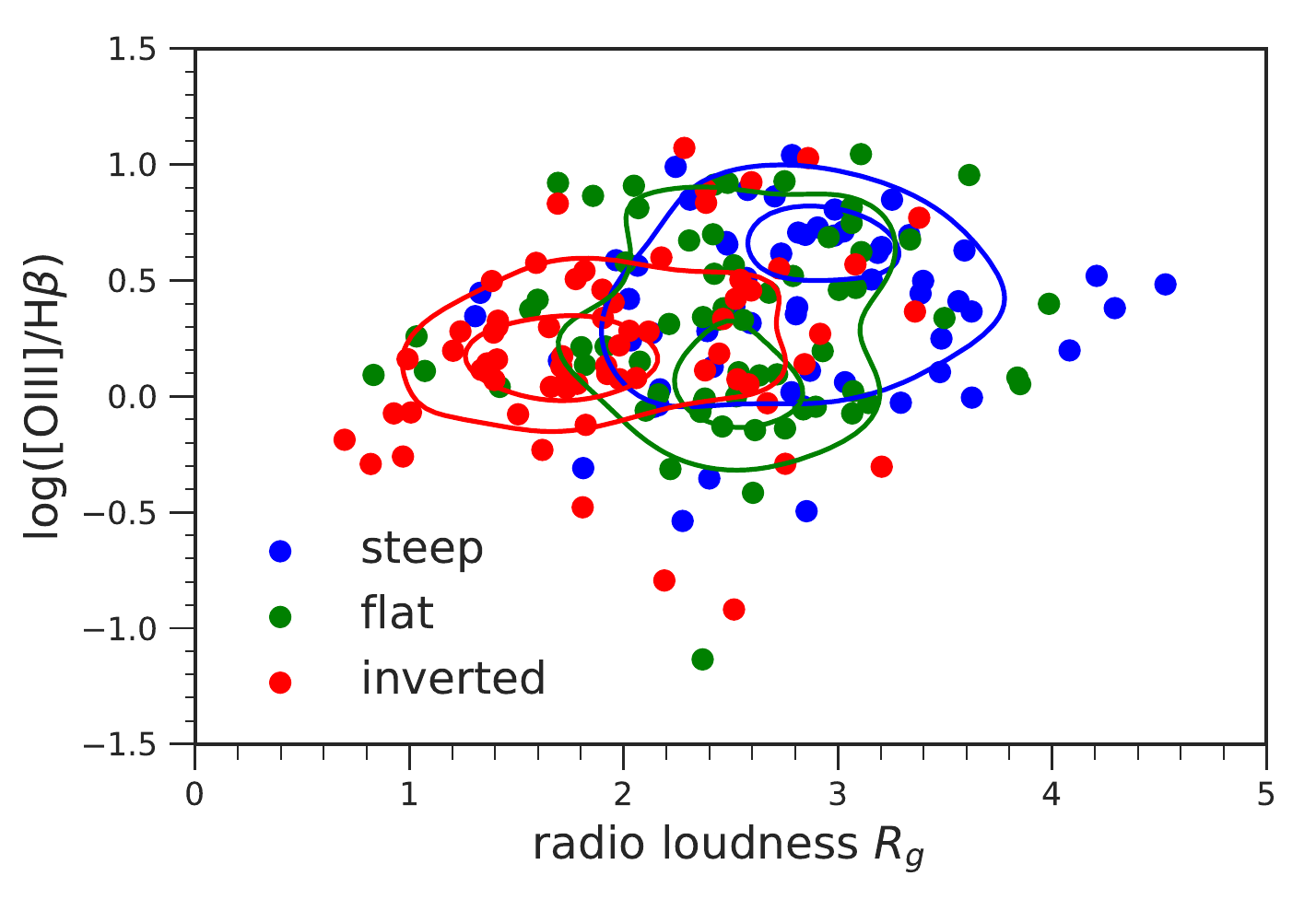}
  \caption{Clustering of the radio sources in the radio loudness--ionization ratio plane. The galaxies are grouped with respect to the spectral index in the following categories: $\alpha_{[4.85-10.45]}<-0.7$, $-0.7\leq\alpha_{[4.85-10.45]}\leq-0.4$, and $\alpha_{[4.85-10.45]}>-0.4$.}
  \label{fig_loudness_oiii}
\end{figure}

Since this vertical movement in the optical diagnostic diagram follows the same direction as the usual division line between Seyfert and LINER galaxies \citep{2006MNRAS.372..961K,2007MNRAS.382.1415S}, this trend will lead to the previously discussed increase of radio spectral index towards LINER sources, namely that steeper sources tend to fall into the Seyfert and more inverted into the LINER category. 
Since we select brighter radio emitters with optical counterparts, it is quite possible that all our sources have a contribution from an AGN to 
some extent. 


In the further search for general trends, we added another parameter, radio loudness, which can directly trace the energetics of AGN and their hosts rather than the purely astronomical division into low- and high-flux sources. Using the flux density at $20\,{\rm cm}$ from FIRST , $F_{1.4}$, we converted the radio flux density into the $AB_{\nu}$-radio magnitude system of \citet{1983ApJ...266..713O} according to \citet{2002AJ....124.2364I},

\begin{equation}
  m_{1.4}=-2.5\log{\left(\frac{F_{1.4}}{3631\,{\rm Jy}}\right)}\,,
  \label{eq_AB_system}
\end{equation}
in which the zero point $3631\,{\rm Jy}$ does not depend on the wavelength. Subsequently, the radio loudness can be calculated as the ratio of the radio flux density to the optical flux density,

\begin{equation}
   R_{\rm g}\equiv \log{\left(\frac{F_{\rm radio}}{F_{\rm optical}} \right)}=0.4(g-m_{1.4})\,,
   \label{eq_radio_loudness}
\end{equation}
where we use the optical $g$-band for each source from the SDSS-DR7 catalogue.

\begin{figure*}[h!]
  \centering
  \includegraphics[width=0.45\textwidth]{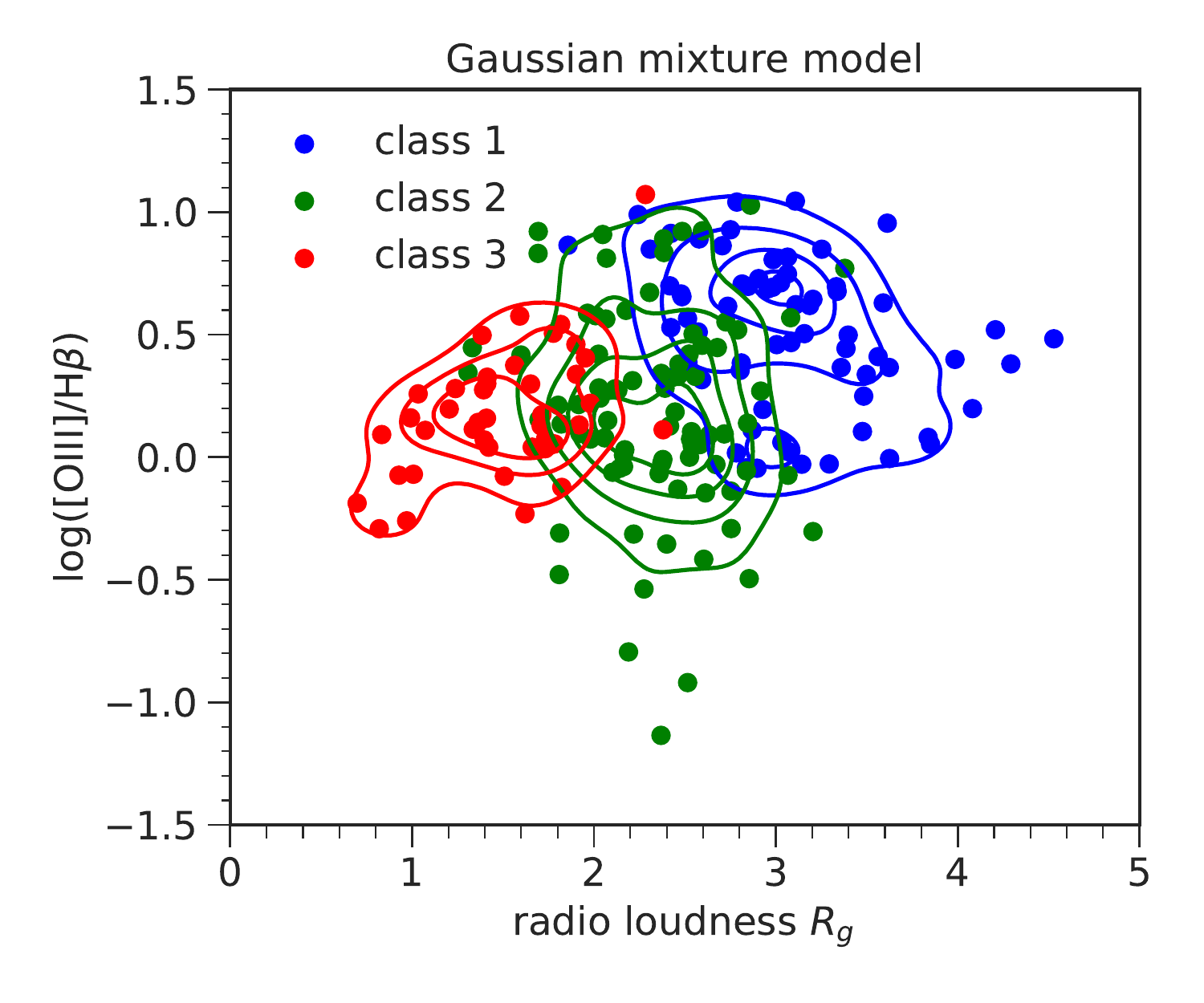}
  \includegraphics[width=0.45\textwidth]{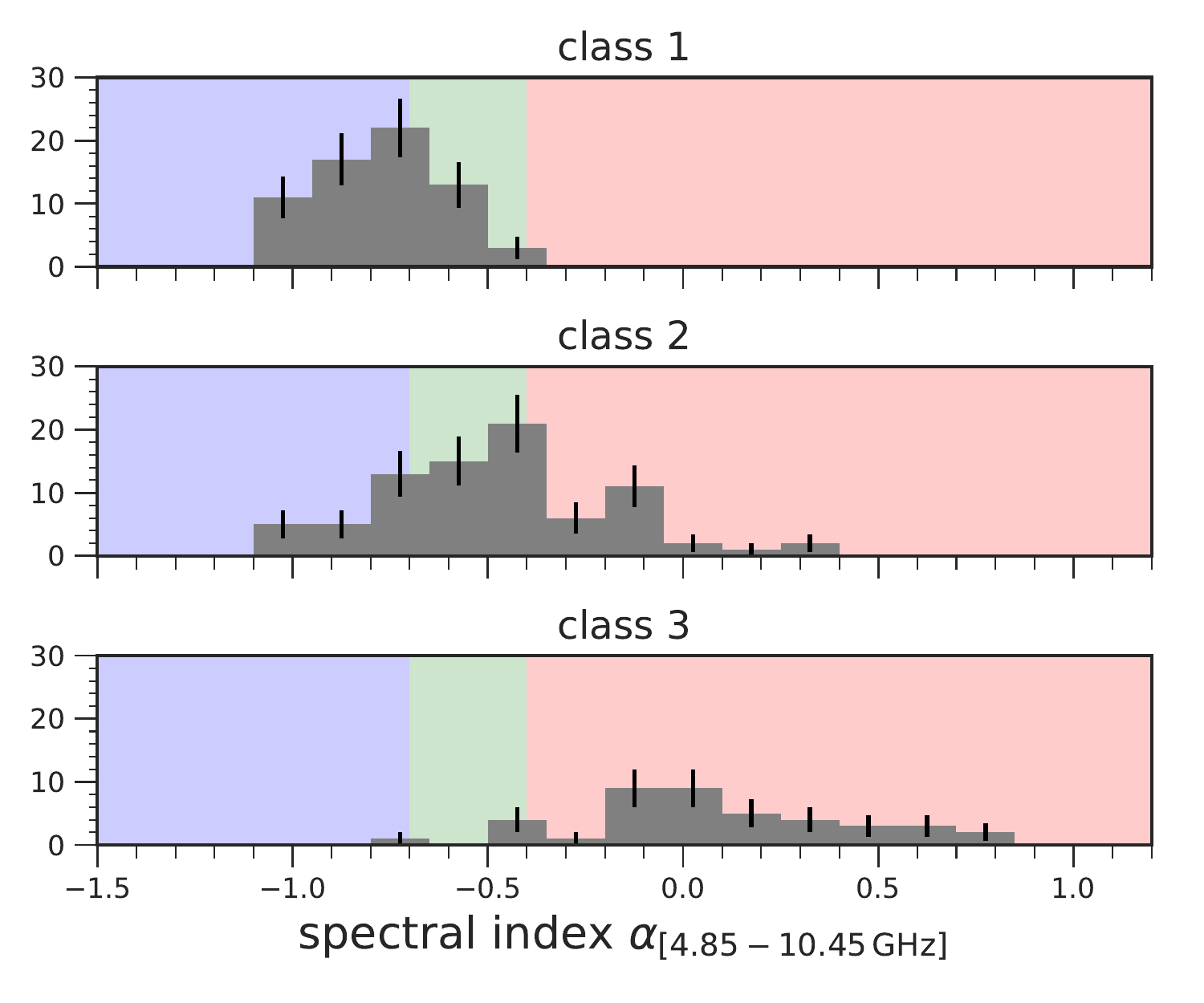}
  \caption{Results of the Gaussian mixture model. \textbf{Left panel:} Radio sources in the radio loudness--ionization ratio plane according to the spectral index divisions as found from the Gaussian fitting. \textbf{Right panel:} The spectral-index histogram with three basic groups that are very similar to the manual cut with these limits: $\alpha_{[4.85-10.45]}<-0.7$, $-0.7\leq\alpha_{[4.85-10.45]}\leq-0.4$, and $\alpha_{[4.85-10.45]}>-0.4$.}
  \label{fig_gaussian_mixture}
\end{figure*}

Here we note that the optical flux density is related to the host galaxy and not to the AGN, since our sample is radio selected. An inspection of the optical spectra also shows that only a handful of sources display the typical power-law continuum shape that is associated to the accretion disk. 
Hence, the radio loudness derived in this way expresses the ratio between the radio power of the AGN to the optical emission of the host galaxy and can be taken as an upper limit of the intrinsic AGN radio loudness. 

Figure~\ref{fig_loudness_oiii} shows the location of the three radio classes, with steep, flat, and inverted radio spectra in the $R_g - \log([\ion{O}{iii}]/\mathrm{H}\beta)$ plane where the radio loudness $R_{\rm g}$ is along the $x$-axis and the low-ionization ratio $\log([\ion{O}{iii}]/\mathrm{H}\beta)$ is along the $y$-axis. We show that sources cluster in well-discriminated regions.

To justify these spectral-index categories, we fit a Gaussian mixture model (GMM), which is an unsupervised machine learning algorithm. For this, we consider the quantities radio loudness, $R_g$, low-ionization ratio, $\log([\ion{O}{iii}]/\mathrm{H}\beta)$, and radio spectral index, $\alpha_{[4.85-10.45]}$ , as a three-dimensional space; this means every galaxy is represented by a vector
\begin{equation}
\vec{x} = \begin{pmatrix}
    R_g \\
    \log([\ion{O}{iii}]/\mathrm{H}\beta) \\
    \alpha_{[4.85-10.45]}.
\end{pmatrix}
.\end{equation}

The GMM assumes that the data points can be described by a superposition of a finite number of multivariate Gaussian distributions with unknown parameters in this parameter space. The model is the probability density function represented as a weighted sum of the Gaussian component densities. Using the model, we can then give probabilities for data points belonging to one of these classes. For the fit, we assume three components and use the expectation maximization technique implemented in the \textsc{Python} library \textsc{Scikit-Learn} \citep{scikit-learn}.

In Fig.~\ref{fig_gaussian_mixture}, we show the three classes in the $R_g - \log([\ion{O}{iii}]/\mathrm{H}\beta)$ plane and, to represent the third dimension, we show histograms of the radio spectral index $\alpha_{[4.85-10.45]}$. In accordance with our results based on a manual cut in radio spectral index (Fig.~\ref{fig_loudness_oiii}), we find three distinct classes:
\begin{itemize}
\item[(1)] associated with a steep radio index, high ionization ratio, and high radio loudness;
\item[(2)] associated with a flat radio index, lower ionization ratio, and intermediate radio loudness;
\item[(3)] associated with an inverted radio index, low ionization ratio, and low radio loudness.
\end{itemize}

\begin{figure*}[h!]
  \centering
  \includegraphics[width=\textwidth]{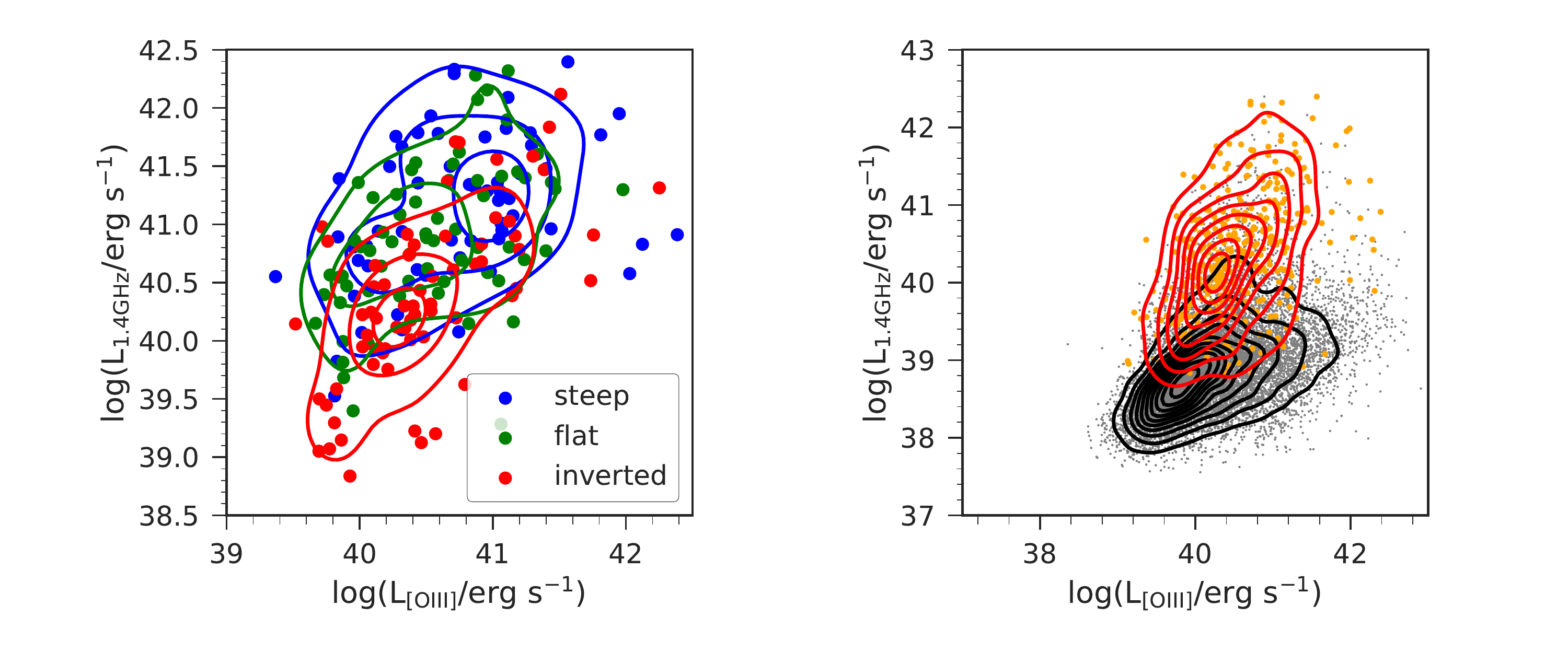}
   \includegraphics[width=\textwidth]{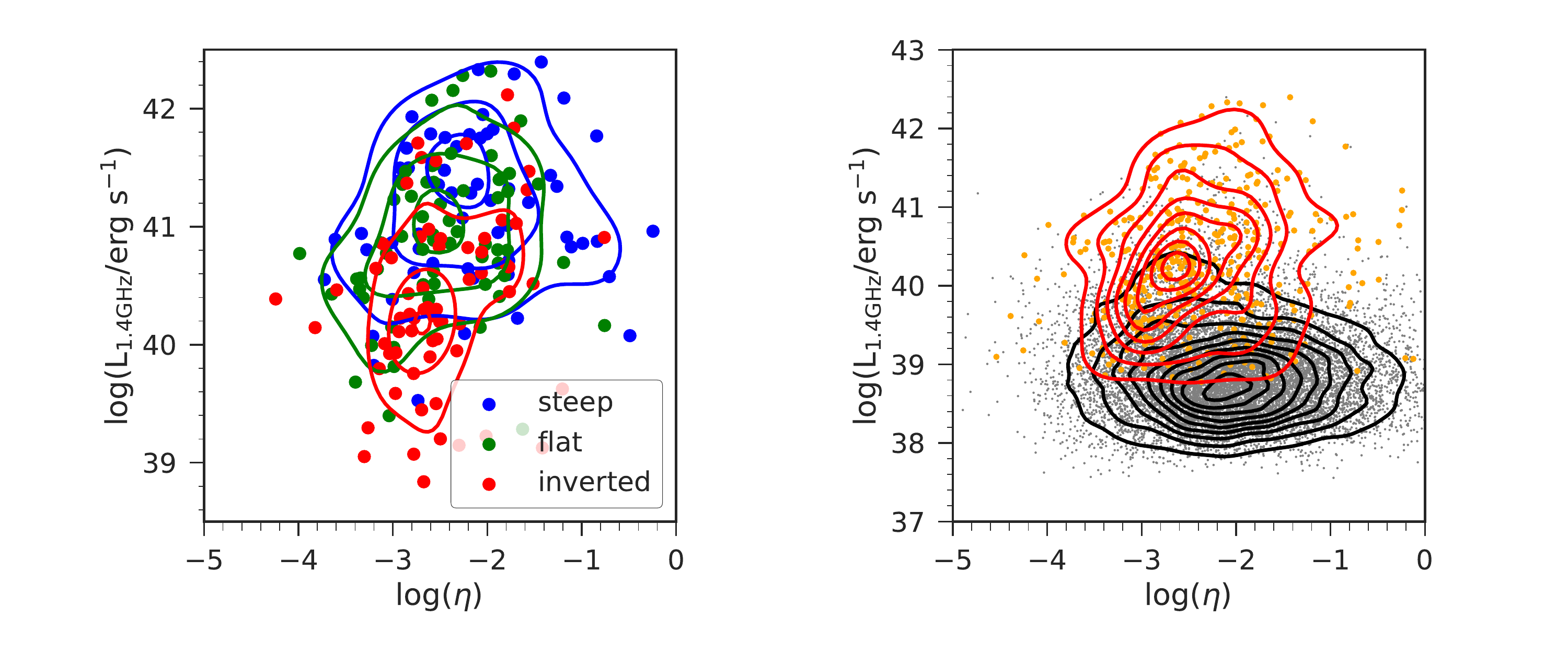}
  \caption{Distribution of the Effelsberg sources with respect to the radio luminosity, luminosity of [$\ion{O}{iii}$] line, and the Eddington ratio. \textbf{Top row}--Left panel: The distribution of the radio spectral index $\alpha_{[4.85-10.45]}$ with respect to the radio luminosity $L_{\rm 1.4GHz}$ and the luminosity of [$\ion{O}{iii}$] line. Right panel: The localization of the Effelsberg sources (low$+$high flux; orange points) in the plane of $L_{\rm 1.4GHz}$ and $L_{[\ion{O}{iii}]}$ with respect to the parent sample (grey points). \textbf{Bottom row}--Left panel: The distribution of steep, flat, and inverted radio spectral indices with respect to the radio luminosity $L_{\rm 1.4GHz}$ and the Eddington ratio $\eta$. Right panel: The distribution of the Effelsberg sources (orange points) in the same plane as in the left figure with respect to the parent sample (grey points).}
  \label{fig_1_4_luminosity_OIIIlum}
\end{figure*}

\section{Interpretation of the results}
\label{interpretation}

 \subsection{General trends in radio-optical properties}

By measuring flux densities with the Effelsberg radio telescope, calculating spectral indices, and analysing their distributions in the optical diagnostic diagrams, we can recover these basic trends in the radio-optical properties of our selected sources,
\begin{itemize}
  \item[(a)] the radio luminosity increases in the direction of increasing low-ionization ratio [$\ion{O}{iii}$]/H$\beta$ (see Fig.~\ref{fig_1_4_luminosity}) with the exception of the LINERs, which show lower radio luminosities and low ionization ratios (plus higher stellar masses) compared to the Seyferts with high radio luminosities;
  \item[(b)] there is a trend of the radio spectral index steepening in the direction of increasing [$\ion{O}{iii}$]/H$\beta$ (see Fig.~\ref{fig:BPT-index-reverse});
  \item[(c)] the radio loudness increases in the same direction, as shown by Fig.~\ref{fig_gaussian_mixture}.
\end{itemize}

Our results are representative for the nearby luminous, active SDSS-FIRST sources, which are predominantly located in the AGN (Seyfert-LINER) region of the optical diagnostic diagram. In comparison with previous studies, the determination of spectral indices allows us to connect the radio luminosity and radio-loudness trends \citep{1989AJ.....98.1195K,2007ApJ...658..815S} with the radio-morphological structures, such as the activity of radio primary components (cores), jet components, and radio lobes, as is known from the studies of quasar and blazar studies \citep{1986A&A...168...17E}. 

Considering the bolometric luminosities of AGN sources, for which the luminosity of the emission line [$\ion{O}{iii}$] serves as a proxy, there is a trend of less radio-luminous galaxies being located towards lower $L_{[\ion{O}{iii}]}$ and these sources have progressively flatter to inverted radio spectra (see the top row of Fig.~\ref{fig_1_4_luminosity_OIIIlum}, left panel). On the other hand, more radio-luminous sources have larger $L_{[\ion{O}{iii}]}$ and steeper spectra.

To analyse the trends (a), (b), and (c) in relation with the accretion rate $\dot{M}_{\rm acc}$, we calculate the Eddington ratio, defined as $\eta \equiv L_{\rm bol}/L_{\rm Edd}$, where the bolometric luminosity is derived from the [$\ion{O}{iii}$] luminosity using $L_{\rm bol} = 3500\, L_{[\ion{O}{iii}]}$ \citep{2004ApJ...613..109H} and the Eddington luminosity is $L_{\rm Edd}=4\pi GM_{\bullet}m_{\rm p}c/\sigma_{\rm T}=1.3 \times 10^{38} (M_{\bullet}/M_{\odot})\,{\rm erg\,s^{-1}}$. The black-hole masses for SDSS-FIRST sources can be estimated from the black-hole mass--velocity dispersion $M_{\bullet}-\sigma_{\star}$ correlation, using the relation found by \cite{2009ApJ...698..198G},

\begin{equation}
 \log{(M_{\bullet}/M_{\odot})}=8.12+4.24\log{(\sigma_{\star}/200\,{\rm km\,s^{-1}})}\,.
 \label{eq_mass_dispersion}
\end{equation}
In the bottom row of Fig.~\ref{fig_1_4_luminosity_OIIIlum} (left panel), we depict the distribution of the spectral index $\alpha_{[4.85-10.45]}$ with respect to the radio luminosity $L_{\rm 1.4GHz}$ and the Eddington ratio $\eta$. All sources irrespective of their radio spectral index have very similar Eddington ratios, with $2/3$ of them within $\log{\eta}\sim[-3,-2]$. This similarity in accretion rates is driven by the strong dependency of $\eta$ on the stellar-velocity dispersion ($\eta \propto \sigma^4$) and the characteristics of our sample: the brightest radio galaxies with optical counterparts (dominated by the stellar emission) on the sky are -- mostly -- very massive, $M_{\star} \sim 10^{11-12}\, M_{\odot}$, have a large velocity dispersion, $\sigma_{\star} \sim 180-320\,{\rm km s^{-1}}$, and thereby have a heavy SMBH with $\log(M_{\bullet}/{M_{\odot}}) \sim 8.5$.

\subsection{Relation between radio emission and Eddington ratio}

Our results seem to be at odds with previous findings that the radio loudness anti-correlates with the Eddington ratio \citep{2007ApJ...658..815S,2011MNRAS.417..184B}. Since the ionization ratio $\log{[\ion{O}{iii}]/{\rm H\beta})}$ is proportional to the hardness of the ionization field, which in turn depends on the accretion efficiency expressed by the Eddington ratio, we find that sources with a lower radio loudness correspond to those with lower ionization ratio and their radio spectral indices are inverted, which is indicative of self-absorbed synchrotron emission (see Fig.~\ref{fig_gaussian_mixture}). On the other hand, the radio-louder sources are associated with larger ionization ratio and their spectral indices demonstrate optically thin synchrotron emission.

Although our sample spans shorter ranges in $L_{\rm 1.4GHz}$ and $\log{\eta}$ when compared with extended samples like that of \citet{2007ApJ...658..815S}, it displays similar luminosities to their broad line radio galaxies (BLRGs) and radio loud quasars (RLQs) when using only core radio powers \citep[see][]{2011MNRAS.417..184B} indicating that our sample might represent the narrow line optical counterpart of those objects.

To estimate how our radio-loudness definition and sample selection bias our results, we study integral quantities and compare the sources selected for the Effelsberg observations to the parent sample. One main source of discrepancy could be the use of the host-galaxy optical luminosity in the radio-loudness calculation (see Eq.~\eqref{eq_radio_loudness})\footnote{We refrain from using $L_{[\ion{O}{iii}]}$ to estimate the optical AGN emission to avoid obtaining an artificial trend by comparing $R \propto 1/L_{[\ion{O}{iii}]}$ with $\eta \propto L_{[\ion{O}{iii}]}$}. However, galaxy hosts of Effelsberg sources seem to be quite similar; $\gtrsim 90\%$ of them have elliptical morphologies spanning only around 0.5dex in $g$-band luminosities. Therefore we conclude that the main trend of decreasing radio spectral index with radio loudness is caused by the variations in the core radio luminosities.

The basic explanation of the trend of decreasing spectral indices $\alpha_{[4.85-10.45]}$ with the increasing ionization ratio [$\ion{O}{iii}$]/H$\beta$, which corresponds to the LINER--Seyfert transition in the optical diagnostic diagrams, might be the renewal of AGN activity in the past $\sim 10^5$--$10^7$ years \citep{2016A&ARv..24...10T,2017A&ARv..25....2P}. At least one third of the sources that have larger [$\ion{O}{iii}$]/H$\beta$ and steeper radio spectral indices have extended jet structures, that is, radio emission is  dominated by older radio lobes, where electrons cooled down. Since the timescales for the formation of the radio extended structures and the optical narrow-line region brightening are quite different, the observed trend can be explained by two scenarios: objects in the  Seyfert region must have been optically-active for a long enough period for the radio lobes to develop, or the radio structures formed in the past and their activity has been re-triggered some decades ago.

On the other hand, only $1/8$ of the sources with lower ionization ratios display jet-like structures. They might have started (or re-started) their nuclear activity very recently, which could explain inverted spectral indices corresponding to compact self-absorbed core emission. Since they did not have enough time to develop extended radio lobes, their radio luminosities are also smaller.

Visual inspection of FIRST images and literature research for each object in the Effelsberg sample allowed us to classify them as jetted or non-jetted sources. We find that 2/3 of flat and steep radio spectrum objects do not show extended structures that could be related to jet emission. Thus, we do not confirm the one-to-one relation between morphologies and spectral indices (compact/extended versus inverted/flat-steep) found by \citet{2011MNRAS.412..318M}.

Our results seem to be in accordance with the correlations found for radio-quiet Palomar-Green quasars by \citet{2019MNRAS.482.5513L}. They found an increase in the line ratio \ion{Fe}{ii}/H$\beta$ from inverted, through flat, up to steep radio spectral indices, in the same way as we found for [\ion{O}{iii}]/H$\beta$. They interpret the correlations of the radio spectral index with the Eigenvector 1 parameters using the nuclear-outflow interpretation, which at least partially can be used for our findings as well. Higher excitation for steep sources, which are also radio louder (see Fig.~\ref{fig_loudness_oiii}) could be indicative of a large-scale nuclear outflow, which is a source of optically thin synchrotron emission. On the other hand, sources with lower [\ion{O}{iii}]/H$\beta$ with flat to inverted radio slopes, which are radio weaker, could lack an outflow and the dominant source of radio emission would be the compact nucleus (coronal emission) that emits optically thick synchrotron emission.

\begin{figure*}[h!]
 \centering
 \includegraphics[width=0.45\textwidth]{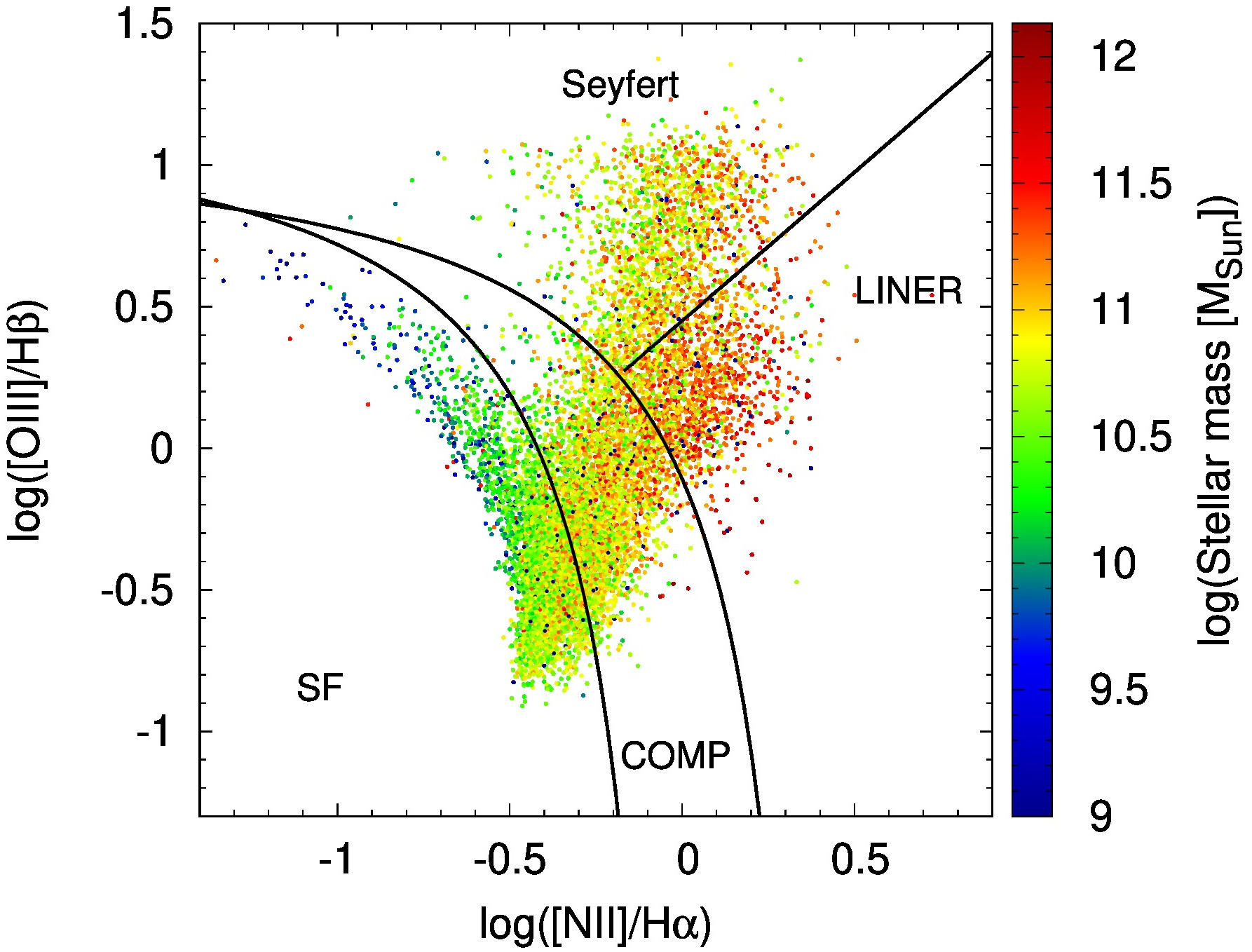}
 \includegraphics[width=0.45\textwidth]{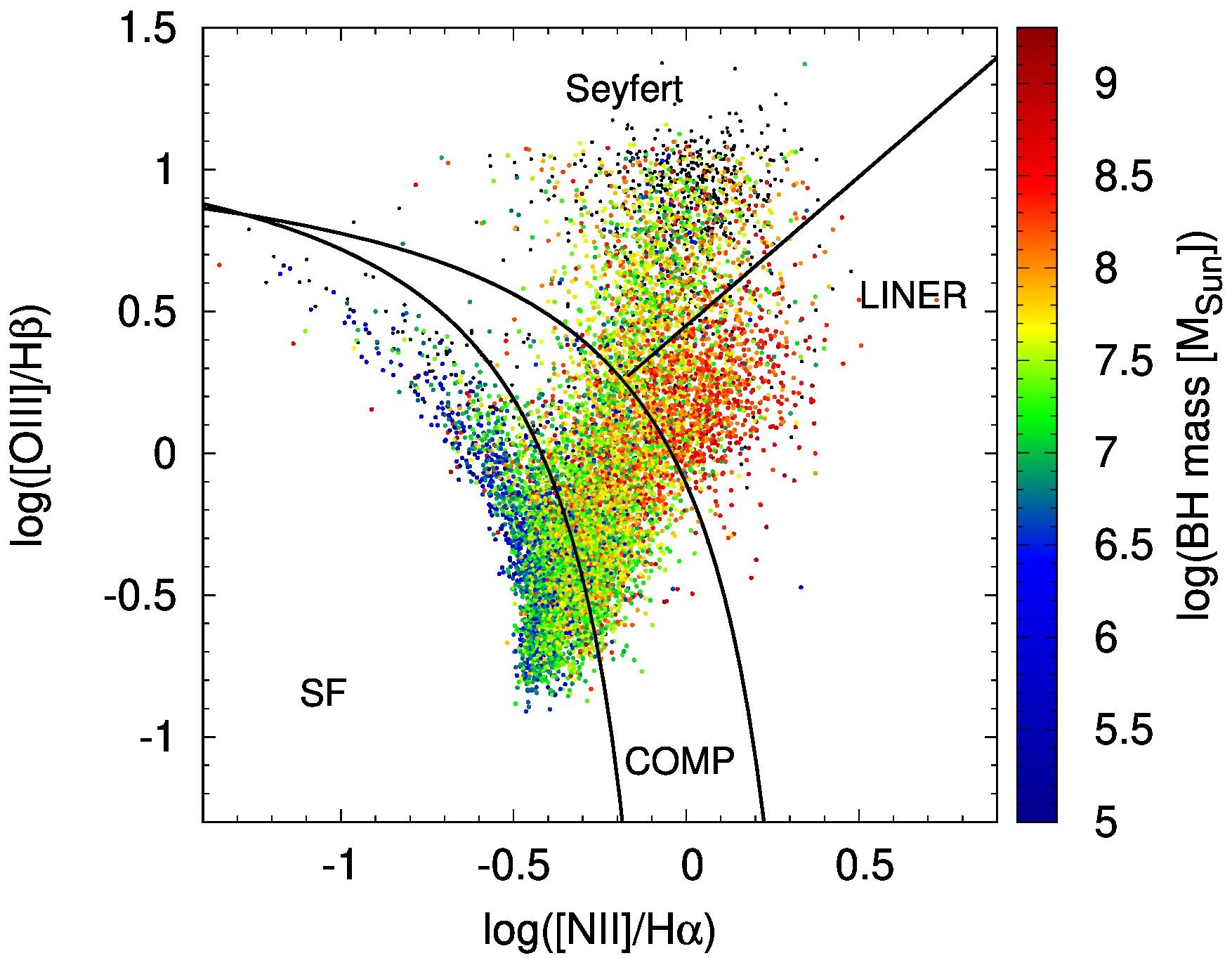}
 \caption{Distribution of stellar and black-hole masses for the SDSS-FIRST parent sample.                    
  \textbf{Left panel:} The SDSS-FIRST distribution of stellar masses in the optical BPT diagram. The colour bar indicates the logarithm of the stellar masses in units of solar masses. \textbf{Right panel:} The distribution of the black hole masses inferred from the measurement of the stellar velocity dispersion in the SDSS-FIRST sources.}
 \label{fig_stellar_bh_masses}
\end{figure*}

\begin{figure*}[h!]
 \centering
 \includegraphics[width=0.9\textwidth]{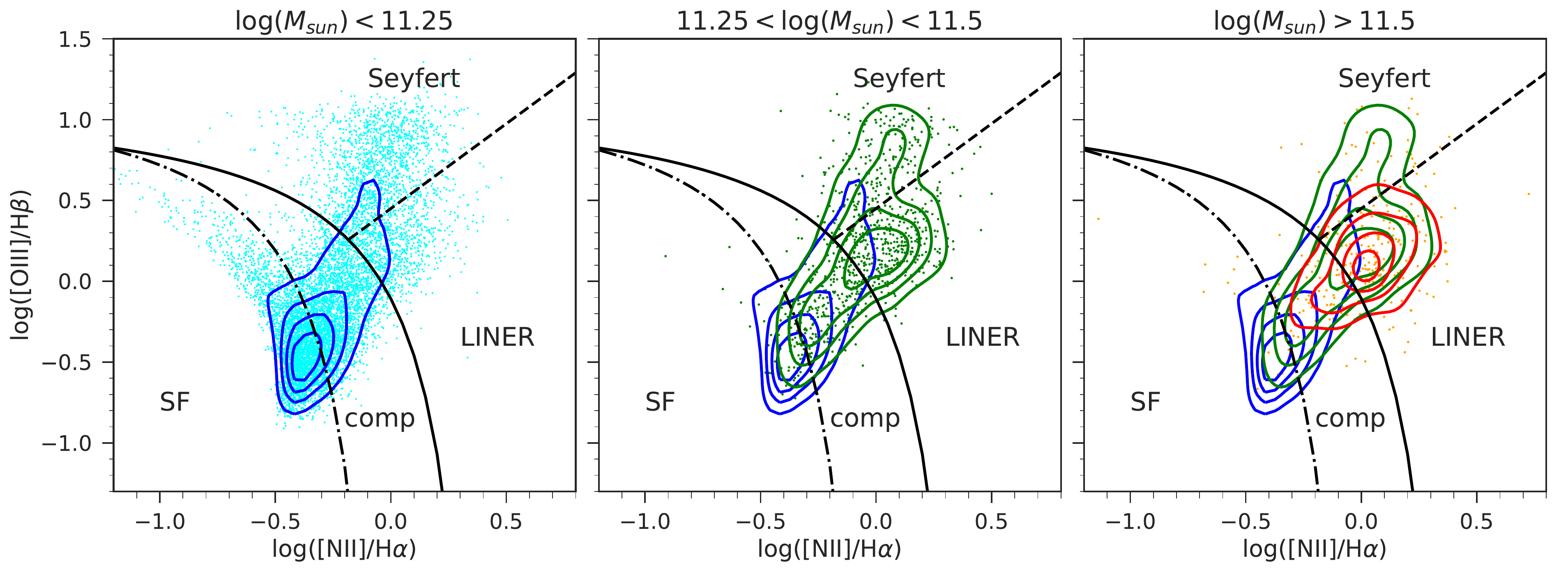}
 \includegraphics[width=0.9\textwidth]{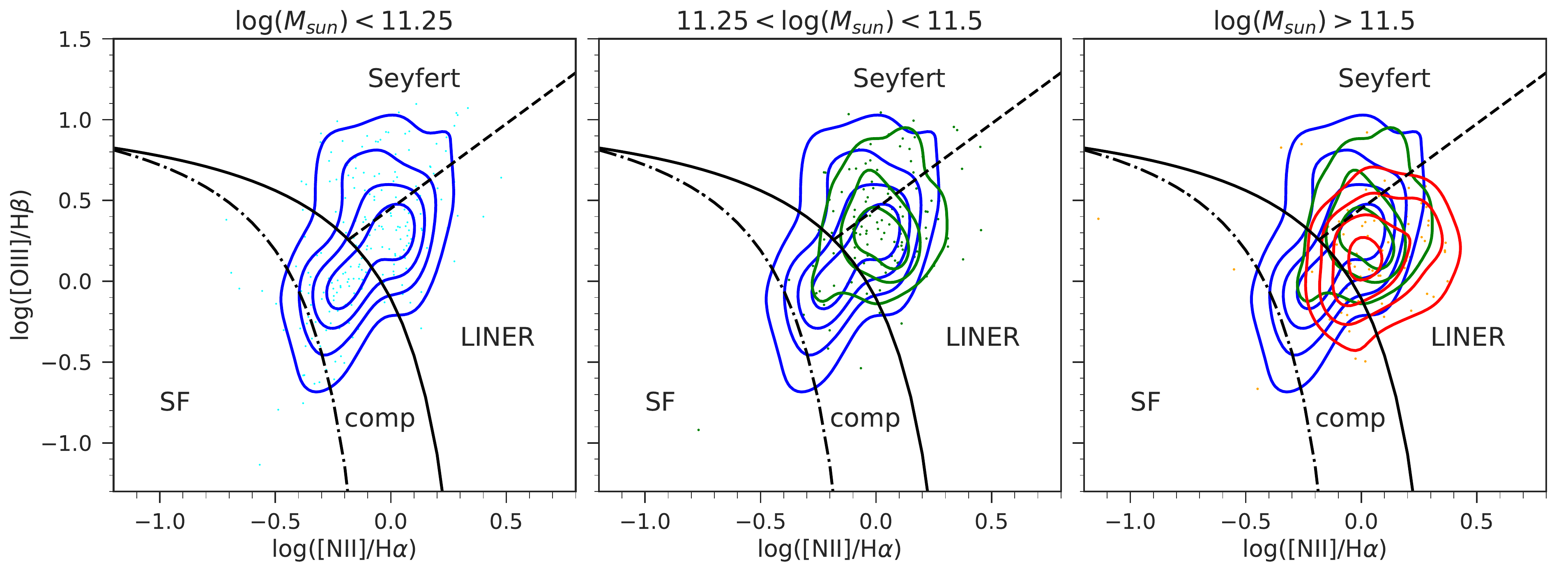}
 \caption{Distribution of the average stellar mass across the parent and the low$+$high flux sample.          
 \textbf{Top panel:} The SDSS-FIRST distribution of stellar masses in the optical BPT diagram for the whole parent sample. From the left to the right figures: galaxies with the average stellar mass in the range $\log(M_{\star}[M_{\odot}])<11.25$, $11.25<\log(M_{\star}[M_{\odot}])<11.5$, $\log(M_{\star}[M_{\odot}])>11.5$, respectively, are represented with density contours. \textbf{Bottom panel:} The distribution of the average stellar mass across the BPT diagram for the same mass bins as in the top panel, but for the combined low$+$high flux sample.}
 \label{fig_stellar_bh_masses}
\end{figure*}

\subsection{Flat and inverted spectral index towards LINERs: Implications for their character}
The interpretation and the nature of LINERs remains still unclear and several recent studies have attempted to shed more light on their characteristics and the potential effect of the environment \citep{2013A&A...558A..43S,2017MNRAS.467.3338C,2018arXiv180300946C}. The fact that $\gtrsim 50\%$ of LINERS in both frequency ranges have flat to inverted spectral indices points to the activity of the nucleus and the dominant contribution of the core and the jet radio emission to the overall radio emission of LINER galaxies.

In our study of the radio-optical properties of SDSS-FIRST sources, LINERs are characterized by a lower ionization ratio [$\ion{O}{iii}$]/H$\beta$ in comparison with Seyfert AGN sources. Keeping in mind the colour-stellar mass sequence \citep[][]{2009AIPC.1201...17S}, LINERs have the largest stellar and black hole masses (see Fig.~\ref{fig_stellar_bh_masses} for the distribution of stellar and black hole masses across the parent sample). In addition, they are associated with redder colours and smaller star formation rates than other optical spectral classes (SF, COMP, Seyfert)  \citep{2016MNRAS.455L..82L}.

The optical low-ionization emission could be explained by the presence of the extended population of hot and old post-asymptotic giant branch (AGB) stars, as is indicated by studying the radial brightness distribution by \citet{2013A&A...558A..43S}, which is consistent with a spatially extended ionizing source (stars) rather than a point source (nucleus). However, in the radio domain, the mean spectral indices $\overline{\alpha}_{[1.4-4.85]}$ and $\overline{\alpha}_{[4.85-10.45]}$ indicate the increase in the mean radio spectral slope in comparison with Seyferts (see also the trend of the increasing spectral index in Fig.~\ref{fig:BPT-index-reverse}). This trend can be explained by the dominant contribution of the jet emission (the overall flat radio spectrum due to the collective contribution of self-absorbed structures) and the nuclear activity (inverted, optically thick synchrotron source).

Therefore, we cannot support the claim of \citet{2013A&A...558A..43S} that LINERs are defined by the lack of AGN activity, at least not for our subsample of radio-emitting LINERs. The AGN ionizing field is likely not dominant in the optical domain, but according to our statistical study, the nuclear and jet activity must be present. This is also in agreement with the studies of \citet{2017MNRAS.467.3338C,2018arXiv180300946C} where they find that LINERs are redder and older than the control sample of galaxies in environments of different density. They also show that LINERs are likely to populate low-density regions in spite of their elliptical morphology, that is, their occurrence in low-density galaxy groups is two times higher than the occurrence of the control, non-LINER galaxies. The fact that LINERs are more likely to be found in low-density regions points towards nuclear activity, since active galaxies typically do not follow the morphology-density relation \citep{2006A&A...460L..23P,2009MNRAS.399...88C,2010MNRAS.409..936P,2014MNRAS.437.1199C}. This also indicates the relevance of major mergers in the galaxy evolution, since the low-density galaxy groups favour major mergers due to lower velocity dispersion among members. Major mergers can restart the nucleus as has been found in several studies \citep{2006A&A...460L..23P,2009MNRAS.399...88C,2007MNRAS.375.1017A}. In addition, \citet{2015ApJ...806..147C} found that major mergers are a trigger for radio-loud AGN and the launching of relativistic jets. Using the luminosity-hardness diagram, which is applied to stellar black hole binaries, \citet{2005A&A...435..521N} argue that LINERs seem to occupy a ``low/hard'' state (geometrically thick and optically thin, hot accretion flows, low Eddington ratio, launching collimated jets), while low-luminosity Seyfert sources are in a ``high" state (geometrically thin and optically thick, cold accretion discs, high Eddington ratios, incapable of launching collimated jets). In this picture, LINERs would be characterized by radiatively inefficient flows with recently (re)started nuclear and jet activity, which could explain their overall lower [OIII]/H$\beta$ ratio in the optical diagnostic diagrams and the increasing spectral index on the transition between Seyfert and LINER sources.

\section{Summary}
\label{conclusions}
We studied the radio-optical properties of selected cross-matched SDSS-FIRST sources, with a particular focus on the spectral index trends in the optical diagnostic diagrams. 
Combining the high-flux sample ($S_{1.4}\ge 100\,{\rm mJy}$; Vitale et al. 2015a) and the low-flux sample presented in this paper 
($10\,{\rm mJy} \leq S_{1.4} \leq 100\,{\rm mJy}$), we cover a total of 417 star-forming, composite, Seyfert, and LINER sources based on the standard spectral classification using the emission-line ratios. For a total of 209 sources (90 from the low-flux sample and 119 from the high-flux sample) we have flux density measurements at 10.45~GHz.

First, we searched for potential trends of the radial spectral index between the classical optical spectral classes of galaxies. Second, we looked at how the different ranges of the radio spectral index are positioned in the optical diagnostic diagrams. While the first approach yielded basic statistics, the second approach turned out to be more appropriate for our sample in the context of radio-optical trends.

We find a scenario that is largely consistent with models in which
the source population shows a dichotomy that reflects
a switch between radiatively efficient and radiatively inefficient accretion modes
at similarly (compared to QSOs and quasars) low accretion rates.
The location of radio sources in the narrow emission-line diagnostic (BPT) diagrams
shifts with the increasing importance of a radio-loud AGN away from 
galaxies dominated by radio emission powered by star formation.
Hence, the radio weakness dominates the radio loudness over stellar mass estimate
(for the stellar mass, see Fig.~\ref{fig_stellar_bh_masses})  and
leads to a clear separation from the radio-loud objects.
This is seen in the diagnostic diagrams in this paper and has also been put forward
by similar investigations (e.g. Fig. A1 by \cite{2012MNRAS.421.1569B}).

Comparing the $\alpha_{1.4/4.85}$ values of all the sources in the low-flux density sample
with each other, one finds that while the high-flux density sources are dominated by the        
steep spectral index components and steepen towards higher frequencies,
the low-flux density sample is significantly influenced by
the increasing presence of flat-spectrum components.
A detailed investigation of the spectral-index distributions of the high and low-flux density samples 
shows that in both samples the presence of flat-spectrum components
implies a higher excitation in the optical diagnostic diagrams.
In particular, the fainter sources that contain a significant contribution by a 
compact flat-spectrum component can be investigated through the present low-flux density sample.

If we turn to the sources from both samples that have 10.45~GHz measurements and hence a known $\alpha_{[4.85-10.45]}$ value, we find the
following:
considering steep ($\alpha_{[4.85-10.45]} > -0.7$), flat ($-0.7 < \alpha_{[4.85-10.45]} < -0.4$), and inverted radio spectra ($\alpha_{[4.85-10.45]} < -0.4$), we recovered three basic classes with respect to the radio loudness and ionization ratio [$\ion{O}{iii}$]/H$\beta$:

\begin{itemize}
\item[(1)] sources with a steep radio index, high ionization ratio, and high radio loudness;
\item[(2)] sources with a flat radio index, lower ionization ratio, and intermediate radio loudness;
\item[(3)] sources with an inverted radio index, low ionization ratio, and low radio loudness.
\end{itemize}

In the optical diagnostic diagrams, these three classes correspond to the transition from Seyfert to LINER classification in terms of the ionization line ratios. Seyfert sources with higher ionization ratio are dominated by older, optically thin radio emission. Towards the lower ionization ratio, LINERs exhibit a flat to inverted radio spectral index, which is indicative of the compact, self-absorbed core and the jet emission. In the local Universe, these trends may result from re-triggered nuclear and jet activity.

\begin{acknowledgements} 
We are grateful to the referee, Brent Groves, for very constructive comments that helped to improve the manuscript. We thank Stefanie Komossa (MPIfR), Pavel Kroupa (University of Bonn), Thomas Krichbaum (MPIfR), Madeleine Yttergren (University of Cologne), Bozena Czerny (CFT PAN), Mary Loli Martinez-Aldama (CFT PAN), and Swayamtrupta Panda (CFT PAN)  for very helpful discussions and input. 

This work was done with the financial support of SFB956 -- ``Conditions and Impact of Star Formation: Astrophysics, Instrumentation and Laboratory Research '' at the Universities of Cologne and Bonn and MPIfR, in which M.Z., G.B., A.E., N.F., S.B., J.-A. Z., K. H. are members of sub-group A2 -- ``Conditions for Star Formation in Nearby AGN and QSO Hosts'' and A1 --``Understanding Galaxy Assembly'', with M. V. and L. F. being its former members. Michal Zaja\v{c}ek acknowledges the financial support by National Science Centre, Poland, grant No. 2017/26/A/ST9/00756 (Maestro 9).
 
\end{acknowledgements}

\bibliographystyle{aa} 
\bibliography{zajacek} 

\begin{thebibliography}{87}
\expandafter\ifx\csname natexlab\endcsname\relax\def\natexlab#1{#1}\fi

\bibitem[{{Abazajian} {et~al.}(2009){Abazajian}, {Adelman-McCarthy},
  {Ag{\"u}eros}, {Allam}, {Allende Prieto}, {An}, {Anderson}, {Anderson},
  {Annis}, {Bahcall}, \& et~al.}]{2009ApJS..182..543A}
{Abazajian}, K.~N., {Adelman-McCarthy}, J.~K., {Ag{\"u}eros}, M.~A., {et~al.}
  2009, \apjs, 182, 543

\bibitem[{{Agudo} {et~al.}(2015){Agudo}, {Boettcher}, {Falcke},
  {Georganopoulos}, {Ghisellini}, {Giovannini}, {Giroletti}, {Gurvits},
  {G{\'o}mez}, {Laing}, {Lister}, {Mart{\'{\i}}}, {Meyer}, {Mizuno},
  {O'Sullivan}, {Padovani}, {Paragi}, {Perucho}, {Schleicher}, {Stawarz},
  {Vlahakis}, \& {Wardle}}]{2015aska.confE..93A}
{Agudo}, I., {Boettcher}, M., {Falcke}, H.~D.~E., {et~al.} 2015, Advancing
  Astrophysics with the Square Kilometre Array (AASKA14), 93

\bibitem[{{Alonso} {et~al.}(2007){Alonso}, {Lambas}, {Tissera}, \&
  {Coldwell}}]{2007MNRAS.375.1017A}
{Alonso}, M.~S., {Lambas}, D.~G., {Tissera}, P., \& {Coldwell}, G. 2007,
  \mnras, 375, 1017

\bibitem[{{Baldry} {et~al.}(2004){Baldry}, {Glazebrook}, {Brinkmann},
  {Ivezi{\'c}}, {Lupton}, {Nichol}, \& {Szalay}}]{2004ApJ...600..681B}
{Baldry}, I.~K., {Glazebrook}, K., {Brinkmann}, J., {et~al.} 2004, \apj, 600,
  681

\bibitem[{{Baldwin} {et~al.}(1981){Baldwin}, {Phillips}, \&
  {Terlevich}}]{1981PASP...93....5B}
{Baldwin}, J.~A., {Phillips}, M.~M., \& {Terlevich}, R. 1981, \pasp, 93, 5

\bibitem[{{Balogh} {et~al.}(2004){Balogh}, {Baldry}, {Nichol}, {Miller},
  {Bower}, \& {Glazebrook}}]{2004ApJ...615L.101B}
{Balogh}, M.~L., {Baldry}, I.~K., {Nichol}, R., {et~al.} 2004, \apjl, 615, L101

\bibitem[{{Becker} {et~al.}(1995){Becker}, {White}, \&
  {Helfand}}]{1995ApJ...450..559B}
{Becker}, R.~H., {White}, R.~L., \& {Helfand}, D.~J. 1995, \apj, 450, 559

\bibitem[{{Bennert} {et~al.}(2006{\natexlab{a}}){Bennert}, {Jungwiert},
  {Komossa}, {Haas}, \& {Chini}}]{2006A&A...459...55B}
{Bennert}, N., {Jungwiert}, B., {Komossa}, S., {Haas}, M., \& {Chini}, R.
  2006{\natexlab{a}}, \aap, 459, 55

\bibitem[{{Bennert} {et~al.}(2006{\natexlab{b}}){Bennert}, {Jungwiert},
  {Komossa}, {Haas}, \& {Chini}}]{2006A&A...456..953B}
{Bennert}, N., {Jungwiert}, B., {Komossa}, S., {Haas}, M., \& {Chini}, R.
  2006{\natexlab{b}}, \aap, 456, 953

\bibitem[{{Best} \& {Heckman}(2012)}]{2012MNRAS.421.1569B}
{Best}, P.~N. \& {Heckman}, T.~M. 2012, \mnras, 421, 1569

\bibitem[{{Binette} {et~al.}(1997){Binette}, {Wilson}, {Raga}, \&
  {Storchi-Bergmann}}]{1997A&A...327..909B}
{Binette}, L., {Wilson}, A.~S., {Raga}, A., \& {Storchi-Bergmann}, T. 1997,
  \aap, 327, 909

\bibitem[{{Boroson} \& {Green}(1992)}]{1992ApJS...80..109B}
{Boroson}, T.~A. \& {Green}, R.~F. 1992, \apjs, 80, 109

\bibitem[{{Broderick} \& {Fender}(2011)}]{2011MNRAS.417..184B}
{Broderick}, J.~W. \& {Fender}, R.~P. 2011, \mnras, 417, 184

\bibitem[{{Bruzual} \& {Charlot}(2003)}]{2003MNRAS.344.1000B}
{Bruzual}, G. \& {Charlot}, S. 2003, \mnras, 344, 1000

\bibitem[{{Busch}(2016)}]{2016arXiv161107872B}
{Busch}, G. 2016, ArXiv e-prints [\eprint[arXiv]{1611.07872}]

\bibitem[{{Chiaberge} {et~al.}(2015){Chiaberge}, {Gilli}, {Lotz}, \&
  {Norman}}]{2015ApJ...806..147C}
{Chiaberge}, M., {Gilli}, R., {Lotz}, J.~M., \& {Norman}, C. 2015, \apj, 806,
  147

\bibitem[{{Cirasuolo} {et~al.}(2003{\natexlab{a}}){Cirasuolo}, {Celotti},
  {Magliocchetti}, \& {Danese}}]{2003MNRAS.346..447C}
{Cirasuolo}, M., {Celotti}, A., {Magliocchetti}, M., \& {Danese}, L.
  2003{\natexlab{a}}, \mnras, 346, 447

\bibitem[{{Cirasuolo} {et~al.}(2003{\natexlab{b}}){Cirasuolo}, {Magliocchetti},
  {Celotti}, \& {Danese}}]{2003MNRAS.341..993C}
{Cirasuolo}, M., {Magliocchetti}, M., {Celotti}, A., \& {Danese}, L.
  2003{\natexlab{b}}, \mnras, 341, 993

\bibitem[{{Coldwell} {et~al.}(2018){Coldwell}, {Alonso}, {Duplancic}, \&
  {Mesa}}]{2018arXiv180300946C}
{Coldwell}, G.~V., {Alonso}, S., {Duplancic}, F., \& {Mesa}, V. 2018, ArXiv
  e-prints [\eprint[arXiv]{1803.00946}]

\bibitem[{{Coldwell} {et~al.}(2014){Coldwell}, {Gurovich}, {D{\'{\i}}az Tello},
  {S{\"o}chting}, \& {Lambas}}]{2014MNRAS.437.1199C}
{Coldwell}, G.~V., {Gurovich}, S., {D{\'{\i}}az Tello}, J., {S{\"o}chting},
  I.~K., \& {Lambas}, D.~G. 2014, \mnras, 437, 1199

\bibitem[{{Coldwell} {et~al.}(2009){Coldwell}, {Lambas}, {S{\"o}chting}, \&
  {Gurovich}}]{2009MNRAS.399...88C}
{Coldwell}, G.~V., {Lambas}, D.~G., {S{\"o}chting}, I.~K., \& {Gurovich}, S.
  2009, \mnras, 399, 88

\bibitem[{{Coldwell} {et~al.}(2017){Coldwell}, {Pereyra}, {Alonso}, {Donoso},
  \& {Duplancic}}]{2017MNRAS.467.3338C}
{Coldwell}, G.~V., {Pereyra}, L., {Alonso}, S., {Donoso}, E., \& {Duplancic},
  F. 2017, \mnras, 467, 3338

\bibitem[{{Condon} {et~al.}(1998){Condon}, {Cotton}, {Greisen}, {Yin},
  {Perley}, {Taylor}, \& {Broderick}}]{1998AJ....115.1693C}
{Condon}, J.~J., {Cotton}, W.~D., {Greisen}, E.~W., {et~al.} 1998, \aj, 115,
  1693

\bibitem[{{Croom} {et~al.}(2001){Croom}, {Smith}, {Boyle}, {Shanks}, {Loaring},
  {Miller}, \& {Lewis}}]{2001MNRAS.322L..29C}
{Croom}, S.~M., {Smith}, R.~J., {Boyle}, B.~J., {et~al.} 2001, \mnras, 322, L29

\bibitem[{{Czerny} {et~al.}(2009){Czerny}, {Siemiginowska}, {Janiuk},
  {Nikiel-Wroczy{\'n}ski}, \& {Stawarz}}]{2009ApJ...698..840C}
{Czerny}, B., {Siemiginowska}, A., {Janiuk}, A., {Nikiel-Wroczy{\'n}ski}, B.,
  \& {Stawarz}, {\L}. 2009, \apj, 698, 840

\bibitem[{{Davies} {et~al.}(2016){Davies}, {Groves}, {Kewley}, {Dopita},
  {Hampton}, {Shastri}, {Scharw{\"a}chter}, {Sutherland}, {Kharb}, {Bhatt},
  {Jin}, {Banfield}, {Zaw}, {James}, {Juneau}, \&
  {Srivastava}}]{2016MNRAS.462.1616D}
{Davies}, R.~L., {Groves}, B., {Kewley}, L.~J., {et~al.} 2016, \mnras, 462,
  1616

\bibitem[{{Duric} {et~al.}(1988){Duric}, {Bourneuf}, \&
  {Gregory}}]{1988AJ.....96...81D}
{Duric}, N., {Bourneuf}, E., \& {Gregory}, P.~C. 1988, \aj, 96, 81

\bibitem[{{Eckart} {et~al.}(1986){Eckart}, {Witzel}, {Biermann}, {Johnston},
  {Simon}, {Schalinski}, \& {Kuhr}}]{1986A&A...168...17E}
{Eckart}, A., {Witzel}, A., {Biermann}, P., {et~al.} 1986, \aap, 168, 17

\bibitem[{{Faber} {et~al.}(2007){Faber}, {Willmer}, {Wolf}, {Koo}, {Weiner},
  {Newman}, {Im}, {Coil}, {Conroy}, {Cooper}, {Davis}, {Finkbeiner}, {Gerke},
  {Gebhardt}, {Groth}, {Guhathakurta}, {Harker}, {Kaiser}, {Kassin},
  {Kleinheinrich}, {Konidaris}, {Kron}, {Lin}, {Luppino}, {Madgwick},
  {Meisenheimer}, {Noeske}, {Phillips}, {Sarajedini}, {Schiavon}, {Simard},
  {Szalay}, {Vogt}, \& {Yan}}]{2007ApJ...665..265F}
{Faber}, S.~M., {Willmer}, C.~N.~A., {Wolf}, C., {et~al.} 2007, \apj, 665, 265

\bibitem[{{Fender} {et~al.}(2004){Fender}, {Belloni}, \&
  {Gallo}}]{2004MNRAS.355.1105F}
{Fender}, R.~P., {Belloni}, T.~M., \& {Gallo}, E. 2004, \mnras, 355, 1105

\bibitem[{{Ferrarese} \& {Merritt}(2000)}]{2000ApJ...539L...9F}
{Ferrarese}, L. \& {Merritt}, D. 2000, \apjl, 539, L9

\bibitem[{{Gioia} {et~al.}(1982){Gioia}, {Gregorini}, \&
  {Klein}}]{1982A&A...116..164G}
{Gioia}, I.~M., {Gregorini}, L., \& {Klein}, U. 1982, \aap, 116, 164

\bibitem[{{Gregorini} {et~al.}(1984){Gregorini}, {Mantovani}, {Eckart},
  {Biermann}, {Witzel}, \& {Kuhr}}]{1984AJ.....89..323G}
{Gregorini}, L., {Mantovani}, F., {Eckart}, A., {et~al.} 1984, \aj, 89, 323

\bibitem[{{Groves} {et~al.}(2004{\natexlab{a}}){Groves}, {Dopita}, \&
  {Sutherland}}]{2004ApJS..153....9G}
{Groves}, B.~A., {Dopita}, M.~A., \& {Sutherland}, R.~S. 2004{\natexlab{a}},
  \apjs, 153, 9

\bibitem[{{Groves} {et~al.}(2004{\natexlab{b}}){Groves}, {Dopita}, \&
  {Sutherland}}]{2004ApJS..153...75G}
{Groves}, B.~A., {Dopita}, M.~A., \& {Sutherland}, R.~S. 2004{\natexlab{b}},
  \apjs, 153, 75

\bibitem[{{G{\"u}ltekin} {et~al.}(2009){G{\"u}ltekin}, {Richstone}, {Gebhardt},
  {Lauer}, {Tremaine}, {Aller}, {Bender}, {Dressler}, {Faber}, {Filippenko},
  {Green}, {Ho}, {Kormendy}, {Magorrian}, {Pinkney}, \&
  {Siopis}}]{2009ApJ...698..198G}
{G{\"u}ltekin}, K., {Richstone}, D.~O., {Gebhardt}, K., {et~al.} 2009, \apj,
  698, 198

\bibitem[{{Heckman} \& {Kauffmann}(2006)}]{2006NewAR..50..677H}
{Heckman}, T.~M. \& {Kauffmann}, G. 2006, \nar, 50, 677

\bibitem[{{Heckman} {et~al.}(2004){Heckman}, {Kauffmann}, {Brinchmann},
  {Charlot}, {Tremonti}, \& {White}}]{2004ApJ...613..109H}
{Heckman}, T.~M., {Kauffmann}, G., {Brinchmann}, J., {et~al.} 2004, \apj, 613,
  109

\bibitem[{{Ho}(2002)}]{2002ApJ...564..120H}
{Ho}, L.~C. 2002, \apj, 564, 120

\bibitem[{{Ichimaru}(1977)}]{1977ApJ...214..840I}
{Ichimaru}, S. 1977, \apj, 214, 840

\bibitem[{{Ivezi{\'c}} {et~al.}(2002){Ivezi{\'c}}, {Menou}, {Knapp}, {Strauss},
  {Lupton}, {Vanden Berk}, {Richards}, {Tremonti}, {Weinstein}, {Anderson},
  {Bahcall}, {Becker}, {Bernardi}, {Blanton}, {Eisenstein}, {Fan},
  {Finkbeiner}, {Finlator}, {Frieman}, {Gunn}, {Hall}, {Kim}, {Kinkhabwala},
  {Narayanan}, {Rockosi}, {Schlegel}, {Schneider}, {Strateva}, {SubbaRao},
  {Thakar}, {Voges}, {White}, {Yanny}, {Brinkmann}, {Doi}, {Fukugita},
  {Hennessy}, {Munn}, {Nichol}, \& {York}}]{2002AJ....124.2364I}
{Ivezi{\'c}}, {\v Z}., {Menou}, K., {Knapp}, G.~R., {et~al.} 2002, \aj, 124,
  2364

\bibitem[{{Kauffmann} {et~al.}(2003){Kauffmann}, {Heckman}, {Tremonti},
  {Brinchmann}, {Charlot}, {White}, {Ridgway}, {Brinkmann}, {Fukugita}, {Hall},
  {Ivezi{\'c}}, {Richards}, \& {Schneider}}]{2003MNRAS.346.1055K}
{Kauffmann}, G., {Heckman}, T.~M., {Tremonti}, C., {et~al.} 2003, \mnras, 346,
  1055

\bibitem[{{Kellermann} {et~al.}(2016){Kellermann}, {Condon}, {Kimball},
  {Perley}, \& {Ivezi{\'c}}}]{2016ApJ...831..168K}
{Kellermann}, K.~I., {Condon}, J.~J., {Kimball}, A.~E., {Perley}, R.~A., \&
  {Ivezi{\'c}}, {\v Z}. 2016, \apj, 831, 168

\bibitem[{{Kellermann} {et~al.}(1989){Kellermann}, {Sramek}, {Schmidt},
  {Shaffer}, \& {Green}}]{1989AJ.....98.1195K}
{Kellermann}, K.~I., {Sramek}, R., {Schmidt}, M., {Shaffer}, D.~B., \& {Green},
  R. 1989, \aj, 98, 1195

\bibitem[{{Kewley} {et~al.}(2001){Kewley}, {Dopita}, {Sutherland}, {Heisler},
  \& {Trevena}}]{2001ApJ...556..121K}
{Kewley}, L.~J., {Dopita}, M.~A., {Sutherland}, R.~S., {Heisler}, C.~A., \&
  {Trevena}, J. 2001, \apj, 556, 121

\bibitem[{{Kewley} {et~al.}(2003){Kewley}, {Geller}, \&
  {Jansen}}]{2003AAS...20311901K}
{Kewley}, L.~J., {Geller}, M.~J., \& {Jansen}, R.~A. 2003, in Bulletin of the
  American Astronomical Society, Vol.~35, American Astronomical Society Meeting
  Abstracts, 1404

\bibitem[{{Kewley} {et~al.}(2006){Kewley}, {Groves}, {Kauffmann}, \&
  {Heckman}}]{2006MNRAS.372..961K}
{Kewley}, L.~J., {Groves}, B., {Kauffmann}, G., \& {Heckman}, T. 2006, \mnras,
  372, 961

\bibitem[{{Komossa} \& {Schulz}(1997)}]{1997A&A...323...31K}
{Komossa}, S. \& {Schulz}, H. 1997, \aap, 323, 31

\bibitem[{{Kormendy} \& {Ho}(2013)}]{2013ARA&A..51..511K}
{Kormendy}, J. \& {Ho}, L.~C. 2013, \araa, 51, 511

\bibitem[{{Kuehr} {et~al.}(1981){Kuehr}, {Witzel}, {Pauliny-Toth}, \&
  {Nauber}}]{1981A&AS...45..367K}
{Kuehr}, H., {Witzel}, A., {Pauliny-Toth}, I.~I.~K., \& {Nauber}, U. 1981,
  \aaps, 45, 367

\bibitem[{{Laor}(2003)}]{2003ApJ...590...86L}
{Laor}, A. 2003, \apj, 590, 86

\bibitem[{{Laor} {et~al.}(2019){Laor}, {Baldi}, \&
  {Behar}}]{2019MNRAS.482.5513L}
{Laor}, A., {Baldi}, R.~D., \& {Behar}, E. 2019, \mnras, 482, 5513

\bibitem[{{Leslie} {et~al.}(2016){Leslie}, {Kewley}, {Sanders}, \&
  {Lee}}]{2016MNRAS.455L..82L}
{Leslie}, S.~K., {Kewley}, L.~J., {Sanders}, D.~B., \& {Lee}, N. 2016, \mnras,
  455, L82

\bibitem[{{Madau} \& {Dickinson}(2014)}]{2014ARA&A..52..415M}
{Madau}, P. \& {Dickinson}, M. 2014, \araa, 52, 415

\bibitem[{{Magorrian} {et~al.}(1998){Magorrian}, {Tremaine}, {Richstone},
  {Bender}, {Bower}, {Dressler}, {Faber}, {Gebhardt}, {Green}, {Grillmair},
  {Kormendy}, \& {Lauer}}]{1998AJ....115.2285M}
{Magorrian}, J., {Tremaine}, S., {Richstone}, D., {et~al.} 1998, \aj, 115, 2285

\bibitem[{{Marziani} {et~al.}(2001){Marziani}, {Sulentic}, {Zwitter},
  {Dultzin-Hacyan}, \& {Calvani}}]{2001ApJ...558..553M}
{Marziani}, P., {Sulentic}, J.~W., {Zwitter}, T., {Dultzin-Hacyan}, D., \&
  {Calvani}, M. 2001, \apj, 558, 553

\bibitem[{{Massardi} {et~al.}(2011){Massardi}, {Ekers}, {Murphy}, {Mahony},
  {Hancock}, {Chhetri}, {de Zotti}, {Sadler}, {Burke-Spolaor}, {Calabretta},
  {Edwards}, {Ekers}, {Jackson}, {Kesteven}, {Newton-McGee}, {Phillips},
  {Ricci}, {Roberts}, {Sault}, {Staveley-Smith}, {Subrahmanyan}, {Walker}, \&
  {Wilson}}]{2011MNRAS.412..318M}
{Massardi}, M., {Ekers}, R.~D., {Murphy}, T., {et~al.} 2011, \mnras, 412, 318

\bibitem[{{Nagar} {et~al.}(2005){Nagar}, {Falcke}, \&
  {Wilson}}]{2005A&A...435..521N}
{Nagar}, N.~M., {Falcke}, H., \& {Wilson}, A.~S. 2005, \aap, 435, 521

\bibitem[{{Narayan} \& {Yi}(1995)}]{1995ApJ...444..231N}
{Narayan}, R. \& {Yi}, I. 1995, \apj, 444, 231

\bibitem[{{Oke} \& {Gunn}(1983)}]{1983ApJ...266..713O}
{Oke}, J.~B. \& {Gunn}, J.~E. 1983, \apj, 266, 713

\bibitem[{{OMullane} {et~al.}(2005){OMullane}, {Li}, {Nieto-Santisteban},
  {Szalay}, {Thakar}, \& {Gray}}]{2005cs........2072O}
{OMullane}, W., {Li}, N., {Nieto-Santisteban}, M., {et~al.} 2005, eprint
  arXiv:cs/0502072 [\eprint{cs/0502072}]

\bibitem[{{Padilla} {et~al.}(2010){Padilla}, {Lambas}, \&
  {Gonz{\'a}lez}}]{2010MNRAS.409..936P}
{Padilla}, N., {Lambas}, D.~G., \& {Gonz{\'a}lez}, R. 2010, \mnras, 409, 936

\bibitem[{{Padovani} {et~al.}(2017){Padovani}, {Alexander}, {Assef}, {De
  Marco}, {Giommi}, {Hickox}, {Richards}, {Smol{\v c}i{\'c}}, {Hatziminaoglou},
  {Mainieri}, \& {Salvato}}]{2017A&ARv..25....2P}
{Padovani}, P., {Alexander}, D.~M., {Assef}, R.~J., {et~al.} 2017, \aapr, 25, 2

\bibitem[{Pedregosa {et~al.}(2011)Pedregosa, Varoquaux, Gramfort, Michel,
  Thirion, Grisel, Blondel, Prettenhofer, Weiss, Dubourg, Vanderplas, Passos,
  Cournapeau, Brucher, Perrot, \& Duchesnay}]{scikit-learn}
Pedregosa, F., Varoquaux, G., Gramfort, A., {et~al.} 2011, Journal of Machine
  Learning Research, 12, 2825

\bibitem[{{Popesso} \& {Biviano}(2006)}]{2006A&A...460L..23P}
{Popesso}, P. \& {Biviano}, A. 2006, \aap, 460, L23

\bibitem[{{Rees} {et~al.}(1982){Rees}, {Begelman}, {Blandford}, \&
  {Phinney}}]{1982Natur.295...17R}
{Rees}, M.~J., {Begelman}, M.~C., {Blandford}, R.~D., \& {Phinney}, E.~S. 1982,
  \nat, 295, 17

\bibitem[{{Richardson} {et~al.}(2016){Richardson}, {Allen}, {Baldwin},
  {Hewett}, {Ferland}, {Crider}, \& {Meskhidze}}]{2016MNRAS.458..988R}
{Richardson}, C.~T., {Allen}, J.~T., {Baldwin}, J.~A., {et~al.} 2016, \mnras,
  458, 988

\bibitem[{{Scharw{\"a}chter} {et~al.}(2011){Scharw{\"a}chter}, {Dopita},
  {Zuther}, {Fischer}, {Komossa}, \& {Eckart}}]{2011AJ....142...43S}
{Scharw{\"a}chter}, J., {Dopita}, M.~A., {Zuther}, J., {et~al.} 2011, \aj, 142,
  43

\bibitem[{{Schawinski}(2009)}]{2009AIPC.1201...17S}
{Schawinski}, K. 2009, in American Institute of Physics Conference Series, Vol.
  1201, American Institute of Physics Conference Series, ed. S.~{Heinz} \&
  E.~{Wilcots}, 17--20

\bibitem[{{Schawinski} {et~al.}(2007){Schawinski}, {Thomas}, {Sarzi},
  {Maraston}, {Kaviraj}, {Joo}, {Yi}, \& {Silk}}]{2007MNRAS.382.1415S}
{Schawinski}, K., {Thomas}, D., {Sarzi}, M., {et~al.} 2007, \mnras, 382, 1415

\bibitem[{{Schulz} \& {Fritsch}(1994)}]{1994A&A...291..713S}
{Schulz}, H. \& {Fritsch}, C. 1994, \aap, 291, 713

\bibitem[{{Shakura} \& {Sunyaev}(1973)}]{1973A&A....24..337S}
{Shakura}, N.~I. \& {Sunyaev}, R.~A. 1973, \aap, 24, 337

\bibitem[{{Sikora} {et~al.}(2007){Sikora}, {Stawarz}, \&
  {Lasota}}]{2007ApJ...658..815S}
{Sikora}, M., {Stawarz}, {\L}., \& {Lasota}, J.-P. 2007, \apj, 658, 815

\bibitem[{{Singh} {et~al.}(2013){Singh}, {van de Ven}, {Jahnke}, {Lyubenova},
  {Falc{\'o}n-Barroso}, {Alves}, {Cid Fernandes}, {Galbany},
  {Garc{\'{\i}}a-Benito}, {Husemann}, {Kennicutt}, {Marino}, {M{\'a}rquez},
  {Masegosa}, {Mast}, {Pasquali}, {S{\'a}nchez}, {Walcher}, {Wild}, {Wisotzki},
  \& {Ziegler}}]{2013A&A...558A..43S}
{Singh}, R., {van de Ven}, G., {Jahnke}, K., {et~al.} 2013, \aap, 558, A43

\bibitem[{{Stoughton} {et~al.}(2002){Stoughton}, {Lupton}, {Bernardi},
  {Blanton}, {Burles}, {Castander}, {Connolly}, {Eisenstein}, {Frieman},
  {Hennessy}, {Hindsley}, {Ivezi{\'c}}, {Kent}, {Kunszt}, {Lee}, {Meiksin},
  {Munn}, {Newberg}, {Nichol}, {Nicinski}, {Pier}, {Richards}, {Richmond},
  {Schlegel}, {Smith}, {Strauss}, {SubbaRao}, {Szalay}, {Thakar}, {Tucker},
  {Vanden Berk}, {Yanny}, {Adelman}, {Anderson}, {Anderson}, {Annis},
  {Bahcall}, {Bakken}, {Bartelmann}, {Bastian}, {Bauer}, {Berman},
  {B{\"o}hringer}, {Boroski}, {Bracker}, {Briegel}, {Briggs}, {Brinkmann},
  {Brunner}, {Carey}, {Carr}, {Chen}, {Christian}, {Colestock}, {Crocker},
  {Csabai}, {Czarapata}, {Dalcanton}, {Davidsen}, {Davis}, {Dehnen},
  {Dodelson}, {Doi}, {Dombeck}, {Donahue}, {Ellman}, {Elms}, {Evans}, {Eyer},
  {Fan}, {Federwitz}, {Friedman}, {Fukugita}, {Gal}, {Gillespie}, {Glazebrook},
  {Gray}, {Grebel}, {Greenawalt}, {Greene}, {Gunn}, {de Haas}, {Haiman},
  {Haldeman}, {Hall}, {Hamabe}, {Hansen}, {Harris}, {Harris}, {Harvanek},
  {Hawley}, {Hayes}, {Heckman}, {Helmi}, {Henden}, {Hogan}, {Hogg}, {Holmgren},
  {Holtzman}, {Huang}, {Hull}, {Ichikawa}, {Ichikawa}, {Johnston}, {Kauffmann},
  {Kim}, {Kimball}, {Kinney}, {Klaene}, {Kleinman}, {Klypin}, {Knapp},
  {Korienek}, {Krolik}, {Kron}, {Krzesi{\'n}ski}, {Lamb}, {Leger},
  {Limmongkol}, {Lindenmeyer}, {Long}, {Loomis}, {Loveday}, {MacKinnon},
  {Mannery}, {Mantsch}, {Margon}, {McGehee}, {McKay}, {McLean}, {Menou},
  {Merelli}, {Mo}, {Monet}, {Nakamura}, {Narayanan}, {Nash}, {Neilsen},
  {Newman}, {Nitta}, {Odenkirchen}, {Okada}, {Okamura}, {Ostriker}, {Owen},
  {Pauls}, {Peoples}, {Peterson}, {Petravick}, {Pope}, {Pordes}, {Postman},
  {Prosapio}, {Quinn}, {Rechenmacher}, {Rivetta}, {Rix}, {Rockosi}, {Rosner},
  {Ruthmansdorfer}, {Sandford}, {Schneider}, {Scranton}, {Sekiguchi}, {Sergey},
  {Sheth}, {Shimasaku}, {Smee}, {Snedden}, {Stebbins}, {Stubbs}, {Szapudi},
  {Szkody}, {Szokoly}, {Tabachnik}, {Tsvetanov}, {Uomoto}, {Vogeley}, {Voges},
  {Waddell}, {Walterbos}, {Wang}, {Watanabe}, {Weinberg}, {White}, {White},
  {Wilhite}, {Wolfe}, {Yasuda}, {York}, {Zehavi}, \&
  {Zheng}}]{2002AJ....123..485S}
{Stoughton}, C., {Lupton}, R.~H., {Bernardi}, M., {et~al.} 2002, \aj, 123, 485

\bibitem[{{Strateva} {et~al.}(2001){Strateva}, {Ivezi{\'c}}, {Knapp},
  {Narayanan}, {Strauss}, {Gunn}, {Lupton}, {Schlegel}, {Bahcall}, {Brinkmann},
  {Brunner}, {Budav{\'a}ri}, {Csabai}, {Castander}, {Doi}, {Fukugita}, {Gy{\H
  o}ry}, {Hamabe}, {Hennessy}, {Ichikawa}, {Kunszt}, {Lamb}, {McKay},
  {Okamura}, {Racusin}, {Sekiguchi}, {Schneider}, {Shimasaku}, \&
  {York}}]{2001AJ....122.1861S}
{Strateva}, I., {Ivezi{\'c}}, {\v Z}., {Knapp}, G.~R., {et~al.} 2001, \aj, 122,
  1861

\bibitem[{{Strittmatter} {et~al.}(1980){Strittmatter}, {Hill}, {Pauliny-Toth},
  {Steppe}, \& {Witzel}}]{1980A&A....88L..12S}
{Strittmatter}, P.~A., {Hill}, P., {Pauliny-Toth}, I.~I.~K., {Steppe}, H., \&
  {Witzel}, A. 1980, \aap, 88, L12

\bibitem[{{Sulentic} {et~al.}(2000){Sulentic}, {Zwitter}, {Marziani}, \&
  {Dultzin-Hacyan}}]{2000ApJ...536L...5S}
{Sulentic}, J.~W., {Zwitter}, T., {Marziani}, P., \& {Dultzin-Hacyan}, D. 2000,
  \apjl, 536, L5

\bibitem[{{Svoboda} {et~al.}(2017){Svoboda}, {Guainazzi}, \&
  {Merloni}}]{2017A&A...603A.127S}
{Svoboda}, J., {Guainazzi}, M., \& {Merloni}, A. 2017, \aap, 603, A127

\bibitem[{{Tadhunter}(2016)}]{2016A&ARv..24...10T}
{Tadhunter}, C. 2016, \aapr, 24, 10

\bibitem[{{Tremou} {et~al.}(2015){Tremou}, {Garcia-Marin}, {Zuther}, {Eckart},
  {Valencia-Schneider}, {Vitale}, \& {Shan}}]{2015A&A...580A.113T}
{Tremou}, E., {Garcia-Marin}, M., {Zuther}, J., {et~al.} 2015, \aap, 580, A113

\bibitem[{{Veilleux} \& {Osterbrock}(1987)}]{1987ApJS...63..295V}
{Veilleux}, S. \& {Osterbrock}, D.~E. 1987, \apjs, 63, 295

\bibitem[{{Vitale} {et~al.}(2015{\natexlab{a}}){Vitale}, {Fuhrmann},
  {Garc{\'{\i}}a-Mar{\'{\i}}n}, {Eckart}, {Zuther}, \&
  {Hopkins}}]{2015A&A...573A..93V}
{Vitale}, M., {Fuhrmann}, L., {Garc{\'{\i}}a-Mar{\'{\i}}n}, M., {et~al.}
  2015{\natexlab{a}}, \aap, 573, A93

\bibitem[{{Vitale} {et~al.}(2015{\natexlab{b}}){Vitale}, {Fuhrmann},
  {Garcia-Marin}, {Eckart}, {Zuther}, \& {Hopkins}}]{2015yCat..35730093V}
{Vitale}, M., {Fuhrmann}, L., {Garcia-Marin}, M., {et~al.} 2015{\natexlab{b}},
  VizieR Online Data Catalog, 357

\bibitem[{{Vitale} {et~al.}(2012){Vitale}, {Zuther},
  {Garc{\'{\i}}a-Mar{\'{\i}}n}, {Eckart}, {Bremer}, {Valencia-S.}, \&
  {Zensus}}]{2012A&A...546A..17V}
{Vitale}, M., {Zuther}, J., {Garc{\'{\i}}a-Mar{\'{\i}}n}, M., {et~al.} 2012,
  \aap, 546, A17

\bibitem[{{White} {et~al.}(2000){White}, {Becker}, {Gregg},
  {Laurent-Muehleisen}, {Brotherton}, {Impey}, {Petry}, {Foltz}, {Chaffee},
  {Richards}, {Oegerle}, {Helfand}, {McMahon}, \&
  {Cabanela}}]{2000ApJS..126..133W}
{White}, R.~L., {Becker}, R.~H., {Gregg}, M.~D., {et~al.} 2000, \apjs, 126, 133

\bibitem[{{York} {et~al.}(2000){York}, {Adelman}, {Anderson}, {Anderson},
  {Annis}, {Bahcall}, {Bakken}, {Barkhouser}, {Bastian}, {Berman}, {Boroski},
  {Bracker}, {Briegel}, {Briggs}, {Brinkmann}, {Brunner}, {Burles}, {Carey},
  {Carr}, {Castander}, {Chen}, {Colestock}, {Connolly}, {Crocker}, {Csabai},
  {Czarapata}, {Davis}, {Doi}, {Dombeck}, {Eisenstein}, {Ellman}, {Elms},
  {Evans}, {Fan}, {Federwitz}, {Fiscelli}, {Friedman}, {Frieman}, {Fukugita},
  {Gillespie}, {Gunn}, {Gurbani}, {de Haas}, {Haldeman}, {Harris}, {Hayes},
  {Heckman}, {Hennessy}, {Hindsley}, {Holm}, {Holmgren}, {Huang}, {Hull},
  {Husby}, {Ichikawa}, {Ichikawa}, {Ivezi{\'c}}, {Kent}, {Kim}, {Kinney},
  {Klaene}, {Kleinman}, {Kleinman}, {Knapp}, {Korienek}, {Kron}, {Kunszt},
  {Lamb}, {Lee}, {Leger}, {Limmongkol}, {Lindenmeyer}, {Long}, {Loomis},
  {Loveday}, {Lucinio}, {Lupton}, {MacKinnon}, {Mannery}, {Mantsch}, {Margon},
  {McGehee}, {McKay}, {Meiksin}, {Merelli}, {Monet}, {Munn}, {Narayanan},
  {Nash}, {Neilsen}, {Neswold}, {Newberg}, {Nichol}, {Nicinski}, {Nonino},
  {Okada}, {Okamura}, {Ostriker}, {Owen}, {Pauls}, {Peoples}, {Peterson},
  {Petravick}, {Pier}, {Pope}, {Pordes}, {Prosapio}, {Rechenmacher}, {Quinn},
  {Richards}, {Richmond}, {Rivetta}, {Rockosi}, {Ruthmansdorfer}, {Sandford},
  {Schlegel}, {Schneider}, {Sekiguchi}, {Sergey}, {Shimasaku}, {Siegmund},
  {Smee}, {Smith}, {Snedden}, {Stone}, {Stoughton}, {Strauss}, {Stubbs},
  {SubbaRao}, {Szalay}, {Szapudi}, {Szokoly}, {Thakar}, {Tremonti}, {Tucker},
  {Uomoto}, {Vanden Berk}, {Vogeley}, {Waddell}, {Wang}, {Watanabe},
  {Weinberg}, {Yanny}, {Yasuda}, \& {SDSS Collaboration}}]{2000AJ....120.1579Y}
{York}, D.~G., {Adelman}, J., {Anderson}, Jr., J.~E., {et~al.} 2000, \aj, 120,
  1579

\end{thebibliography}

\appendix
\section{Table of the basic radio-optical properties}
\label{appa}

\begin{figure}[tbh]
  \centering
  \includegraphics[width=0.5\textwidth]{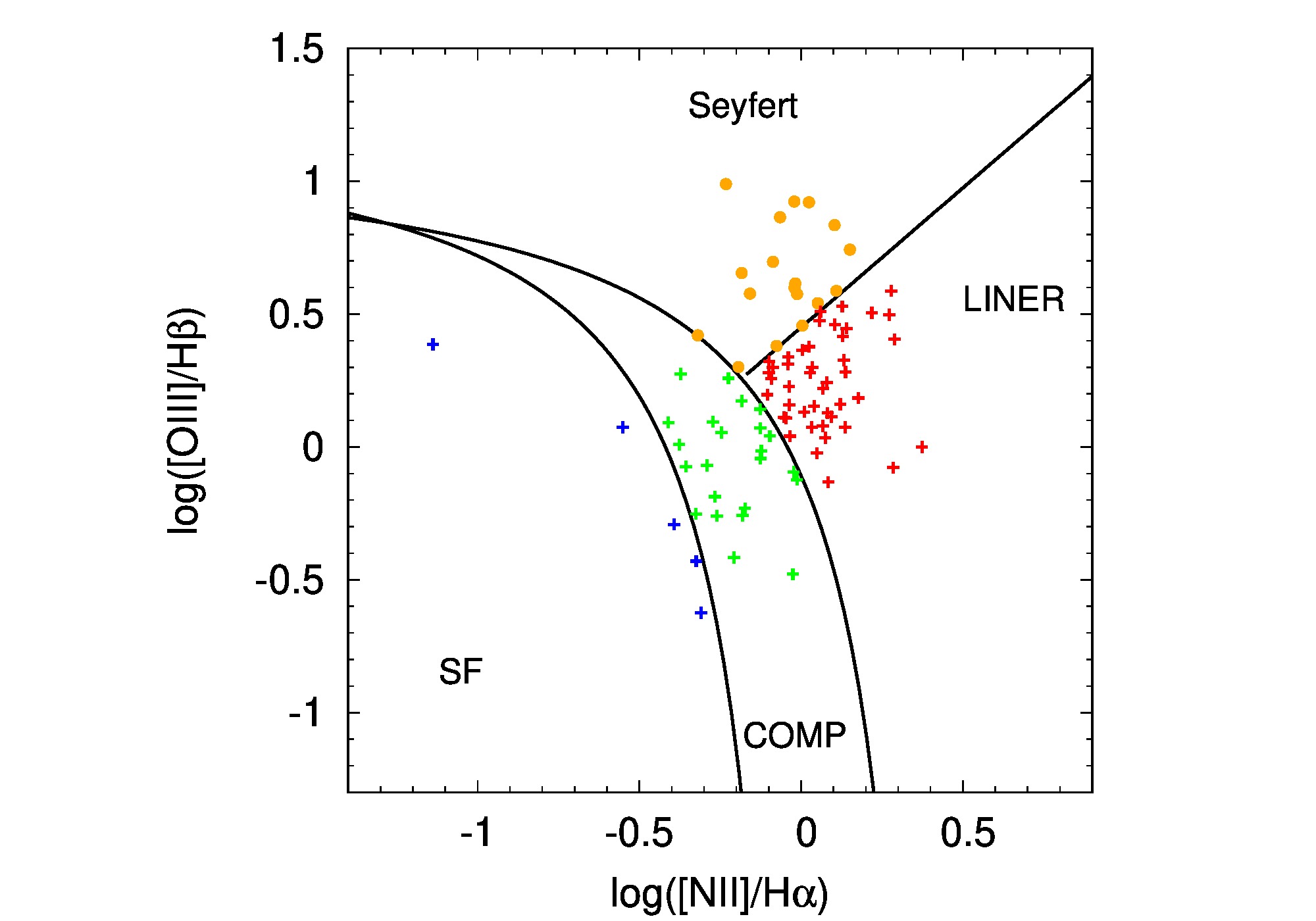}
  \caption{Location of the sources from Table~\ref{tab_three_freq1} in [NII]-based diagnostic diagram.}
  \label{fig_appendix_bpt}
\end{figure}

Here we summarize the information on 90 sources of the low-flux sample, for which we obtained radio flux densities at $4.85\,{\rm GHz}$ and $10.45\,{\rm GHz}$ using the 100-m Effelsberg radio telescope. In combination with the integrated flux of these sources at $1.4\,{\rm GHz}$, we determined the corresponding spectral slopes. 

In Table~\ref{tab_three_freq1}, we include the redshift $z$, right ascension $RA$, declination $DEC$, integrated flux density at $1.4\,{\rm GHz}$ as adopted from the FIRST survey catalogue, the measured flux densities at $4.85\,{\rm GHz}$ and $10.45\,{\rm GHz,}$ and the corresponding errors. The final column denoted as $[NII]_d$ is the galaxy spectral type according to the $[NII]$-based optical diagnostic diagram (BPT diagram): SF denotes star-forming galaxies, COMP composites, SEY Seyfert AGN sources, and LIN stands for LINERs (see also Fig.~\ref{fig_appendix_bpt} for corresponding positions in the $[\ion{N}{ii}]$-based BPT diagram).

\begin{table*}[tbh]
\centering
\caption{Sources with the flux densities $10\,{\rm mJy} \lesssim F_{1.4} \lesssim 100\,{\rm mJy}$ with measured flux densities at two higher frequencies, $4.85\,{\rm GHz}$ and $10.45\,{\rm GHz}$. Using the symbol $[NII]_d$, we denote the diagnostics according to the [NII]-based optical diagnostics diagram, where SF stands for star-forming galaxies, COMP composites, SEY represents Seyfert AGN, and LIN denotes LINER sources.}
\label{tab_three_freq1}
  \begin{tabular}{ccccccccc}
  \hline
  \hline
  $z$ & $RA$ & $DEC$ & $F_{1.4}\,[mJy]$ & $F_{4.85}\,[mJy]$ & Err $[mJy]$ & $F_{10.45}\,[mJy]$ & Err $[mJy]$ & $[NII]_d$\\
  \hline
   $0.054$& $37.92$ & $   -1.1677$&$    14.250$&$      27.5$&$       1.3$&$      30.8$&$       3.3$&COMP                \\
$0.062$&$    116.18$&$   38.6531$&$    19.570$&$      14.9$&$       0.1$&$      10.7$&$       5.1$&COMP                \\
$0.133$&$    119.36$&$   40.9047$&$    12.800$&$      10.8$&$       0.2$&$      13.0$&$       2.4$&COMP                \\
$0.095$&$    129.18$&$   53.5757$&$    17.870$&$      26.2$&$       2.5$&$      22.6$&$       4.6$&LIN                 \\
$0.136$&$    130.40$&$   51.9057$&$    40.580$&$      56.5$&$      18.0$&$      17.3$&$       4.2$&LIN                 \\
$0.127$&$    137.97$&$   56.9317$&$    15.100$&$       7.7$&$       0.6$&$      14.6$&$       5.0$&LIN                 \\
$0.216$&$    119.03$&$   38.5669$&$    63.950$&$      20.0$&$       0.9$&$      12.1$&$       2.9$&LIN                 \\
$0.066$&$    119.49$&$   39.9934$&$    99.200$&$      23.0$&$       0.8$&$      13.5$&$       7.9$&SEY                 \\
$0.098$&$    135.66$&$   52.1874$&$    38.440$&$      18.3$&$       0.8$&$      18.2$&$       4.6$&SEY                 \\
$0.069$&$    235.62$&$   52.9975$&$    28.580$&$      33.6$&$       2.7$&$      64.5$&$       6.3$&LIN                 \\
$0.052$&$    234.90$&$   55.5044$&$    31.490$&$      16.9$&$       1.5$&$      17.5$&$       3.3$&COMP                \\
$0.108$&$    242.15$&$   53.6394$&$    40.340$&$      16.9$&$       0.9$&$      14.5$&$       3.7$&COMP                \\
$0.199$&$    243.92$&$   47.1866$&$    98.080$&$     182.0$&$       1.9$&$     119.9$&$       7.5$&COMP                \\
$0.195$&$    246.83$&$   48.5241$&$    23.590$&$      54.6$&$       1.4$&$      44.4$&$       5.2$&LIN                 \\
$0.191$&$    246.58$&$   50.0678$&$    15.670$&$       9.9$&$       1.0$&$      18.6$&$       5.5$&LIN                 \\
$0.055$&$     10.46$&$   -9.3032$&$    40.200$&$      50.2$&$       2.8$&$      15.9$&$       5.5$&LIN                 \\
$0.077$&$     13.03$&$   -9.4912$&$    29.930$&$      51.6$&$      12.3$&$      18.9$&$       3.3$&LIN                 \\
$0.143$&$     20.94$&$   -9.3843$&$    74.240$&$      31.8$&$       0.4$&$     133.1$&$       6.7$&LIN                 \\
$0.103$&$     20.99$&$   -9.6855$&$    14.700$&$      43.4$&$       3.5$&$      12.0$&$       2.8$&LIN                 \\
$0.049$&$     21.32$&$   -8.8737$&$    11.570$&$     135.9$&$      20.4$&$       7.8$&$       2.8$&COMP                \\
$0.049$&$     21.87$&$   -8.5543$&$    24.250$&$      44.0$&$       5.8$&$       9.2$&$       2.3$&SF                  \\
$0.146$&$     22.93$&$   -8.4370$&$    64.300$&$      32.6$&$       0.8$&$      15.7$&$       3.3$&LIN                 \\
$0.271$&$     23.31$&$   -9.5340$&$    21.170$&$      17.1$&$       1.1$&$       8.7$&$       3.2$&SF                  \\
$0.136$&$     33.21$&$   -9.7125$&$    47.020$&$      22.8$&$       1.3$&$      13.8$&$       2.8$&LIN                 \\
$0.166$&$     31.57$&$   -9.9549$&$    18.720$&$      26.0$&$       1.4$&$      24.9$&$       3.4$&LIN                 \\
$0.081$&$    343.62$&$   -9.2759$&$    27.930$&$      16.2$&$       1.5$&$      14.9$&$       4.1$&LIN                 \\
$0.059$&$    185.77$&$   63.2223$&$    16.500$&$      15.0$&$       1.8$&$      10.9$&$       3.2$&COMP                \\
$0.131$&$    198.60$&$   62.3294$&$    64.520$&$     181.7$&$      10.3$&$      94.8$&$       7.9$&LIN                 \\
$0.099$&$    250.46$&$   37.3093$&$    22.510$&$      27.5$&$       1.6$&$      25.3$&$       3.9$&LIN                 \\
$0.099$&$    130.52$&$   40.8460$&$    24.430$&$      20.7$&$       1.0$&$      30.3$&$       4.0$&LIN                 \\
$0.236$&$    126.72$&$   39.8121$&$    13.870$&$      38.8$&$       1.1$&$      23.3$&$       3.9$&LIN                 \\
$0.147$&$    127.32$&$   40.0373$&$    10.850$&$      14.2$&$       1.0$&$      16.3$&$       3.8$&COMP                \\
$0.167$&$    136.84$&$   46.3384$&$    35.100$&$      23.3$&$       0.9$&$      13.6$&$       3.2$&SEY                 \\
$0.192$&$    139.76$&$   47.4986$&$    13.470$&$       7.7$&$       0.9$&$      18.6$&$       4.6$&COMP                \\
$0.207$&$    145.02$&$   51.0727$&$    16.270$&$      78.2$&$       1.3$&$      58.2$&$       5.9$&SEY                 \\
$0.159$&$    170.12$&$   58.9371$&$    84.050$&$      45.6$&$       3.5$&$      34.1$&$       3.4$&SEY                 \\
$0.234$&$    322.42$&$    0.0892$&$    23.820$&$      12.8$&$       4.8$&$      11.8$&$       5.0$&COMP                \\
$0.074$&$    326.71$&$    0.3542$&$    10.440$&$      18.4$&$       2.6$&$       6.0$&$       2.3$&SF                  \\
$0.133$&$    151.79$&$   50.3990$&$    30.640$&$      24.3$&$       0.8$&$      25.0$&$       5.1$&LIN                 \\
$0.166$&$    157.93$&$   52.4264$&$    83.670$&$     360.8$&$       3.5$&$     178.0$&$       7.0$&SEY                 \\
$0.055$&$    175.82$&$   55.2777$&$    11.720$&$      18.8$&$       2.8$&$      19.1$&$       2.8$&LIN                 \\
$0.091$&$    196.40$&$   54.0284$&$    23.140$&$      34.6$&$       7.4$&$      15.5$&$       3.2$&LIN                 \\
$0.140$&$    200.69$&$   53.0299$&$    22.790$&$     106.1$&$      74.7$&$       8.2$&$       3.2$&SEY                 \\
$0.137$&$    232.51$&$   43.0404$&$    22.650$&$      16.5$&$       0.8$&$      23.6$&$      15.9$&SEY                 \\
$0.119$&$    234.96$&$   41.7237$&$    19.630$&$      24.0$&$       1.0$&$      22.6$&$       5.0$&LIN                 \\
$0.066$&$    241.43$&$   37.1791$&$    28.190$&$       9.8$&$       0.8$&$      14.5$&$       4.1$&SEY                 \\
$0.110$&$    235.44$&$   47.4652$&$    83.380$&$      68.3$&$       1.3$&$      37.1$&$       5.4$&SEY                 \\
$0.042$&$    239.97$&$   44.7090$&$    58.810$&$      26.1$&$       0.7$&$      11.6$&$       4.7$&LIN                 \\
$0.173$&$    188.57$&$   50.9069$&$    52.400$&$      57.9$&$       0.7$&$      45.0$&$       4.3$&SEY                 \\
$0.181$&$    173.13$&$   57.5192$&$    35.560$&$      25.4$&$       1.4$&$      22.3$&$       3.3$&SEY                 \\
$0.117$&$    179.04$&$   56.7764$&$    19.940$&$      25.9$&$       9.9$&$      16.3$&$       3.2$&LIN                 \\
$0.047$&$    218.30$&$   52.9631$&$    15.190$&$      11.6$&$       0.9$&$      15.0$&$       3.7$&COMP                \\
$0.150$&$    243.84$&$   39.9390$&$    28.440$&$      33.1$&$      10.7$&$      12.0$&$       3.3$&COMP                \\
$0.075$&$    252.55$&$   29.2934$&$    10.220$&$      13.9$&$       1.2$&$      13.2$&$       5.5$&LIN                 \\
$0.168$&$    211.91$&$   40.8988$&$    61.280$&$      23.2$&$       2.2$&$      24.0$&$       4.7$&LIN                 \\
$0.173$&$    219.51$&$   39.5710$&$    17.760$&$      19.7$&$       0.9$&$      33.2$&$       4.8$&LIN                 \\
$0.237$&$    231.28$&$   35.5338$&$    17.740$&$      17.8$&$       1.0$&$      12.7$&$       3.2$&COMP                \\
$0.055$&$    231.75$&$   35.9770$&$    13.060$&$      12.3$&$       1.1$&$      15.4$&$       3.7$&COMP                \\
$0.154$&$    239.21$&$   32.9769$&$    18.030$&$      21.1$&$       2.8$&$       7.7$&$       3.2$&LIN                 \\  
\hline
  \end{tabular}
\end{table*}

\begin{table*}[tbh]
\centering
\caption{Table A.1 continued.}

  \begin{tabular}{ccccccccc}
  \hline
  \hline
  $z$ & $RA$ & $DEC$ & $F_{1.4}\,[mJy]$ & $F_{4.85}\,[mJy]$ & Err $[mJy]$ & $F_{10.45}\,[mJy]$ & Err $[mJy]$ & $[NII]_d$\\
  \hline
$0.226$&$    242.04$&$   29.7541$&$    48.500$&$      37.4$&$       6.6$&$      19.5$&$       3.7$&SEY                 \\
$0.112$&$    166.28$&$   46.8885$&$    56.180$&$      45.6$&$       1.2$&$      28.7$&$       3.8$&SEY                 \\
$0.065$&$    205.15$&$   44.8048$&$    82.200$&$      86.0$&$      48.7$&$      43.5$&$       4.5$&SEY                 \\
$0.058$&$    245.20$&$   24.0142$&$    28.520$&$      12.9$&$       1.0$&$      15.0$&$       9.5$&COMP                \\
$0.050$&$    242.09$&$   28.4787$&$    78.160$&$     120.9$&$       0.2$&$      83.9$&$       4.5$&LIN                 \\
$0.201$&$    233.72$&$   29.1555$&$    44.420$&$      32.8$&$       2.3$&$      20.0$&$       8.4$&COMP                \\
$0.108$&$    211.83$&$   50.4603$&$    58.400$&$      29.0$&$       1.4$&$      18.7$&$       3.7$&SEY                 \\
$0.040$&$    229.53$&$   42.7459$&$    30.370$&$      20.2$&$       0.8$&$      18.1$&$       4.6$&COMP                \\
$0.088$&$    113.98$&$   42.2034$&$    17.720$&$      11.0$&$       0.2$&$       6.5$&$       2.3$&LIN                 \\
$0.052$&$    118.18$&$   45.9493$&$    50.150$&$     276.2$&$      27.2$&$      93.7$&$      10.7$&SEY                 \\
$0.044$&$    129.10$&$   56.5974$&$    10.870$&$      17.7$&$       2.3$&$      21.9$&$       4.2$&SF                  \\
$0.084$&$    229.95$&$   27.7727$&$    15.670$&$      28.7$&$       1.1$&$      31.5$&$       8.4$&LIN                 \\
$0.118$&$    238.43$&$   23.8071$&$    64.950$&$     250.5$&$       5.8$&$     168.5$&$       9.3$&SEY                 \\
$0.084$&$    119.20$&$   53.2156$&$    38.980$&$      18.8$&$       0.3$&$      11.1$&$       3.3$&LIN                 \\
$0.122$&$    132.05$&$   60.7737$&$    19.980$&$      34.9$&$       6.5$&$      32.4$&$       3.2$&LIN                 \\
$0.191$&$    137.83$&$   63.1578$&$    13.910$&$      68.4$&$      11.5$&$      11.1$&$       2.8$&COMP                \\
$0.139$&$    208.31$&$   35.1470$&$    42.090$&$      36.5$&$       1.0$&$      30.1$&$       3.9$&LIN                 \\
$0.115$&$    213.01$&$   29.4671$&$    36.970$&$      31.6$&$       1.4$&$      32.4$&$       5.7$&LIN                 \\
$0.126$&$    213.75$&$   26.8312$&$    22.280$&$      12.2$&$       1.3$&$      21.7$&$       4.7$&LIN                 \\
$0.165$&$    222.05$&$   28.8887$&$    42.410$&$      15.7$&$       1.1$&$      26.8$&$       6.5$&LIN                 \\
$0.069$&$    226.00$&$   24.1049$&$    17.260$&$      12.5$&$       1.2$&$      34.5$&$       5.7$&COMP                \\
$0.101$&$    237.75$&$   19.2783$&$    30.570$&$      37.8$&$       1.5$&$      36.8$&$       5.2$&LIN                 \\
$0.110$&$    240.91$&$   15.9007$&$    93.750$&$     199.9$&$       2.5$&$     206.0$&$       9.0$&COMP                \\
$0.229$&$    235.33$&$   14.3696$&$    11.180$&$      26.7$&$       1.0$&$      26.3$&$       4.6$&SF                  \\
$0.149$&$    225.46$&$   16.6183$&$    32.240$&$      26.4$&$       1.4$&$      19.5$&$       4.6$&LIN  \\
$0.139$&$    214.84$&$   21.6037$&$    20.920$&$      14.8$&$       2.4$&$      18.1$&$       3.7$&COMP                \\
$0.138$&$    214.04$&$   22.4760$&$    25.650$&$      14.0$&$       0.9$&$      22.7$&$       6.0$&LIN                 \\
$0.151$&$    225.88$&$   19.6523$&$    91.950$&$      86.7$&$       2.4$&$      76.7$&$       5.4$&LIN                 \\
$0.150$&$    230.43$&$   18.2440$&$    93.270$&$      40.2$&$       4.1$&$      13.7$&$       4.6$&LIN                 \\
$0.057$&$    245.57$&$   50.3720$&$    15.660$&$       7.8$&$       0.8$&$       7.8$&$       3.7$&COMP                \\
$0.104$&$    249.69$&$   27.9109$&$    40.920$&$      26.5$&$       1.4$&$      24.6$&$       4.2$&LIN                 \\              
\hline
  \end{tabular}
\end{table*}

\end{document}